\documentclass{gd-llncs-short}
\newif\ifAnon
\Anonfalse    
\usepackage{cite}
\usepackage{amsmath}
\usepackage{graphicx}
\usepackage{multirow}
\usepackage{pgfplots}
\pgfplotsset{width=9cm,compat=1.9}
\usepackage{subcaption}
\usepackage{microtype} 
\usepackage[linesnumbered,ruled,vlined,nosemicolon]{algorithm2e}
\usepackage{color}
\usepackage[pdfencoding=auto]{hyperref}

\renewcommand{\emph}[1]{\textit{\textbf{#1}}}
\let\epsilon\varepsilon

\begin{document}
\title{Manipulating Weights
to Improve Stress-Graph Drawings of 3-Connected Planar Graphs}
\ifAnon
\author{Anonymous Author(s)}
\institute{Short Paper --- Experimental Track}
\else
\author{
Alvin Chiu 
\and
David Eppstein
\and
Michael T. Goodrich
}
\institute{Dept. of Computer Science, Univ.~of California, Irvine, USA}
\fi

\maketitle
\begin{abstract}
We study methods to manipulate weights in stress-graph
embeddings to improve convex straight-line planar drawings
of 3-connected planar graphs.  
Stress-graph embeddings are weighted
versions of Tutte embeddings, where solving a linear system places
vertices at a minimum-energy configuration for a system of springs.
A major drawback of the unweighted Tutte embedding
is that it often results in drawings with exponential area.
We present
a number of approaches for choosing better weights.
One approach constructs weights (in linear time) that
uniformly spread all vertices in a chosen direction, such as parallel
to the $x$- or $y$-axis.  
A second approach morphs $x$- and $y$-spread
drawings to produce a more aesthetically pleasing and uncluttered drawing.
We further explore a ``kaleidoscope'' paradigm for this $xy$-morph
approach, where we rotate the coordinate axes so as to find the best spreads
and morphs.
A third approach chooses the weight of each edge according
to its depth in a spanning tree rooted at the outer vertices,
such as a Schnyder wood or BFS tree,
in order to pull vertices closer to the boundary.

\renewcommand\and{$\cdot$ }
\keywords{Tutte embedding \and convex drawing \and vertex spreading.}
\end{abstract}

\pagestyle{plain}

\section{Introduction}

Sixty years ago, Tutte provided what is arguably one of
the first graph drawing algorithms~\cite{tutte1963draw}\ifAnon\else\footnote{
   Proofs of F{\'a}ry's Theorem, that any simple, planar graph
   can be embedded in the plane without crossings so each edge
   is drawn as a straight line segment, came 
   earlier~\cite{wagner1936bemerkungen,istvan1948straight,stein1951convex},
   but these proofs do not give specific coordinates for the vertices; hence,
   it is not clear they can be called ``graph drawing algorithms.''}\fi.
Given a simple, undirected 3-connected planar graph,
$G$, Tutte's algorithm produces a 
straight-line, planar drawing of $G$ such that each face is convex.
Tutte's algorithm produces such a drawing of $G$ by solving a set
of linear equations that determine the $x$- and $y$-coordinates of
points to which the vertices of $G$ are assigned.
Intuitively, the equations are based on ``pinning'' the vertices of
the outer face of $G$ to the vertices of a convex polygon, and then
considering all the edges of $G$ to be springs with an idealized length
of $0$. Solving the set of equations amounts to finding a minimum-energy
configuration for the springs given the pinned vertices of the outer 
face~\cite{battista1999graph,kobourov2012spring}.

\begin{figure}[t]

\centering
\begin{subfigure}{.3\textwidth}
  \includegraphics[width=\linewidth]{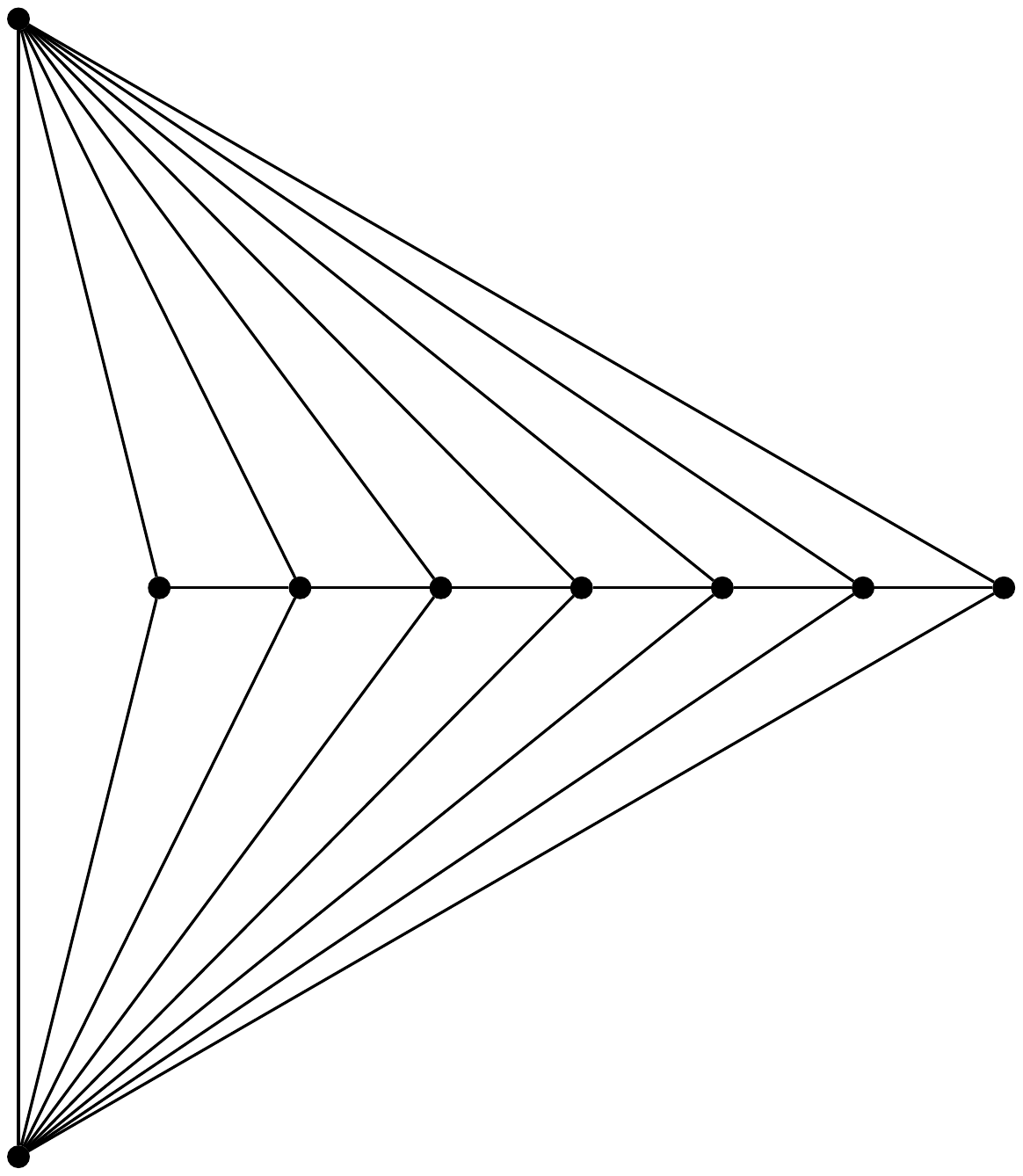}

  \caption{Input graph}
  \label{fig:worst}
\end{subfigure}%
\begin{subfigure}{.3\textwidth}
  \centering
  \includegraphics[width=\linewidth]{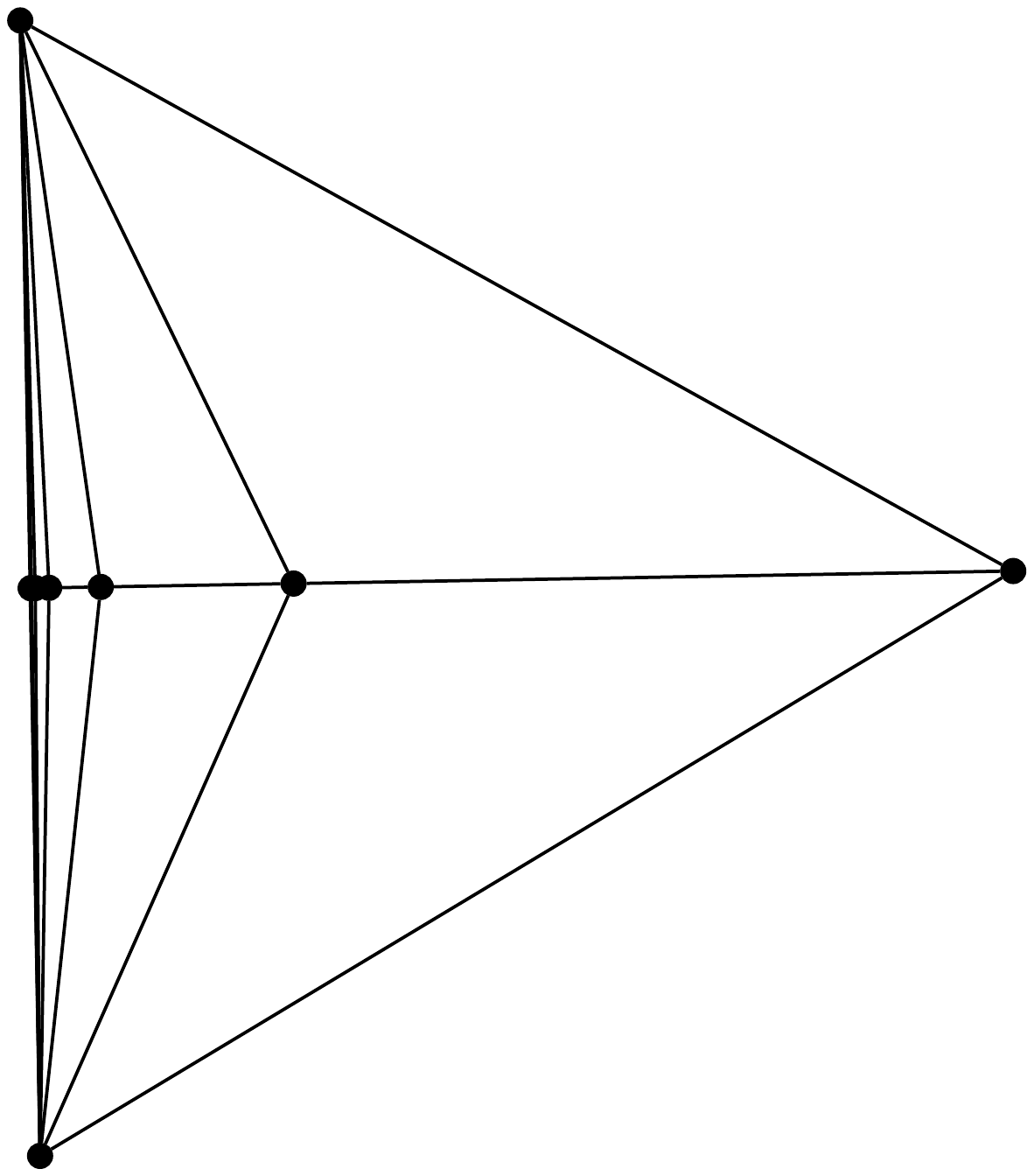}

  \caption{Tutte}
  \label{fig:worst-tutte}
\end{subfigure}
\begin{subfigure}{.3\textwidth}
  \centering
  \includegraphics[width=\linewidth]{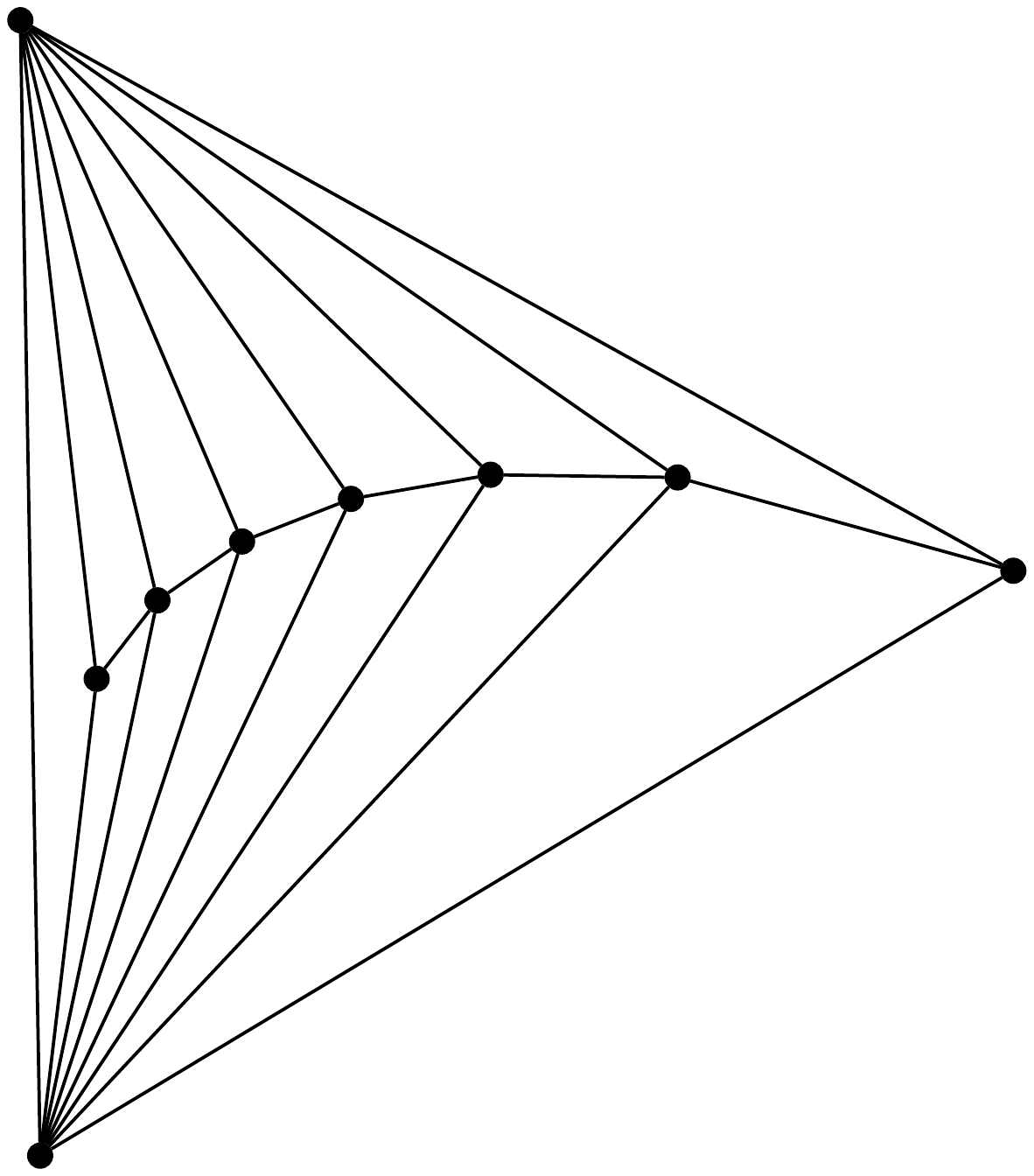}

  \caption{$xy$-morph}
  \label{fig:worst-xy}
\end{subfigure}

\caption{Tutte drawings can have exponential area.}
\label{fig:worst-case}
\end{figure}

One unfortunate drawback of Tutte's algorithm is that
it can produce drawings 
with exponential area or exponentially small edge lengths,
depending on the normalization of coordinates.
Indeed, Eades and Garvan~\cite{eades96} show that this undesirable 
result occurs even for the planar graphs formed by connecting two outer vertices to each vertex of a simple path and to each other, as shown in 
Figs.~\ref{fig:worst} and~\ref{fig:worst-tutte}.
Intuitively, the 
idealized springs representing graph edges have
equal stress, which, in turn, ``pull''
groups of springs into unsightly vertex clusters.

Hopcroft and Kahn~\cite{hopcroft1992paradigm} generalize Tutte's algorithm to spring systems with different stress weights. 
In this framework, which we explore in this paper,
we assign a stress weight, $w_{u,v}$, to each edge, $(u,v)$, of 
$G$.\footnote{
   Tutte's approach can be viewed as being for the case when 
   $w_{u,v}=1$ for each edge.}
We begin as in the Tutte framework
by ``pinning'' the vertices of an outer face, $f$, to
be the vertices of a convex polygon, and 
we then formulate two linear equations for each internal vertex, $u$,
of $G$, as follows:
\begin{align}
    \sum_{(u,v)\in E} w_{u,v}(x_u - x_v) = 0,
    \mbox{~~~and~~~}
    \sum_{(u,v)\in E} w_{u,v}(y_u - y_v) = 0,
\end{align}

where $p_v=(x_v,y_v)$ is the point to which vertex $v$ is assigned.
Note that for a vertex, $v$, on the outer face, we pin
$p_v=(x_v^*,y_v^*)$; hence, 
$x_v=x_v^*$ and $x_v=y_v^*$ are constants in our linear system.
As Hopcroft and Kahn~\cite{hopcroft1992paradigm}, as well
as Floater~\cite{Floater}, show,
if the stresses, $w_{u,v}$, are all positive,
except possibly for the edges of the outer face,
then the resulting drawing is a planar straight-line drawing with 
each face being convex.
In this paper, we experimentally explore the aesthetic improvements to a Tutte
embedding that can be achieved by manipulating the stresses
in such stress-graph drawings of 3-connected planar graphs.

\paragraph{\textbf{Related Prior Results.}}
We are not familiar with any prior work on the manipulation of the weights
in stress-graph drawings strictly for the purpose of improving 
the aesthetic qualities.
Nevertheless,
the general technique of manipulating stresses in stress-graph drawings
is not without precedent.
For example, 
Hopcroft and Kahn~\cite{hopcroft1992paradigm} 
and
Eades and Garvan~\cite{eades96} 
give conditions for stresses
so that the resulting drawing is the projection of a 3-dimensional convex
polyhedron onto the plane.
%
Chrobak, Goodrich, and Tamassia~\cite{chrobak} further explore this approach,
claiming to produce a 3-dimensional realization of a 3-connected planar graph as
the 1-skeleton of a 3-dimensional convex polyhedron with vertex resolution $\Omega(1)$ and with linear volume.\footnote{However, their proof is only valid for polyhedra that have a triangle face.}
Indeed, their approach comes close to ours, in that they first compute weights for a weighted Tutte drawing with good vertex resolution (using a flow-based approach) and then
apply the Maxwell--Cremona correspondence to lift this drawing to a convex polyhedron.
Their method does not necessarily result in aesthetically
pleasing drawings or convex polyhedra, despite the good spacing for the $x$-coordinates.
Researchers have also explored interpolating between stress-graph 
drawings to morph from one layout to another.
For example, Floater and Gotsman~\cite{FLOATER1999117} use 
interpolation of the weights for two convex embeddings 
to morph between them, albeit with vertex movements that are represented implicitly. They also devise a method to obtain weights that will produce a given drawing.
Erickson and Lin~\cite{erickson} morph between two convex
via unidirectional morphs, where vertices move parallel to the direction of an edge. 
Kleist et al.~\cite{kleist2019convexity} turn drawings of planar 3-connected graphs into strictly convex planar drawings with similar morphs. 

\begin{figure}

\begin{subfigure}{.24\textwidth}
  \centering
  \includegraphics[width=\linewidth]{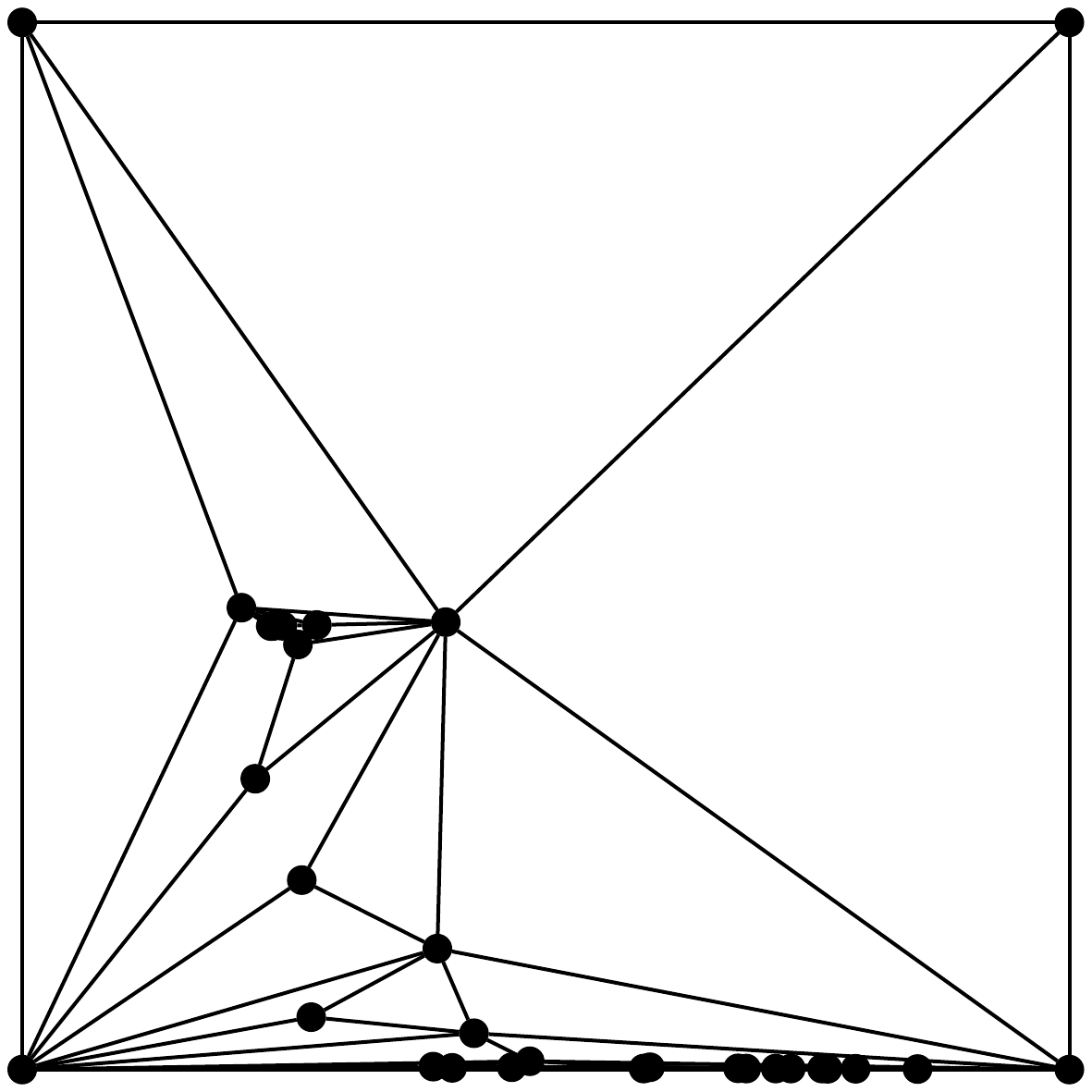}

  \caption{Tutte}
  \label{fig:30-90-90tutte}
\end{subfigure}%
\begin{subfigure}{.24\textwidth}
  \centering
  \includegraphics[width=\linewidth]{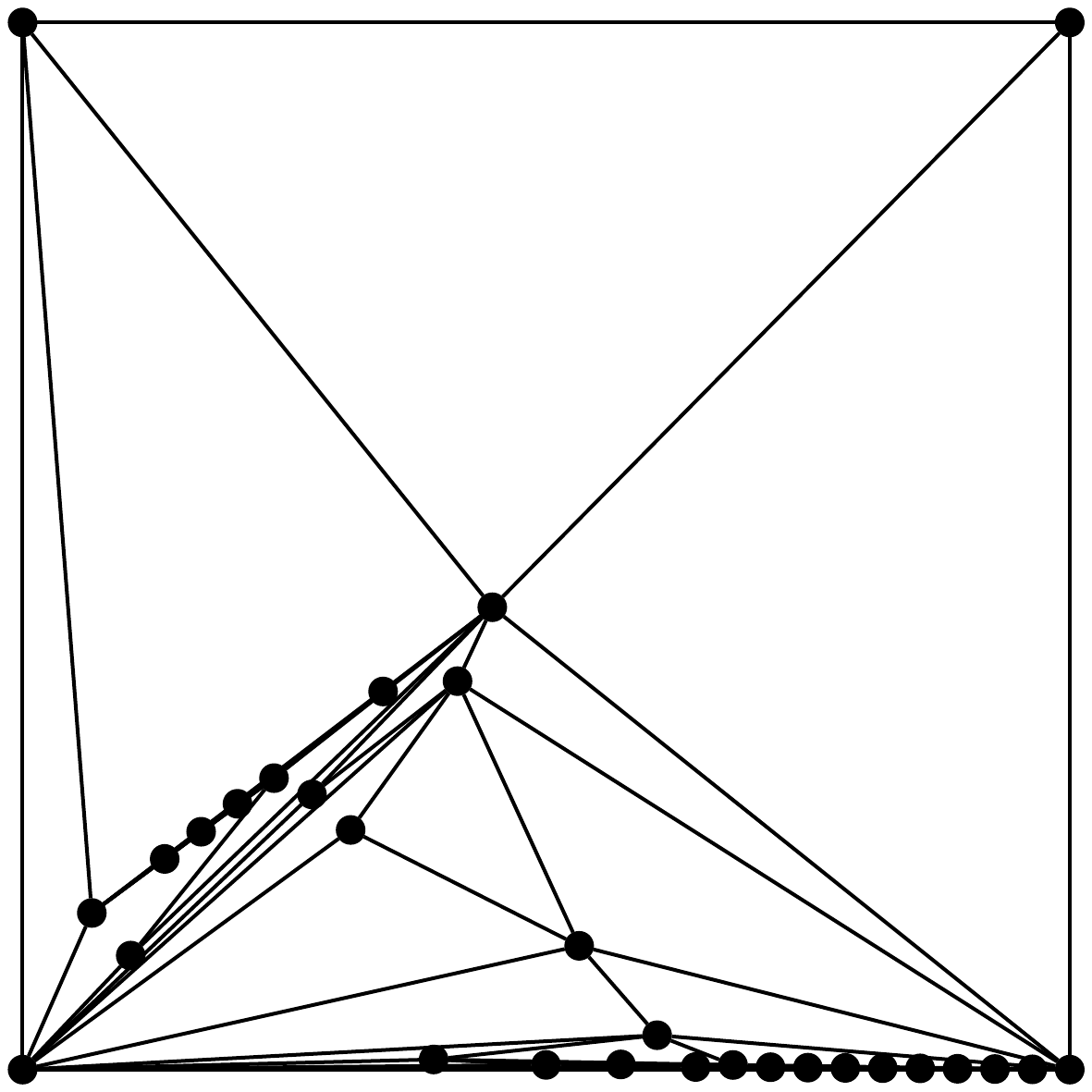}

  \caption{$x$-spread}
  \label{fig:30-90-90x}
\end{subfigure}
\begin{subfigure}{.24\textwidth}
  \centering
  \includegraphics[width=\linewidth]{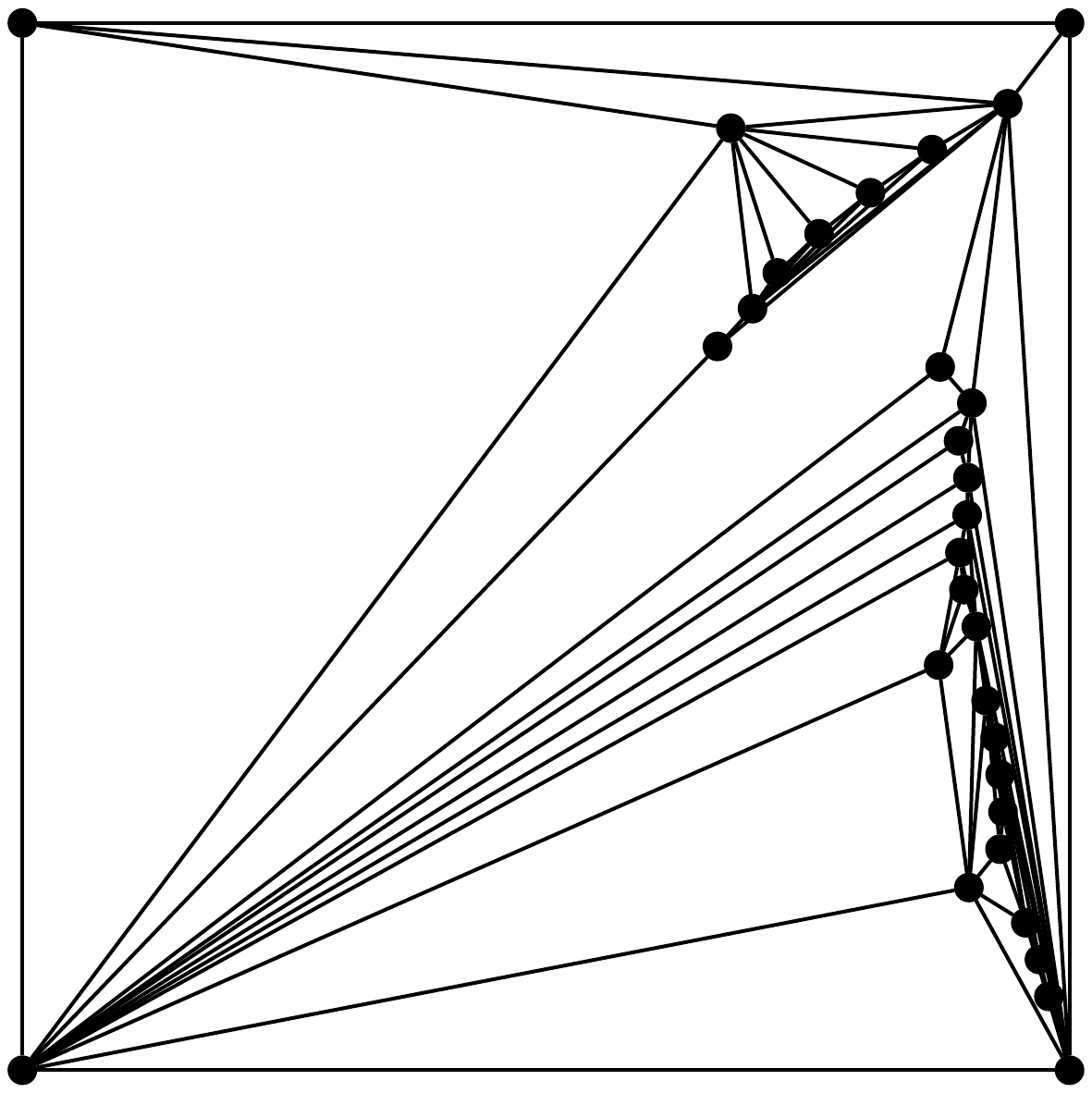}

  \caption{$y$-spread}
  \label{fig:30-90-90y}
\end{subfigure}
\begin{subfigure}{.24\textwidth}
  \centering
  \includegraphics[width=\linewidth]{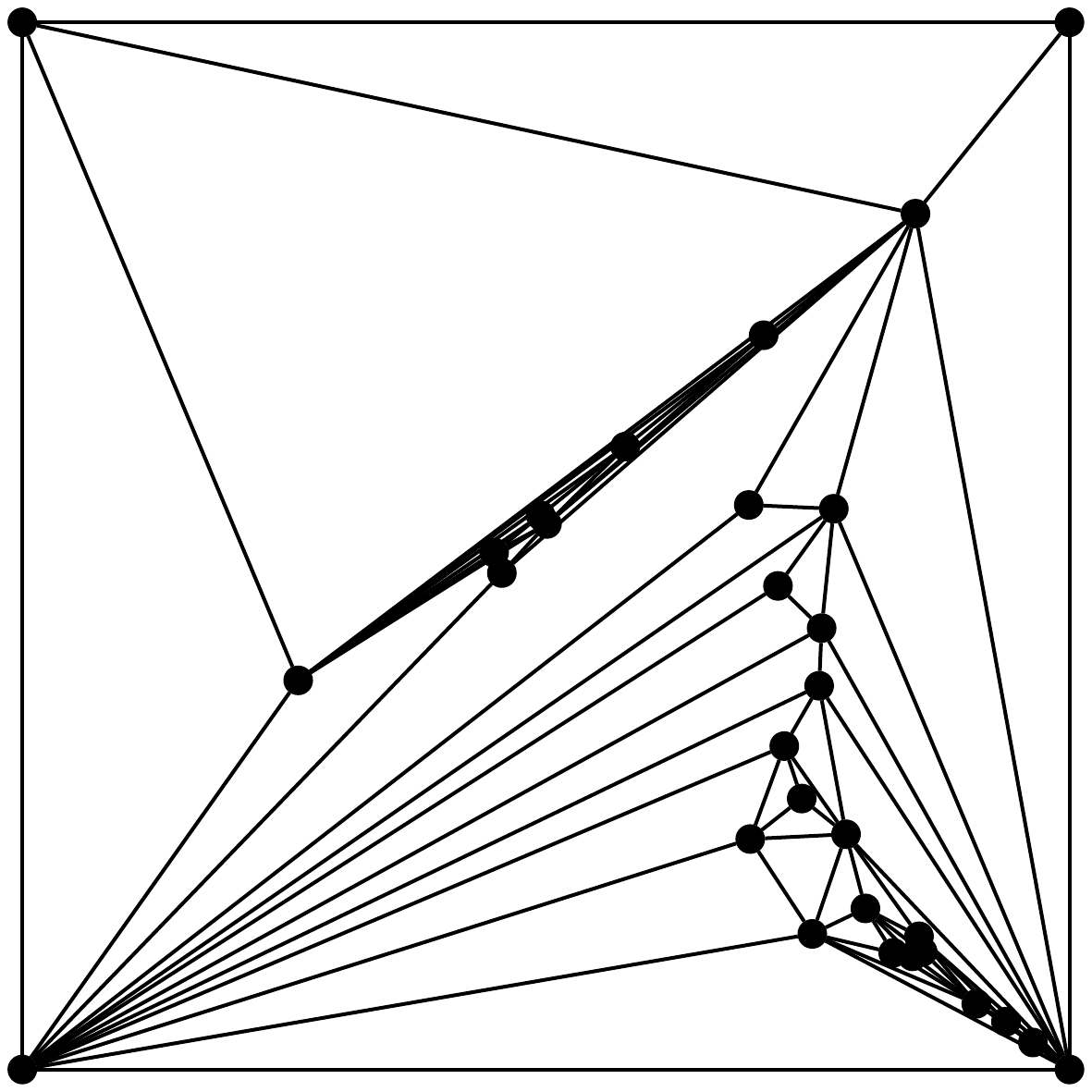}

  \caption{$xy$-morph}
  \label{fig:30-90-90xy}
\end{subfigure}

\caption{Drawings of a planar graph with 30 vertices and 80 edges, $G(30,80)$.}
\label{fig:span-tree}
\end{figure}

\paragraph{\textbf{Our Results.}}
We propose several methods of weight manipulation. In the first, we
simplify (and correct) the approach of Chrobak, Goodrich, and
Tamassia ~\cite{chrobak} for finding drawings in which vertices have
uniformly distributed coordinates.
Instead of using iterated flows,
we find suitable weights in linear time by counting
certain paths in an $st$-orientation of the graph.
Our implementation fixes the outer face as a regular polygon;
in an appendix we show that an alternative choice allows
all vertices, including outer face vertices, to have uniform $x$-coordinates.  We
experiment with a modified version of this method that produces two
planar straight-line drawings that evenly
spread the $x$-coordinates and the $y$-coordinates, respectively. 
We then construct a morph that averages the weights of the $x$- and $y$-spread drawings.  
The idea is that this morph will have fairly
even spacing on both directions, e.g., as shown in Fig.~\ref{fig:worst-xy}
and Figs.~\ref{fig:30-90-90tutte} to~\ref{fig:30-90-90xy}.
We also explore a ``kaleidoscope'' version of this approach,
where we rotate the coordinate axes to find the best spreads.
%
In another approach, we weight edges based on depth in spanning trees rooted at the outer vertices. Edges closer to the outer vertices will have higher weight and thus more ``pull", spreading the internal vertices away from the center of the outer face in a manner that preserves the general structure.
We explore two types of spanning trees: BFS and 
(for fully triangulated graphs) Schnyder woods~\cite{bonichon2007convex,felsner2004lattice,schnyder1990embedding}.


\section{Algorithms}

\paragraph{\textbf{Weight Manipulation to Spread Vertices Uniformly.}}
To find weights whose stress-graph embedding spreads vertices evenly,
we first begin with an unweighted Tutte drawing, rotating it if necessary so no edge is vertical. We sort the vertices by $x$-coordinates in this drawing, and orient edges from left to right, producing an $st$-orientation: an acyclic orientation in which each vertex $v_i$ with $1<i<n$ has both incoming and outgoing edges. Next, we choose new $x$-coordinates $x_i$ for the interior vertices that are as evenly spaced as possible under the constraint that they respect the sorted $x$-ordering of all the vertices. (The same constraint is also present in the flow-based method of Chrobak et al.~\cite{chrobak})
We can choose new positive edge weights for the Tutte drawing to produce the chosen $x$-coordinates in linear time. 
Conceptually, we gradually increase weights along a sequence of paths in the graph, starting with all weights zero. For each edge $e$, we find a directed path from $v_1$ to $v_n$ through $e$, and increase weights on the edges of this path.

Along a single path through consecutive vertices $v_i, v_j, v_k$, the spacing between the vertex placements should be in the proportion $x_j-x_i:x_k-x_j$, which can be achieved by giving edges $v_iv_j$ and $v_jv_k$ the weights $1/(x_j-x_i)$ and $1/(x_k-x_j)$ respectively. Because these weights do not depend on the other edges of the path, we can use this weight for each edge in all of the paths that it belongs to and preserve the $x$-equilibrium. In total, the weight of any edge $v_iv_j$ in the whole graph (summing its weights for each path it appears in) will be $n_{ij}/(x_j-x_i)$, where $n_{ij}$ is the number of paths containing edge $v_iv_j$. 

To calculate these numbers efficiently, we compute two spanning trees in the oriented graph: tree $T_1$ directed out of $v_1$, and tree $T_n$ directed into $v_n$ (shortest-path trees via BFS were used for the implementation). For each edge $v_iv_j$, include a path that follows $T_1$ from $v_1$ to $v_i$, then edge $v_iv_j$, then follow $T_n$ from $v_j$ to $v_n$. We can count the number of these paths that use $v_iv_j$ as follows:

\begin{itemize}
\item There is one path defined in this way from $v_iv_j$.
\item Let $D_j$ be the set of descendants of $v_j$ in $T_1$ (including $v_j$ itself) and $d^+(v_k)$ be the number of outgoing edges from $v_k$. If $v_iv_j$ belongs to $T_1$, then $\sum_{v_k\in D_j} d^+(v_k)$ paths pass through $v_iv_j$ in $T_1$ before crossing to $T_n$. 
\item Let $A_i$ be the set of descendants of $v_i$ in $T_n$ and $d^-(v_k)$ be the number of incoming edges at $v_k$. If $v_iv_j$ belongs to $T_n$, then symmetrically $\sum_{v_k\in A_i} d^-(v_k)$ paths pass through $v_iv_j$ in $T_n$ after crossing to $T_n$.
\end{itemize}

The sums of descendant out-degrees in $T_1$, and of descendant in-degrees in $T_n$, can be computed in linear time by a simple bottom-up tree traversal, after which we can calculate the weight $n_{ij}/(x_j-x_i)$ of all edges in linear time.
A weighted Tutte drawing with positive weights and convex outer face cannot introduce crossings, so we get a convex drawing with spread out $x$-coordinates using these new weights.
To spread by a different direction, we can 
rotate the initial unweighted Tutte drawing
before doing the spread.
Indeed, as we explore experimentally, we consider a number of distinct
rotation angles, producing drawings similar to the way a kaleidoscope
produces patterns as it is turned.

Moreover, we can produce
an ``$xy$-morph'' drawing of the input graph.
Let a weighted Tutte drawing be represented by $\Gamma = (\Lambda, \mathcal{P})$, where $\Lambda$ is the coefficient matrix containing the edge weights and $\mathcal{P}$ is the convex polygon chosen to be the outer face. One can morph between the $x$-coordinate spread drawing $\Gamma_0 = (\Lambda_0, \mathcal{P})$ and $y$-coordinate spread drawing $\Gamma_1 = (\Lambda_1, \mathcal{P})$ to obtain a more balanced graph drawing $\Gamma_{1/2}$. Intuitively,
this is like stopping halfway in
Floater and Gotsman's morphing algorithm~\cite{FLOATER1999117}, where we construct $\Gamma_{1/2} = (\Lambda_{1/2}, \mathcal{P})$ where $\Lambda_{1/2} = \frac{1}{2} \cdot \Lambda_0 + \frac{1}{2}\cdot \Lambda_1$. 
(See Fig.~\ref{fig:span-tree}.)

\paragraph{\textbf{Weight Manipulation via Spanning Tree Depth.}}
In our spanning-tree approach,
we first do a Tutte drawing, then we find a set of edge-covering spanning trees, $T$, for the graph rooted at the outer vertices, such as BFS trees or 
Schnyder woods~\cite{bonichon2007convex,felsner2004lattice,schnyder1990embedding}.
Next, we assign weights to the edges of each tree, $T$, in a top-down manner according to its depth in the spanning tree. With these new weights, we do another stress-graph drawing.

\begin{figure}

\centering
\begin{subfigure}{.3\textwidth}
  \centering
  \includegraphics[width=\linewidth]{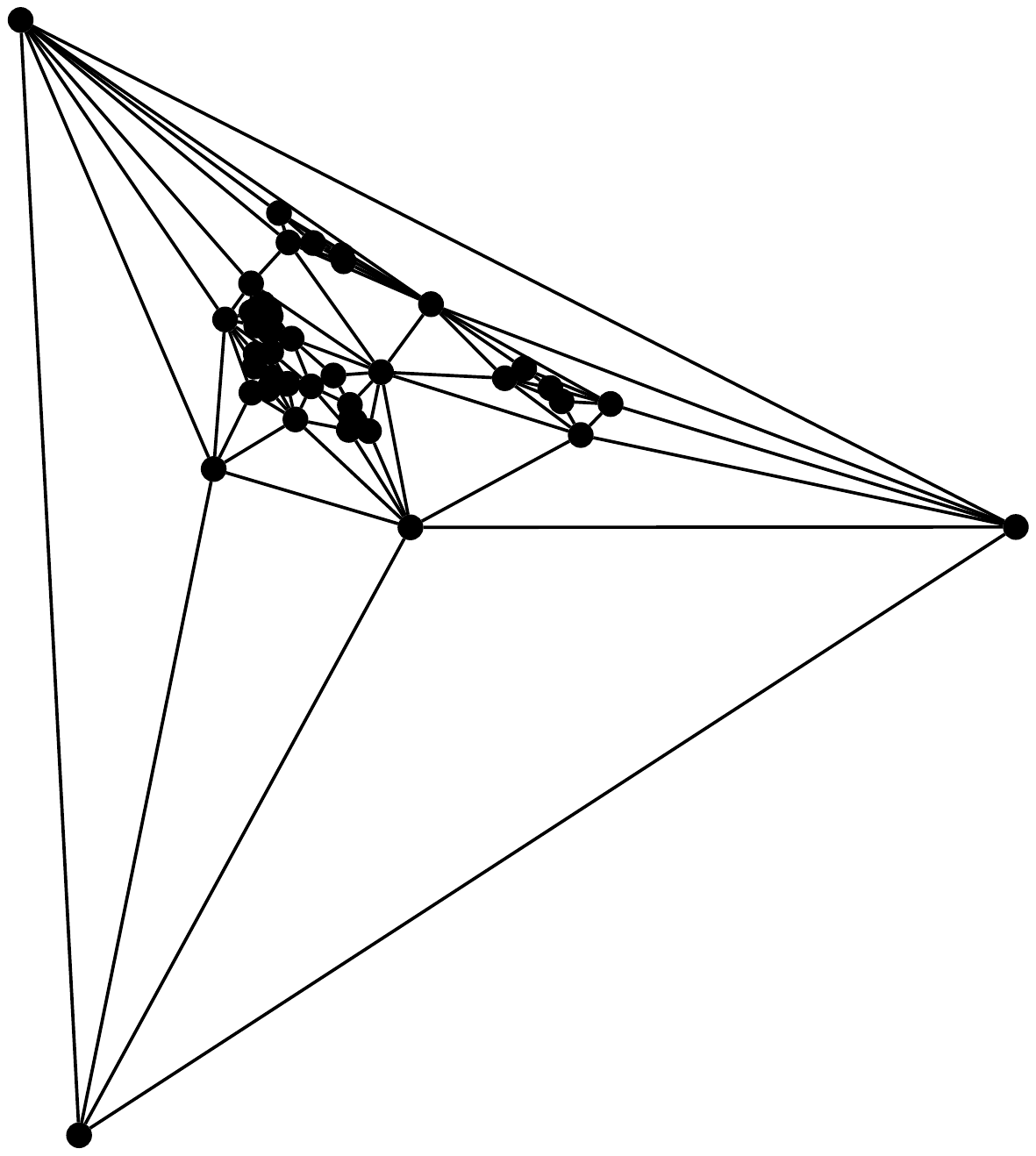}

  \caption{Tutte}
  \label{fig:50-144tutte}
\end{subfigure}%
\begin{subfigure}{.31\textwidth}
  \centering
  \includegraphics[width=\linewidth]{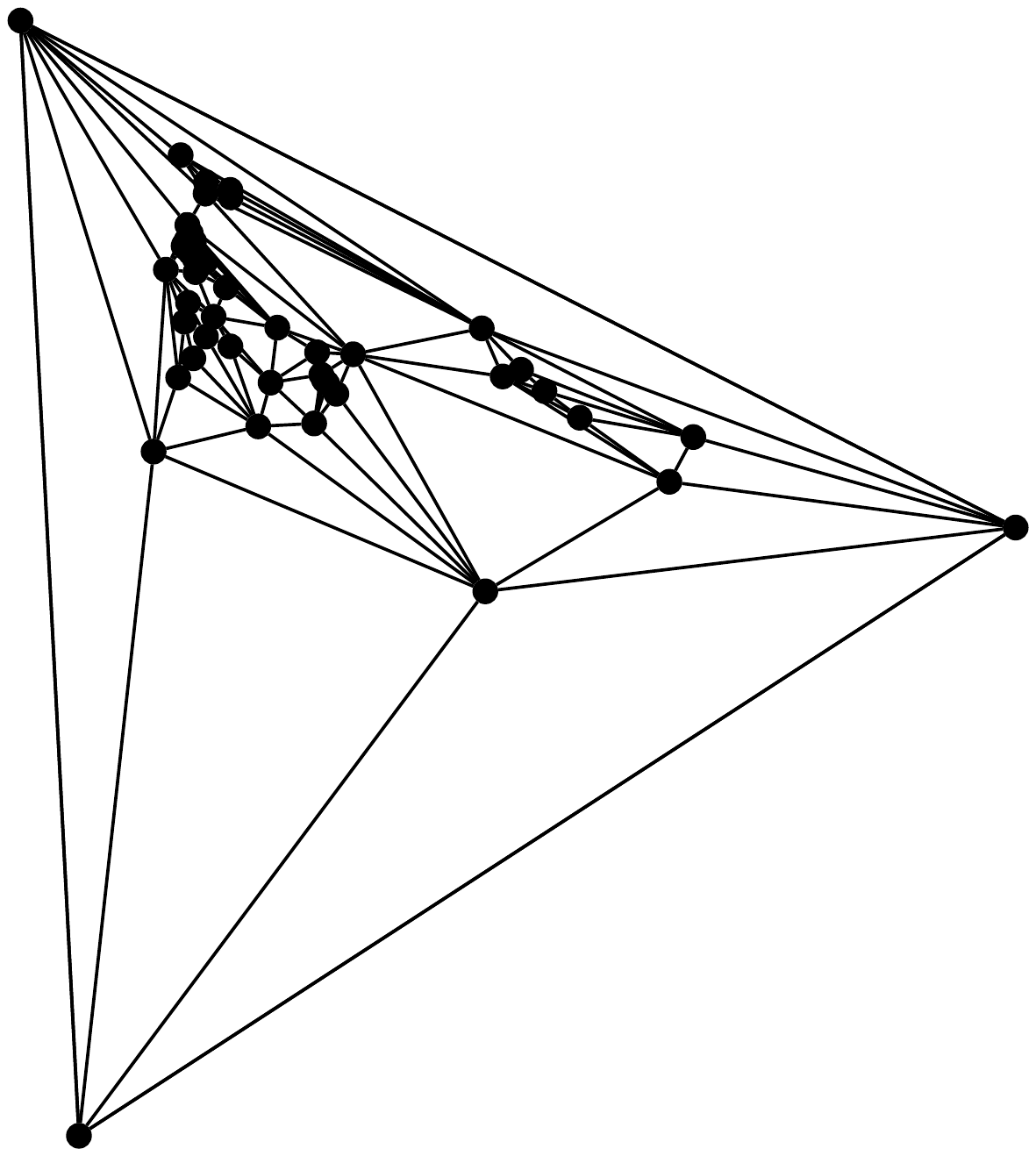}

  \caption{Schnyder-spread, $r=5$}
  \label{fig:50-144sch}
\end{subfigure}
\begin{subfigure}{.3\textwidth}
  \centering
  \includegraphics[width=\linewidth]{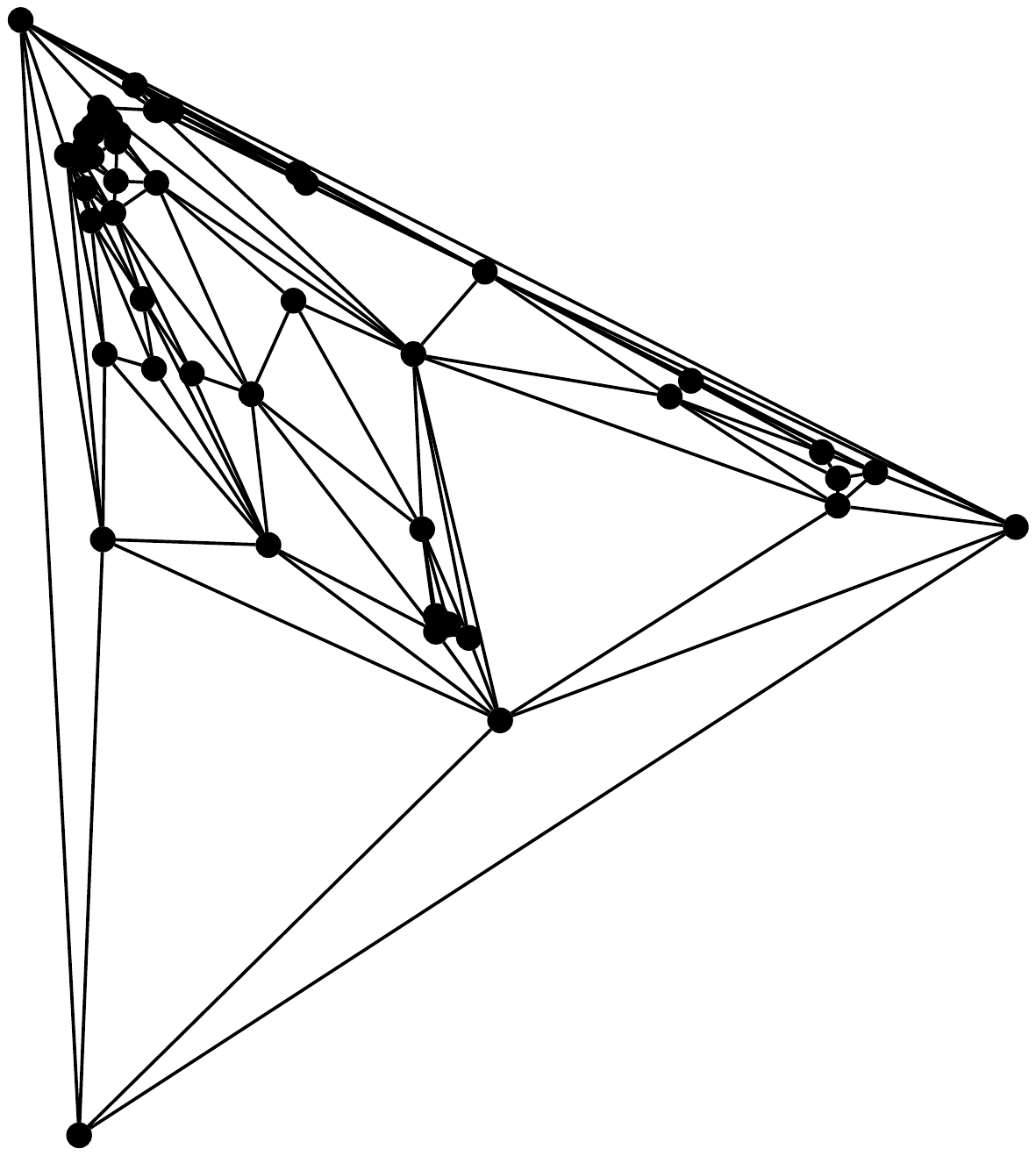}

  \caption{BFS-spread, $r=5$}
  \label{fig:50-144bfs}
\end{subfigure}

\caption{Drawings of a pseudorandom graph, $G(50,144)$.}
\label{fig:flow-diagrams}
\end{figure}

Let the \textit{depth} of an edge in a tree be the number of edges from the root to the edge plus one (to include the edge itself). Then we assign an edge at depth $i$ with weight $a/r^i$, where $a$ is some initial constant and $r$ is a scaling parameter.
When using BFS to find the shortest-path tree $T_v$ rooted at an outer vertex $v$, we assign weights to an edge according to its lowest depth from any of the outer vertices. To do this, we create a dummy ``super"-vertex, $v_s$, connected to all the outer vertices and run BFS from $v_s$, which is akin to running BFS on all the outer vertices simultaneously.
For the case when the outer face is a triangle, we also consider 
Schnyder woods,
which form an edge-covering set of three spanning trees that have nice
``flow'' 
properties~\cite{bonichon2007convex,felsner2004lattice,schnyder1990embedding}.
(See Fig.~\ref{fig:flow-diagrams}.)

\section{Experiments}
Our experimental setup modifies the Open Graph Drawing 
Framework (OGDF) C++ library~\cite{ogdf}. 
%
One of our goals is to compare our weight manipulation methods
against Tutte's algorithm, which often produces exponentially
small edge lengths. Thus, the main metric we use is the \textit{edge-length ratio}
$\rho(\Gamma)$ of drawing $\Gamma$, which is the longest edge length divided by the smallest edge length
in the drawing.
In Table~\ref{tab:gallery}, we compare the edge-length ratios of the Tutte embeddings of several pseudorandom
planar graphs against the $x$-spread, the $y$-spread, the
$xy$-morph between the previous two, and the BFS-spread. For the BFS-spread, we choose
the parameter~$r$ to be the integer that minimizes
the edge-length ratio $p(\Gamma)$.
We do not show the results for the Schnyder-spread, as they were almost
always worse than the BFS-spread.

Not surprisingly,
our testing demonstrates that the $x$-spread and $y$-spread drawings
achieve edge-length ratio close to the number of vertices, $n$, because of the
uniform vertex spacing that they produce.
Nevertheless, optimizing exclusively for edge-length ratio can
result in vertices that cluster close to a straight line as can be seen in
Table~\ref{tab:gallery}. In constrast, the $xy$-spread drawing often is more aesthetically pleasing, as it tends
to have better symmetry visualization than either the $x$- or $y$-spread drawings without clustering.
However, it usually results in higher edge-length ratio than either of the two drawings it morphs. It may even have a higher edge-length ratio than its corresponding Tutte drawing, as seen by the
$xy$-morph for $G(50,130)$ in Table~\ref{tab:gallery}.

The edge-length ratio of BFS-spread
drawings tends to be smaller than Tutte embeddings, while still
preserving those drawings' general structure and symmetry visualization.

\begin{table}[pht]
  \caption{Drawing Gallery. $\rho(\Gamma)$ is the edge-length ratio, $r$ is the scaling parameter.} \label{tab:gallery}
  \begin{tabular}
        {cccccc} \hline & Tutte & $x$-spread & $y$-spread & $xy$-morph & BFS-spread \\
        \hline 
        \multirow{-10}{*}{\rotatebox[origin=c]{90}{$G(60,150)$}} &
        \includegraphics[width=.175\linewidth]{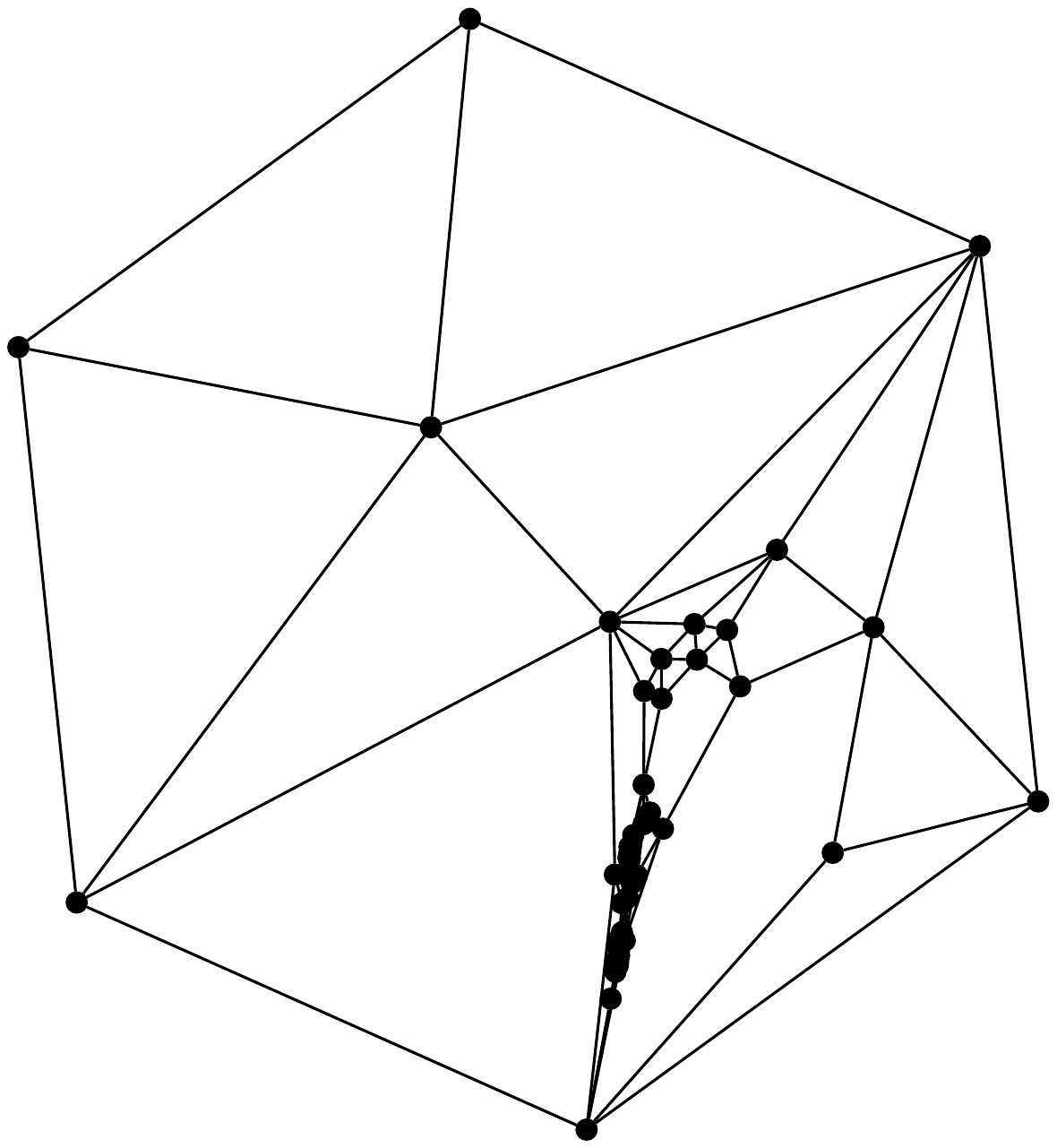} &
        \includegraphics[width=.175\linewidth]{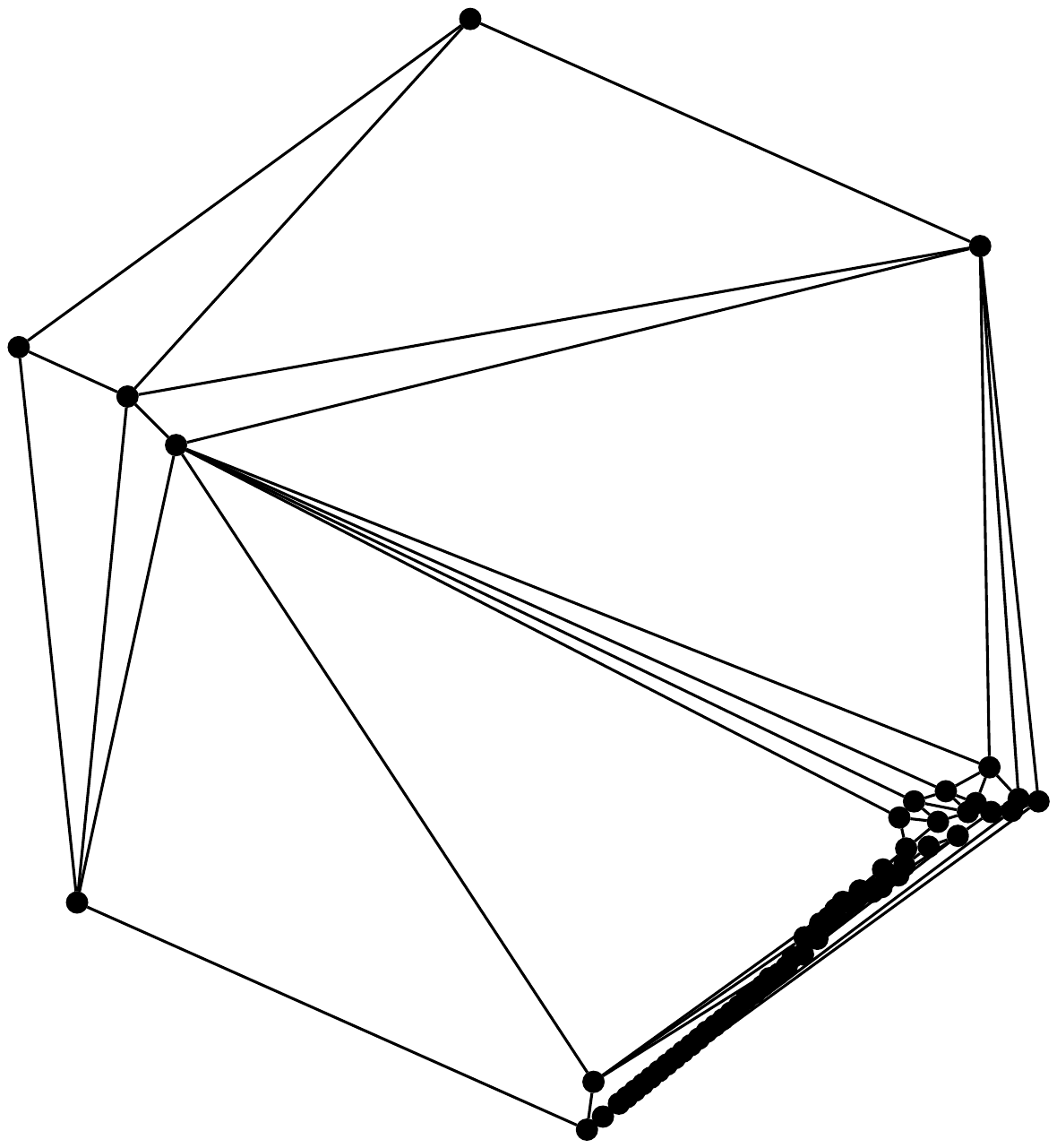} &  
        \includegraphics[width=.175\linewidth]{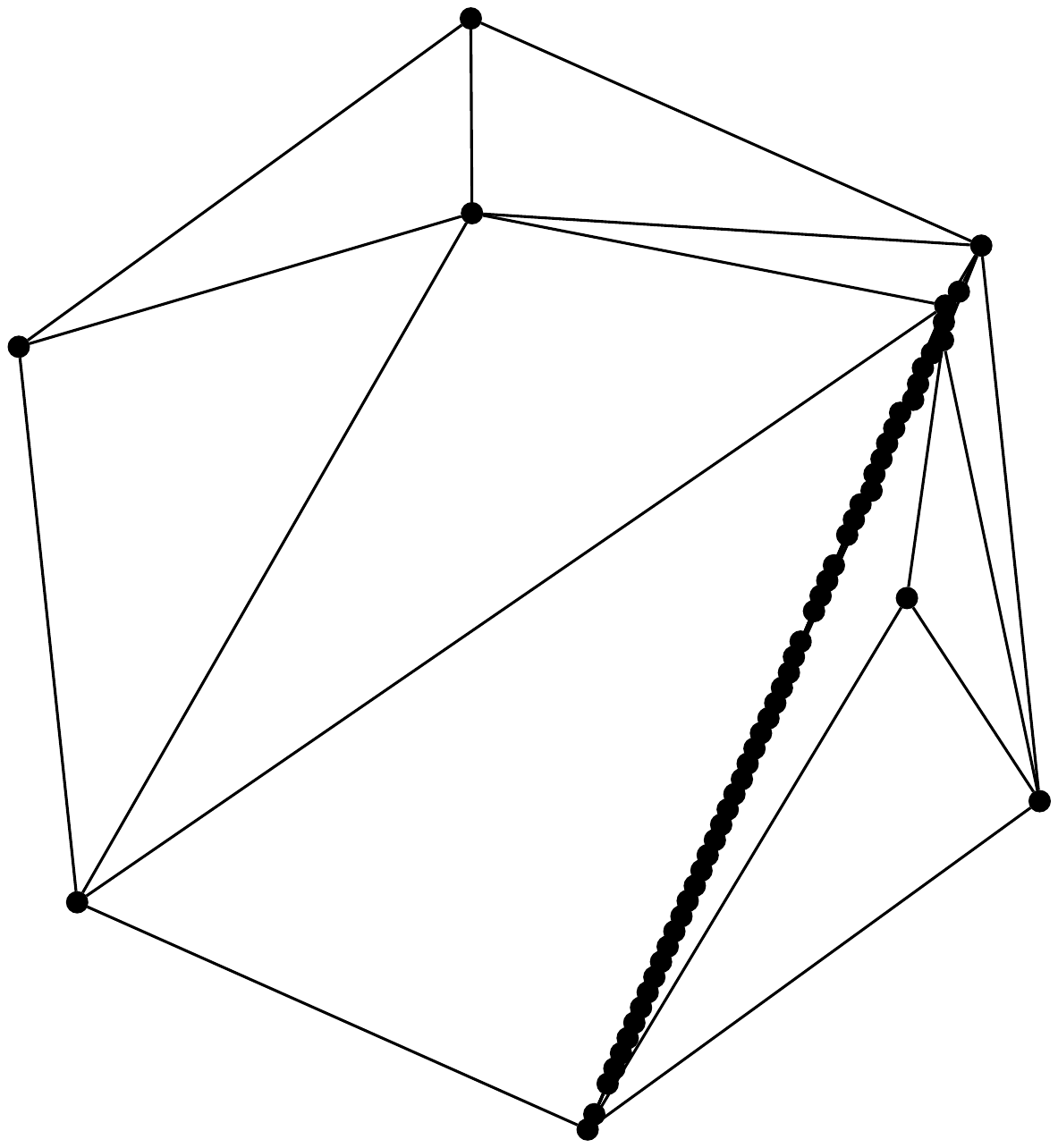} & 
        \includegraphics[width=.175\linewidth]{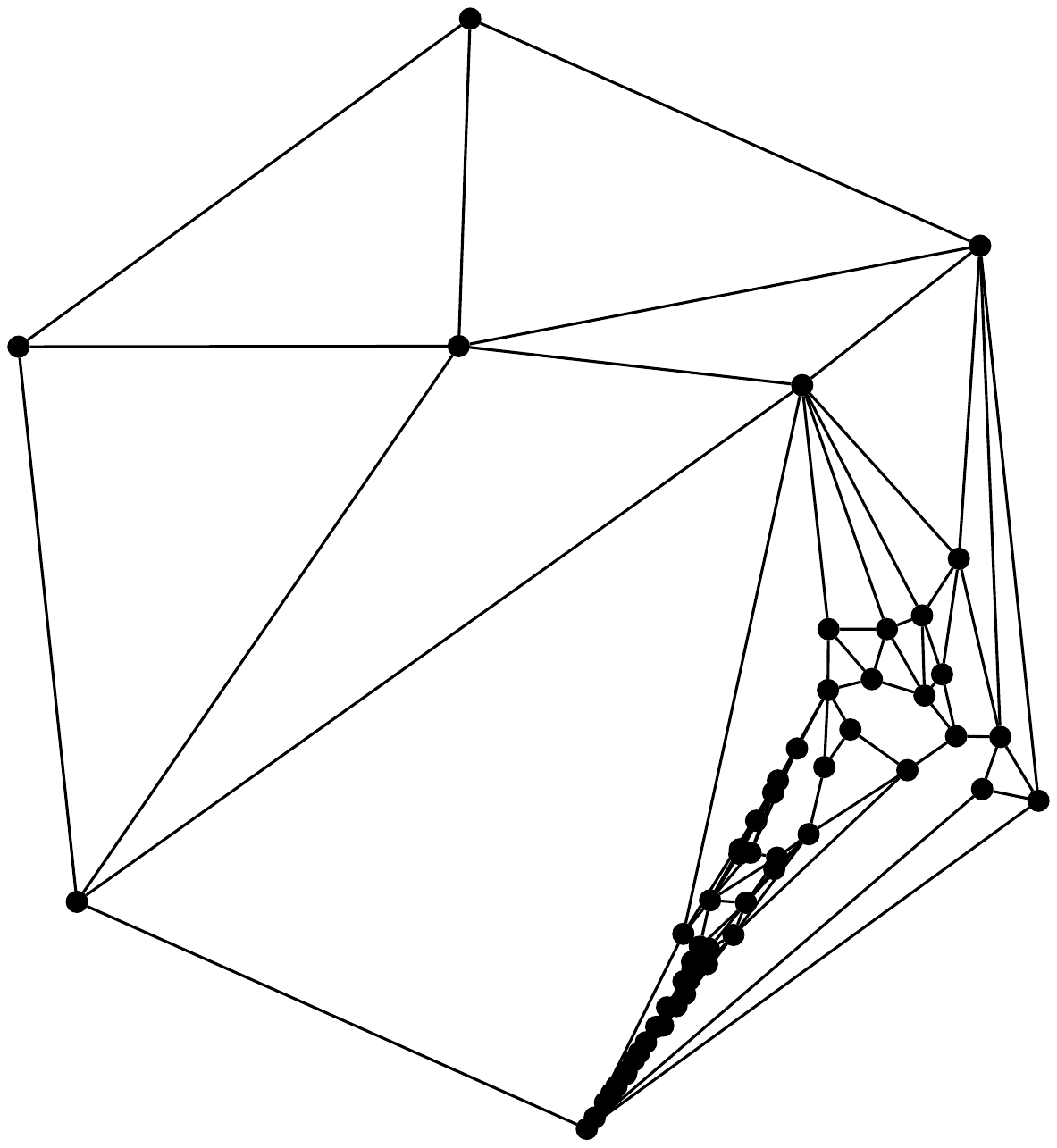} & 
        \includegraphics[width=.175\linewidth]{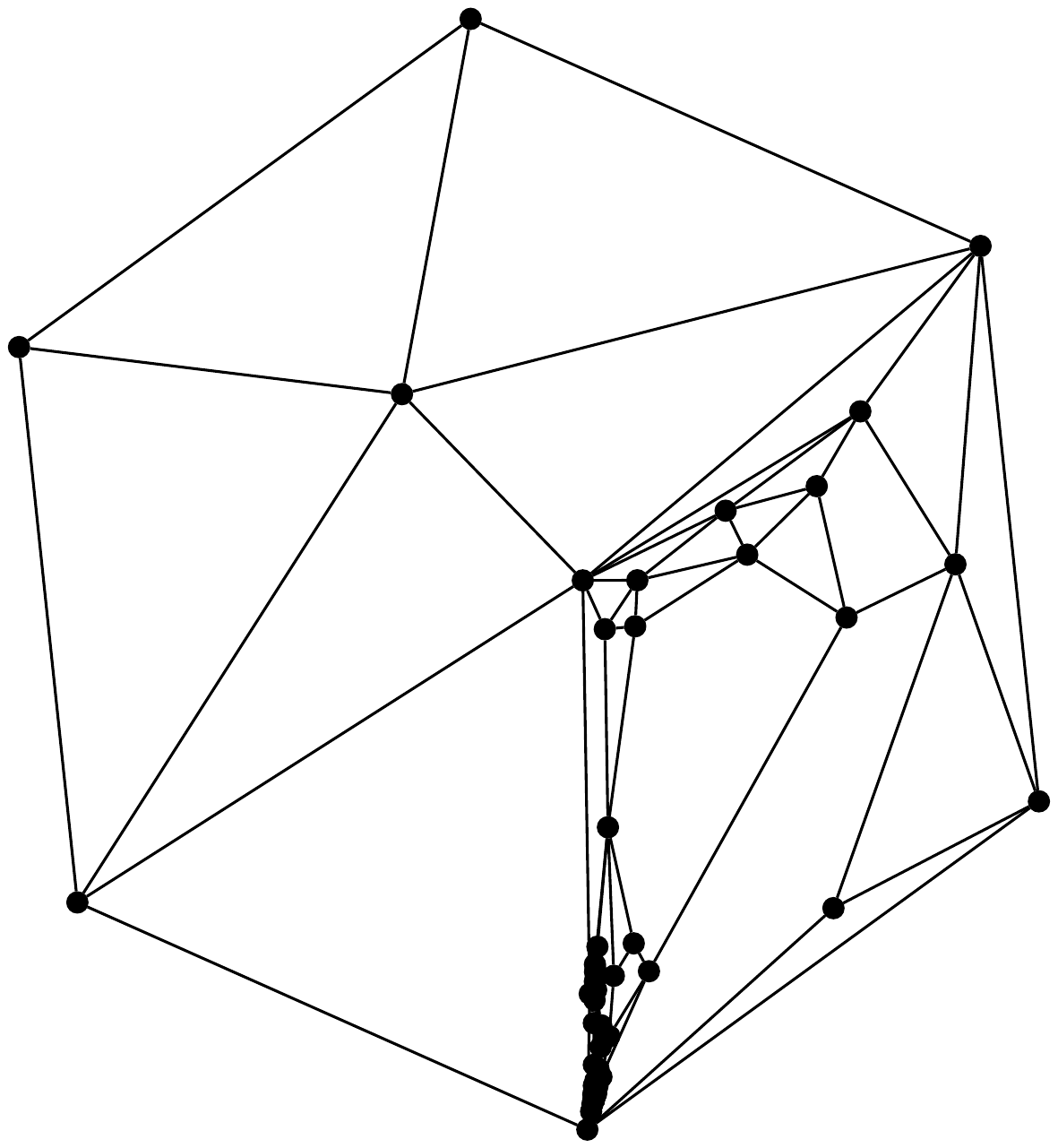} \\
        \hline
        & $\rho(\Gamma) =52717$ & $\rho(\Gamma) =86$ & $\rho(\Gamma) =63$ & $\rho(\Gamma) =2074$ & $\rho(\Gamma) = 16156, r=3$ \\
        \hline
        \multirow{-10}{*}{\rotatebox[origin=c]{90}{$G(100,200)$}} &
        \includegraphics[width=.175\linewidth]{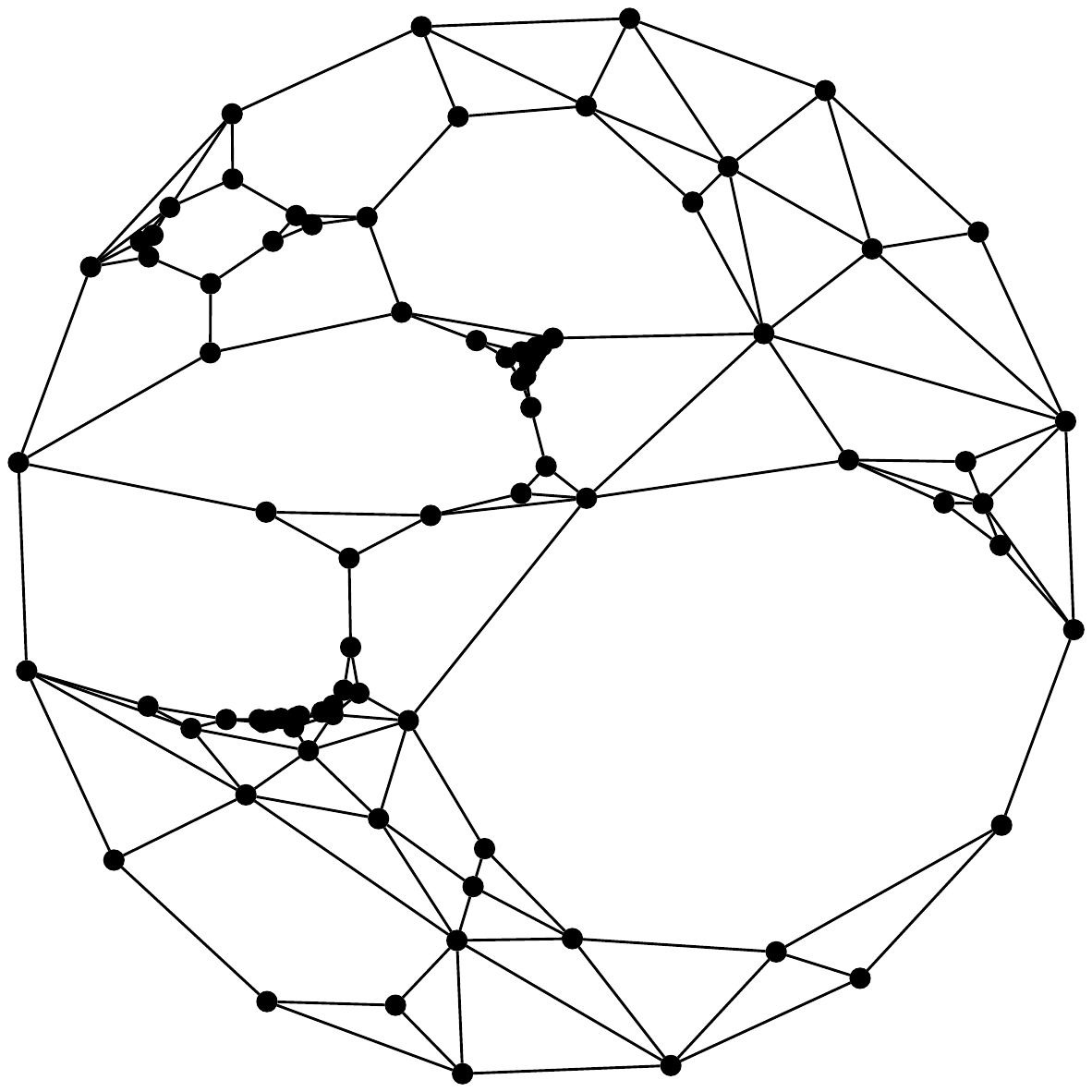} & 
        \includegraphics[width=.175\linewidth]{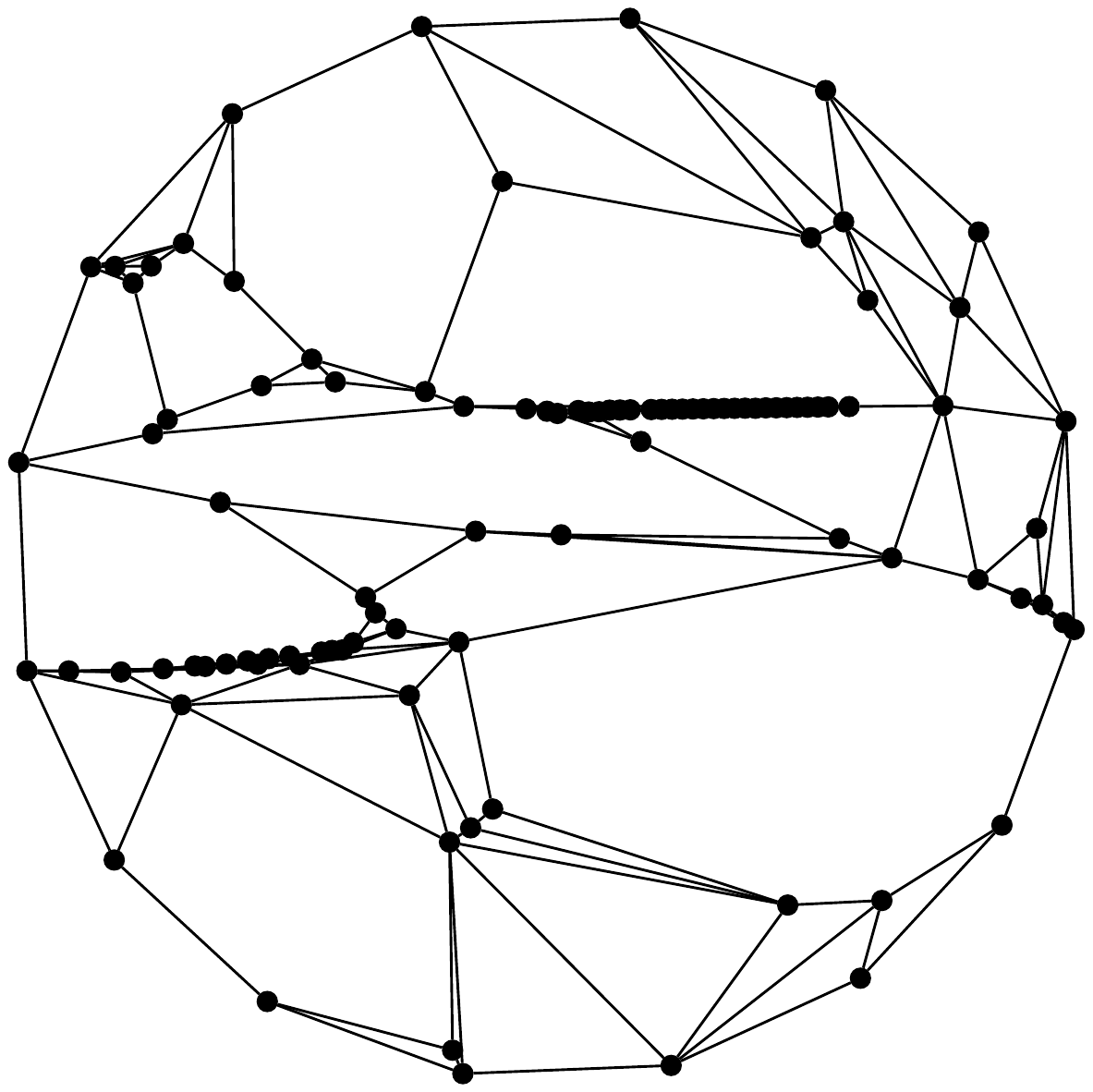} &  
        \includegraphics[width=.175\linewidth]{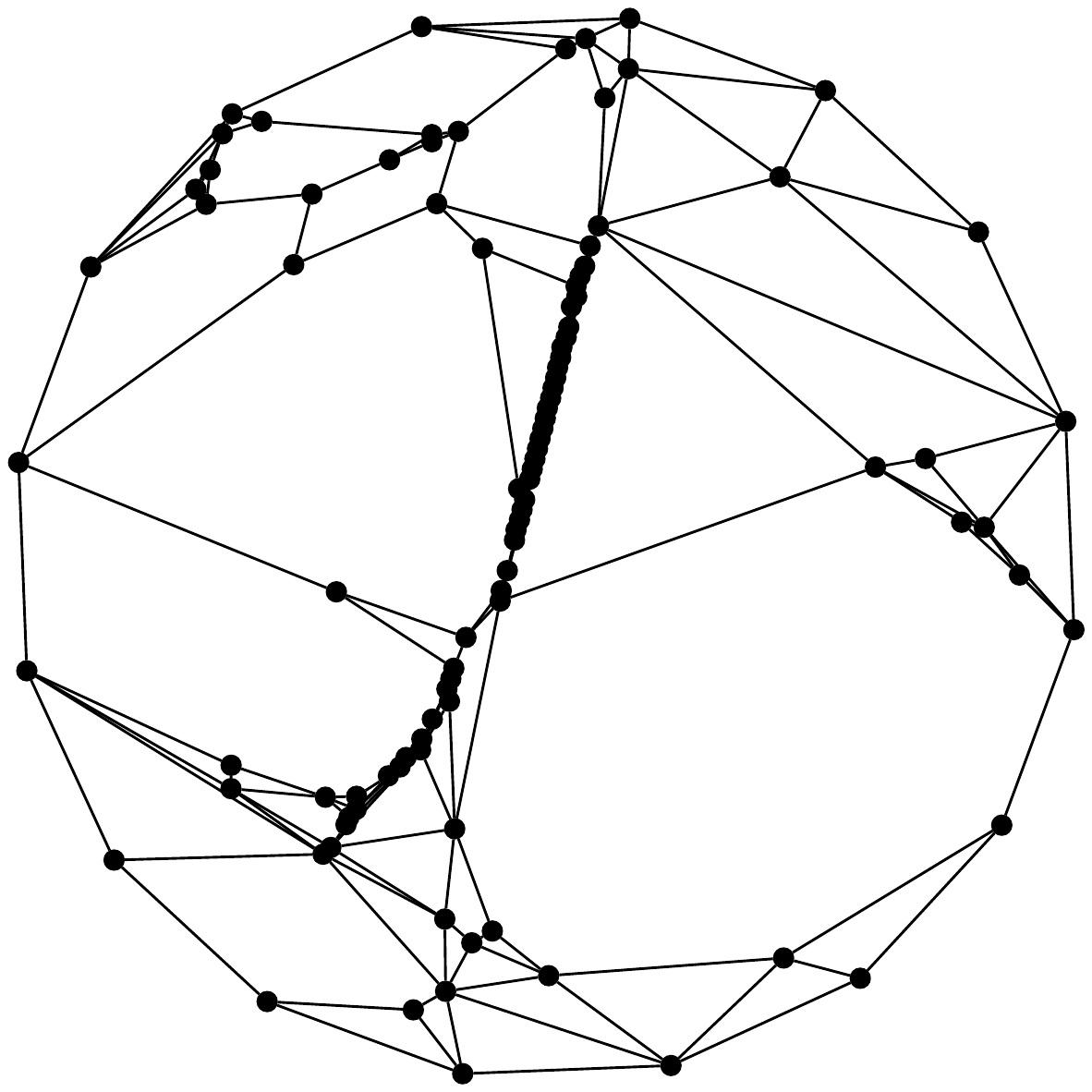} & 
        \includegraphics[width=.175\linewidth]{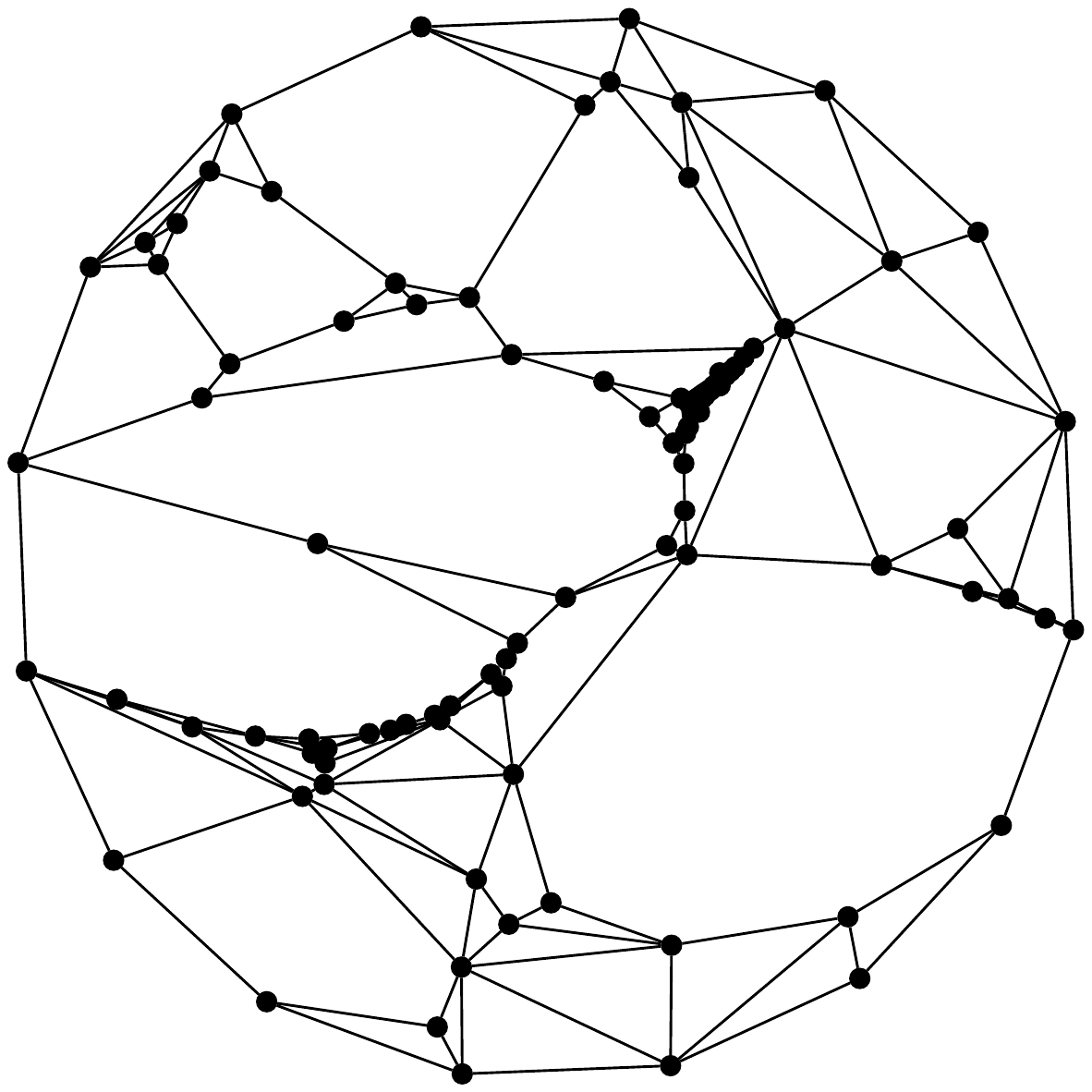} & 
        \includegraphics[width=.175\linewidth]{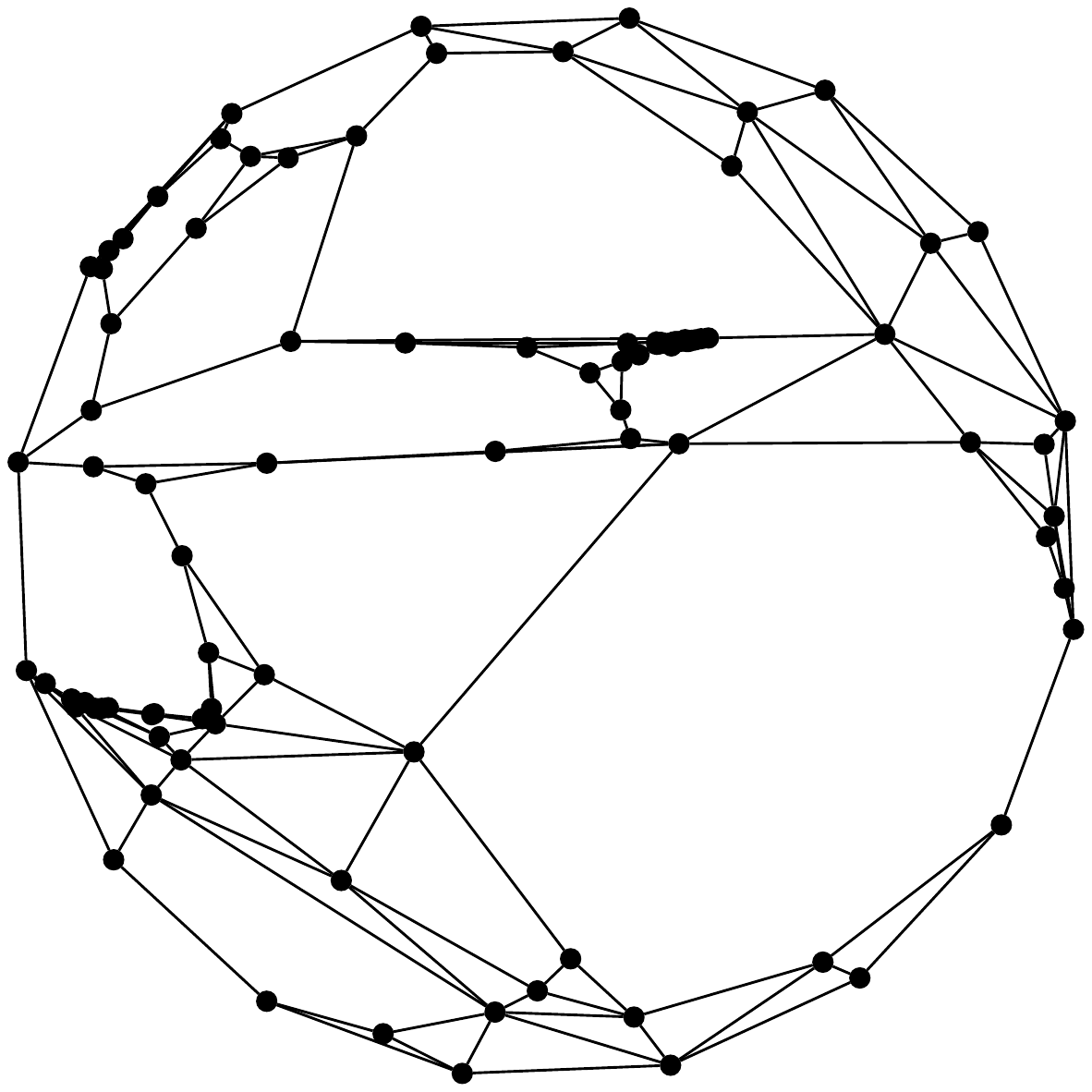} \\
        \hline
        & $\rho(\Gamma) =3973$ & $\rho(\Gamma) =43$ & $\rho(\Gamma) =71$ & $\rho(\Gamma) =229$ & $\rho(\Gamma) = 380, r=3$ \\
        \hline
        \multirow{-10}{*}{\rotatebox[origin=c]{90}{$G(70,200)$}} &
        \includegraphics[width=.175\linewidth]{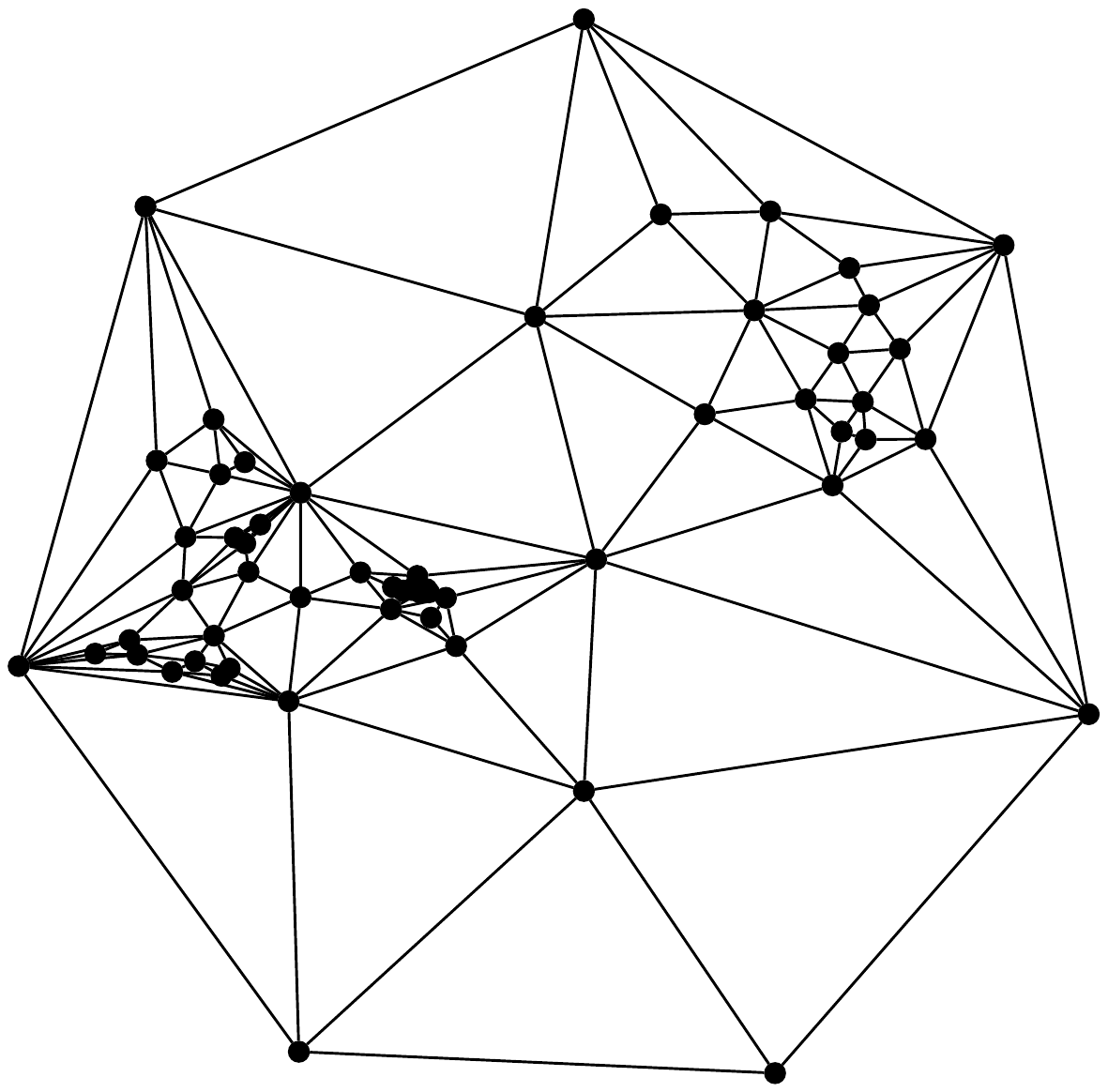} & 
        \includegraphics[width=.175\linewidth]{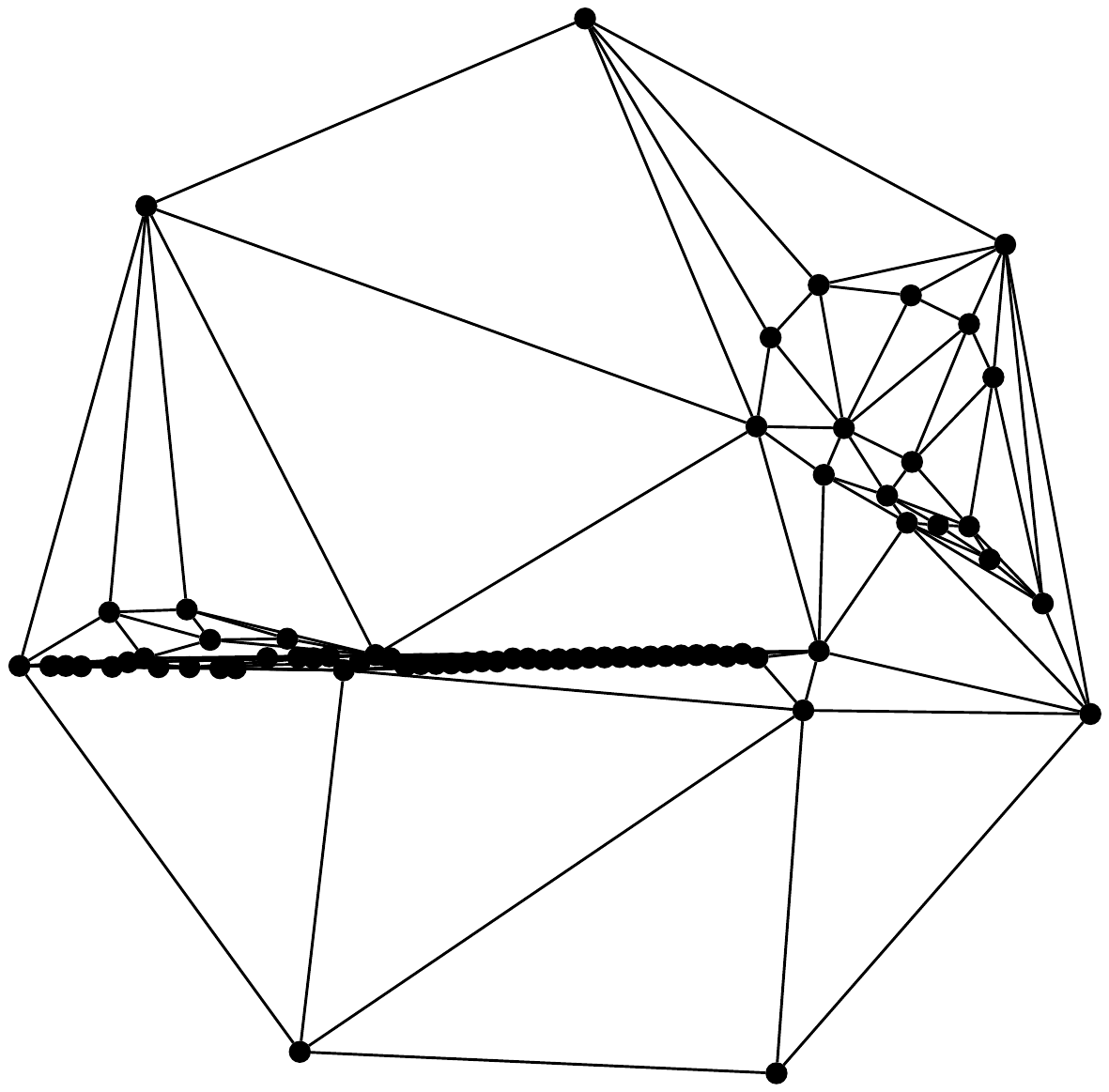} &  
        \includegraphics[width=.175\linewidth]{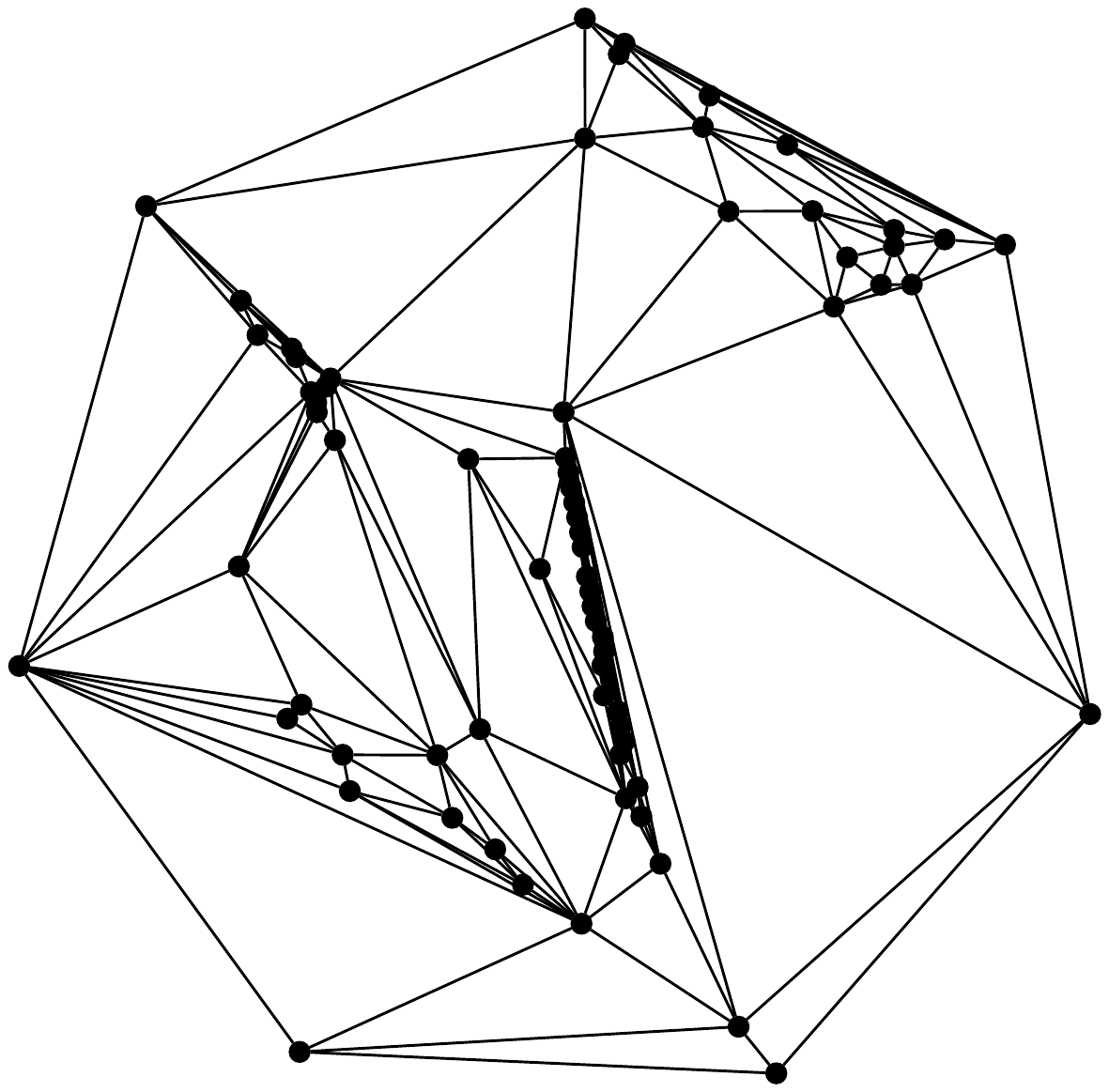} & 
        \includegraphics[width=.175\linewidth]{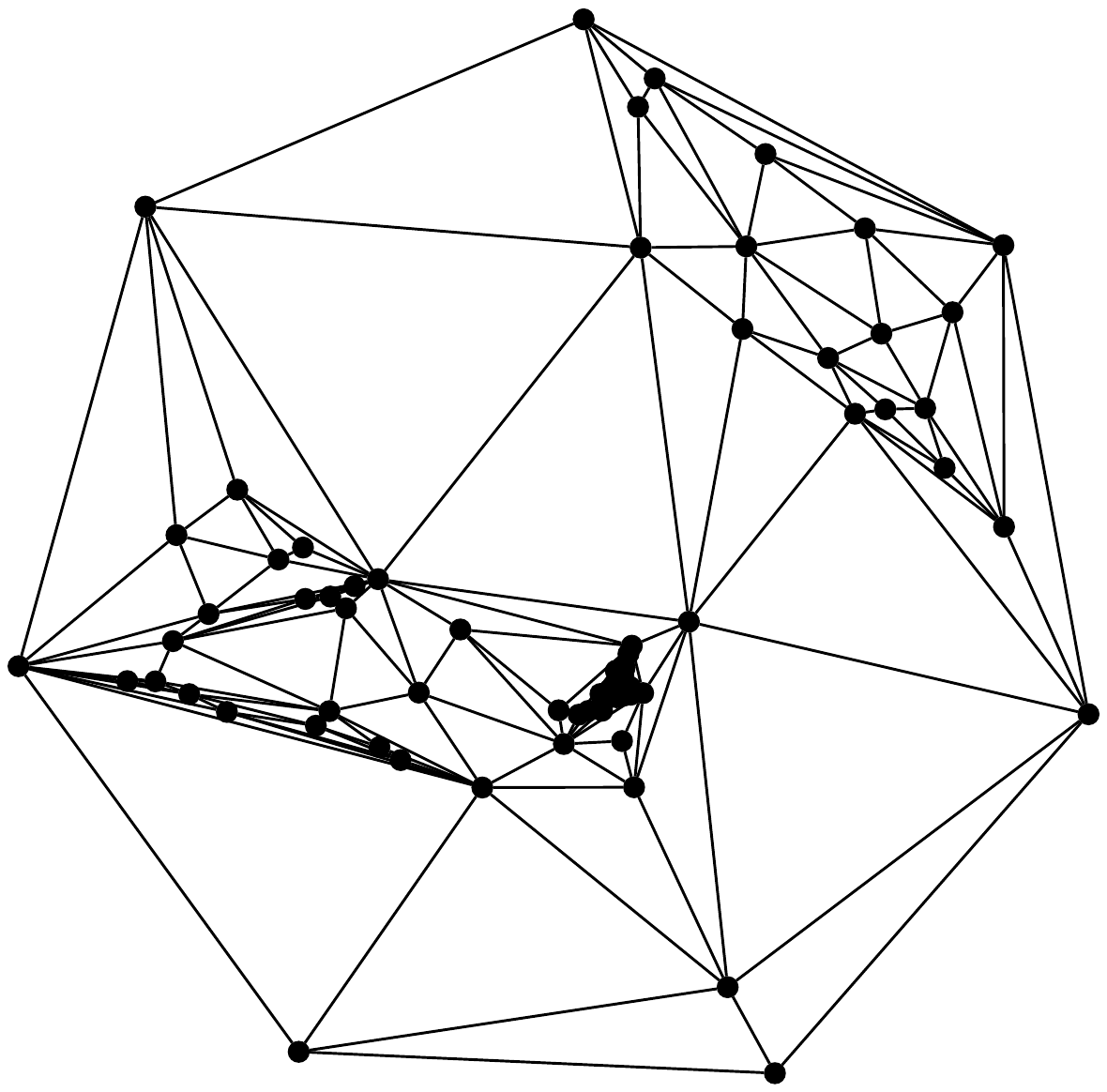} & 
        \includegraphics[width=.175\linewidth]{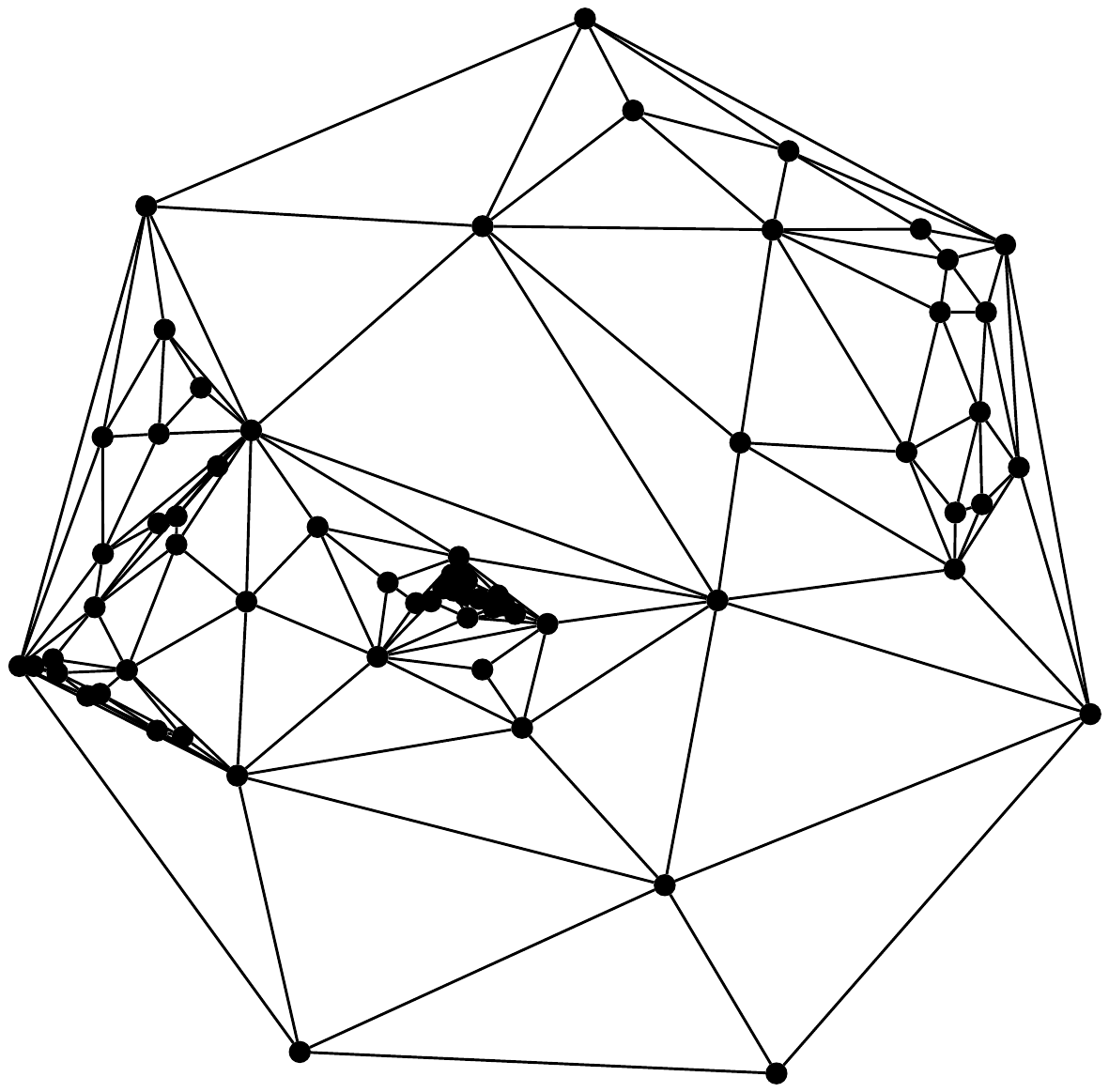} \\
        \hline
        & $\rho(\Gamma) =1165$ & $\rho(\Gamma) =42$  & $\rho(\Gamma) =69$ & $\rho(\Gamma) =216$ & $\rho(\Gamma) = 283, r=3$ \\
        \hline
        \multirow{-10}{*}{\rotatebox[origin=c]{90}{$G(50,130)$}} &
        \includegraphics[width=.175\linewidth]{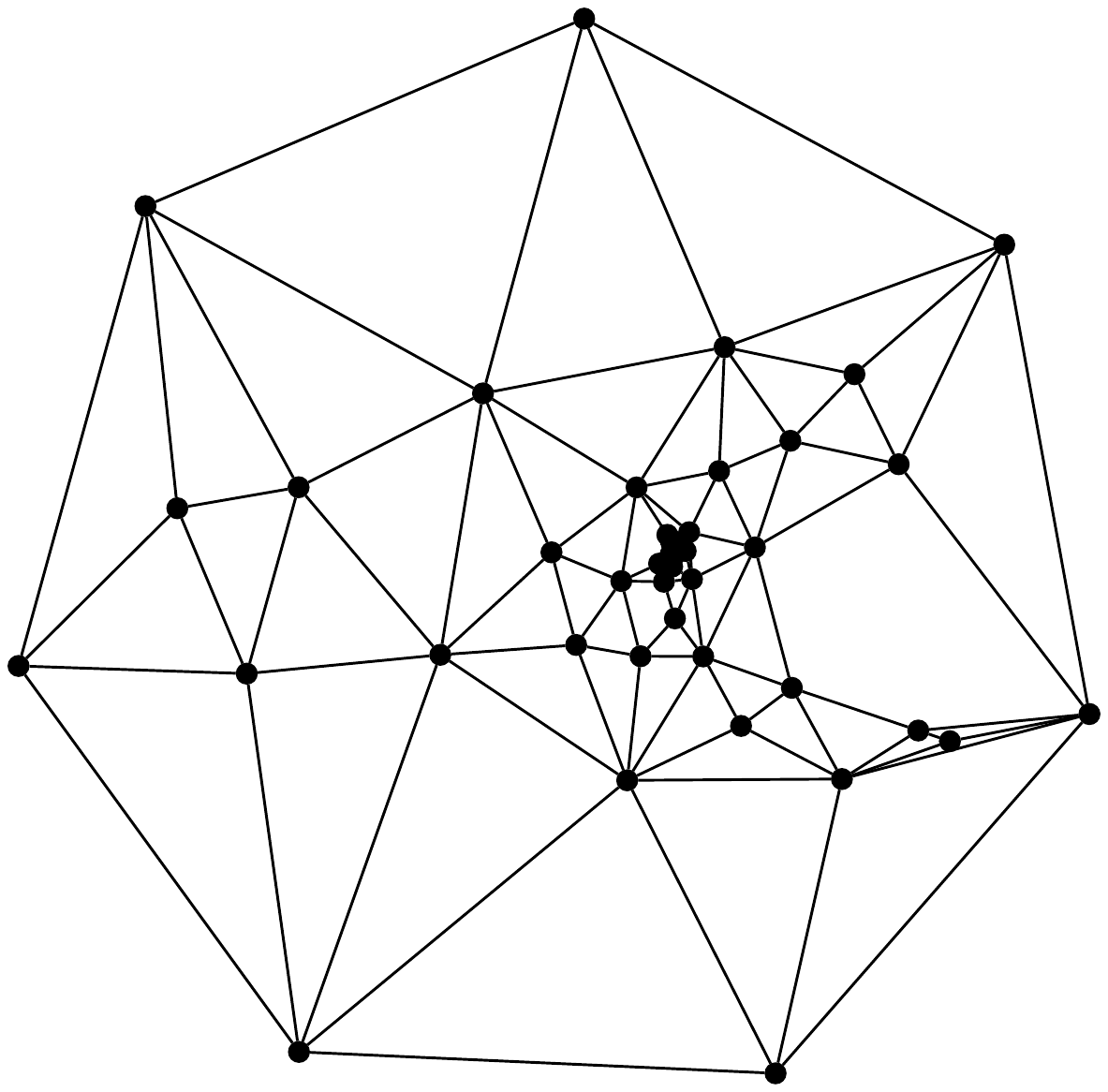} & 
        \includegraphics[width=.175\linewidth]{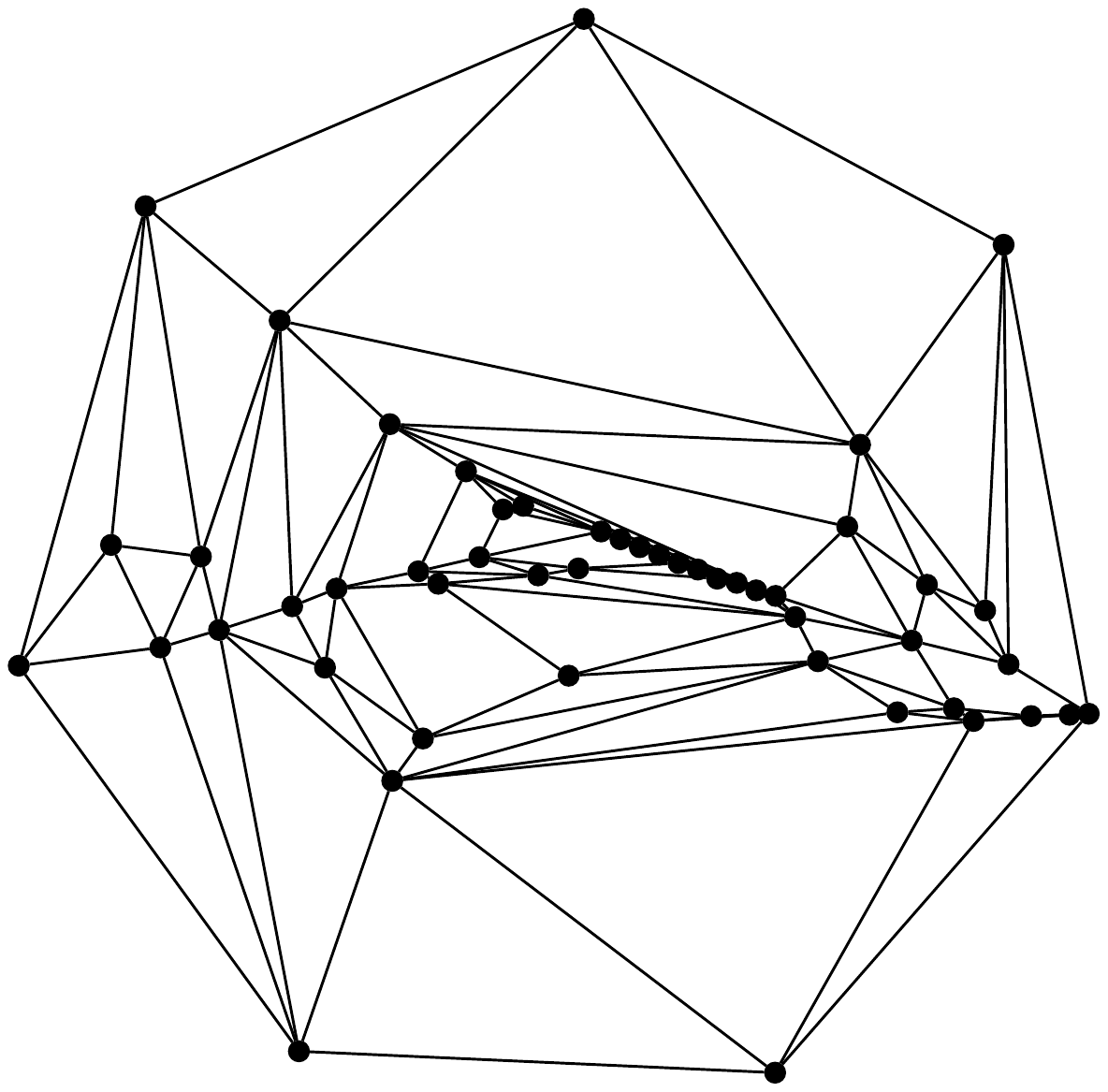} &  
        \includegraphics[width=.175\linewidth]{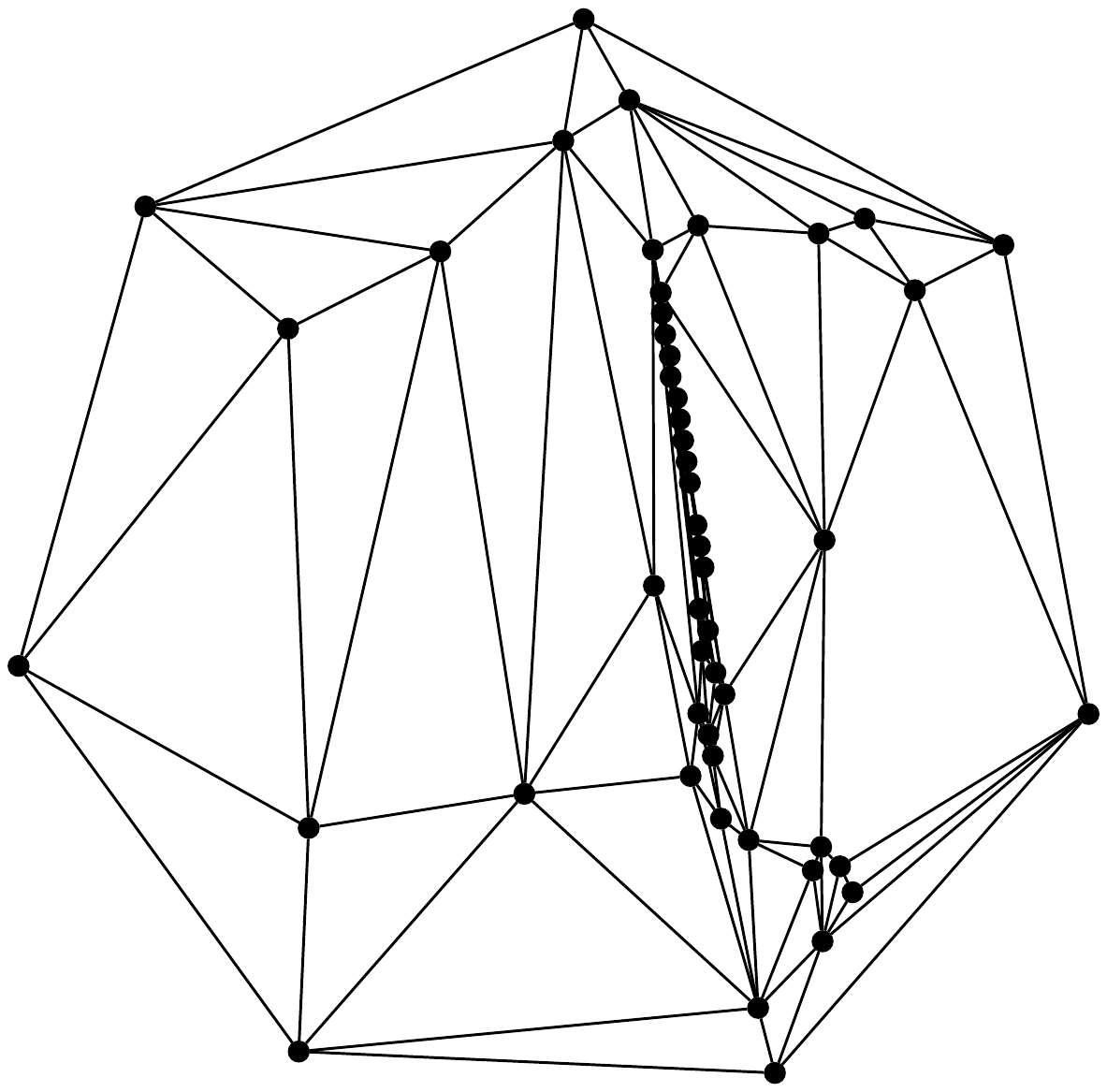} & 
        \includegraphics[width=.175\linewidth]{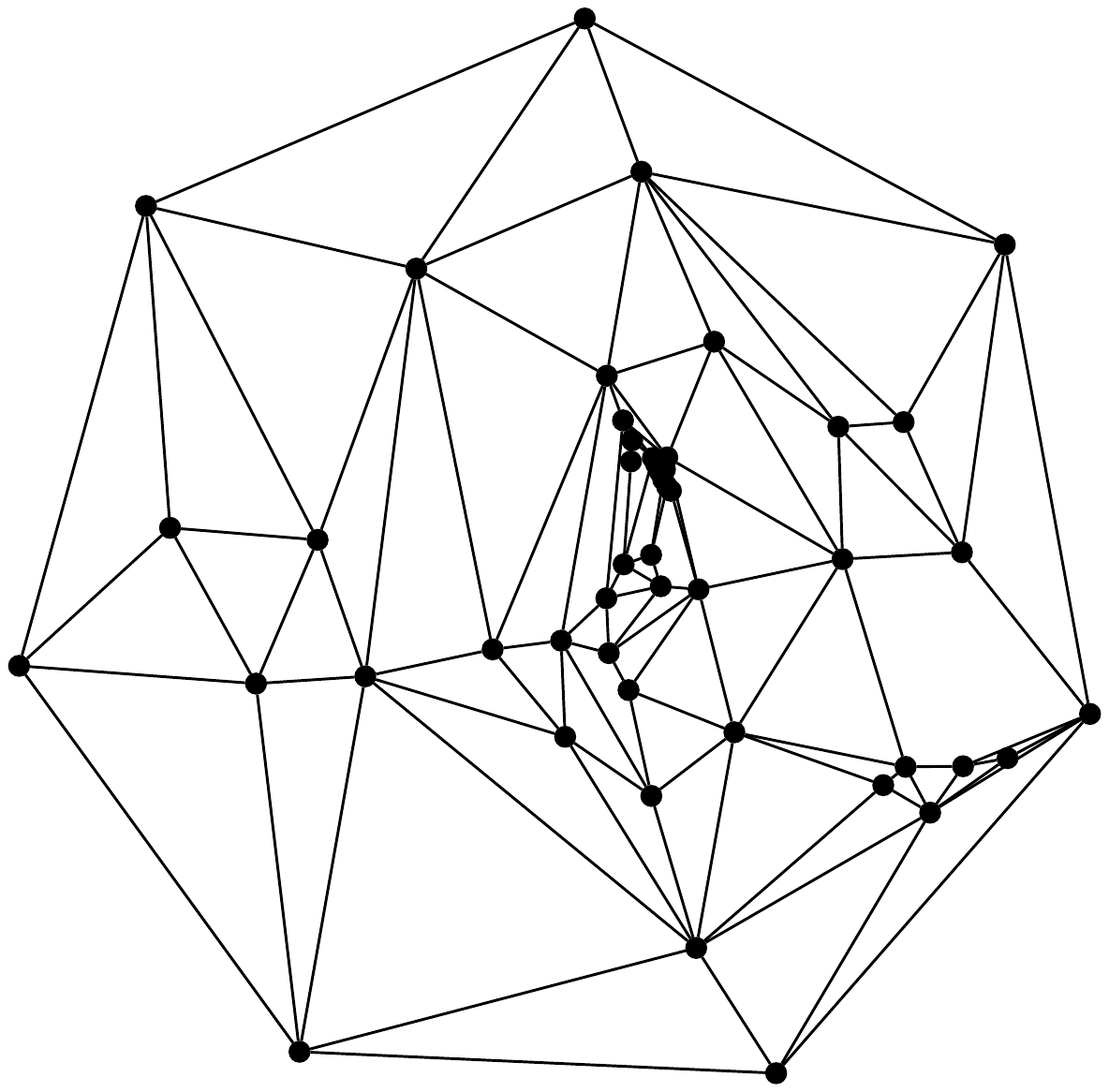} & 
        \includegraphics[width=.175\linewidth]{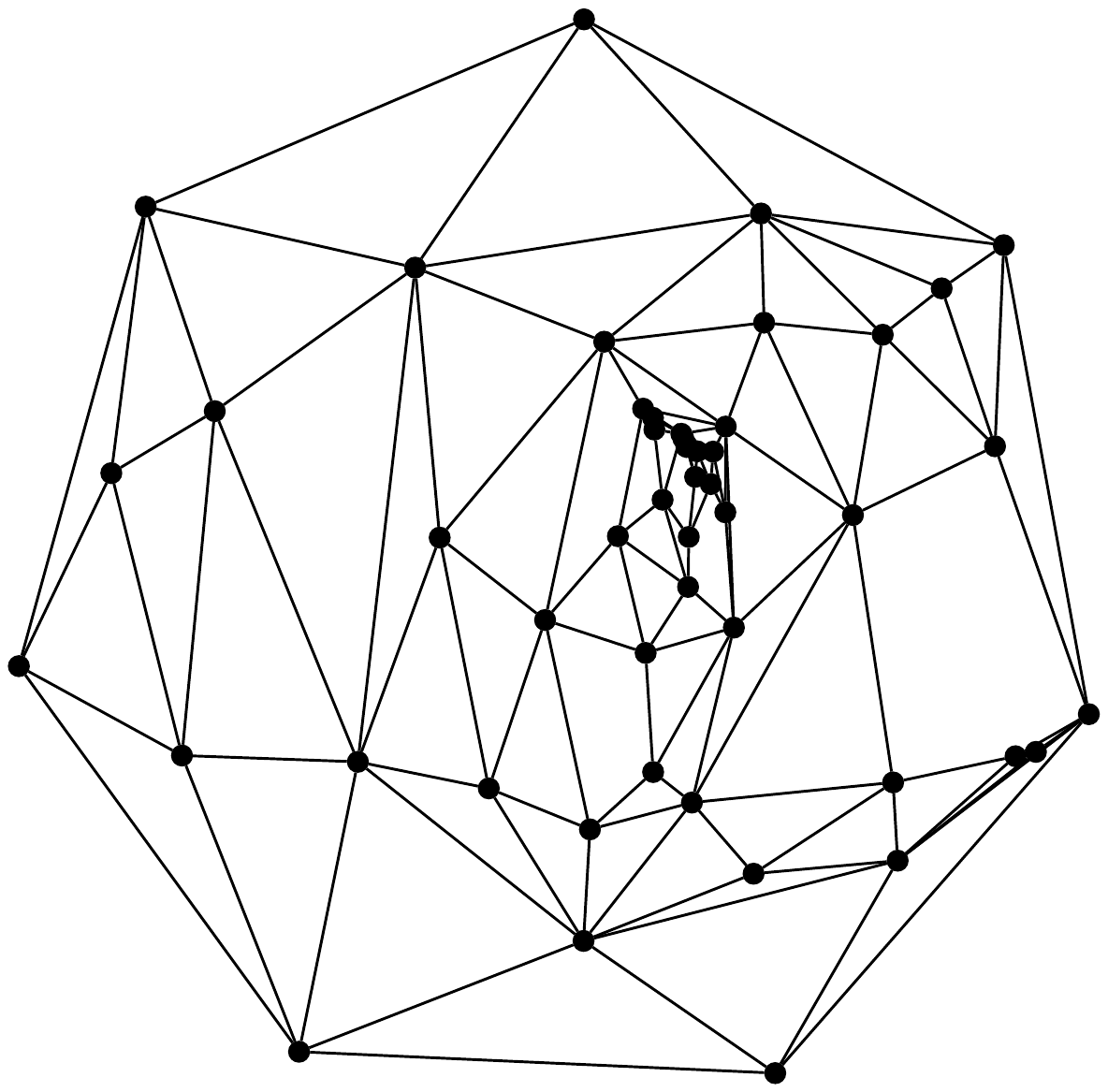} \\
        \hline
        & $\rho(\Gamma) =766$ & $\rho(\Gamma) =31$ & $\rho(\Gamma) =31$ & $\rho(\Gamma) =1020$ & $\rho(\Gamma) = 272, r=3$ \\
        \hline
        \multirow{-10}{*}{\rotatebox[origin=c]{90}{$G(400,1100)$}} &
        \includegraphics[width=.175\linewidth]{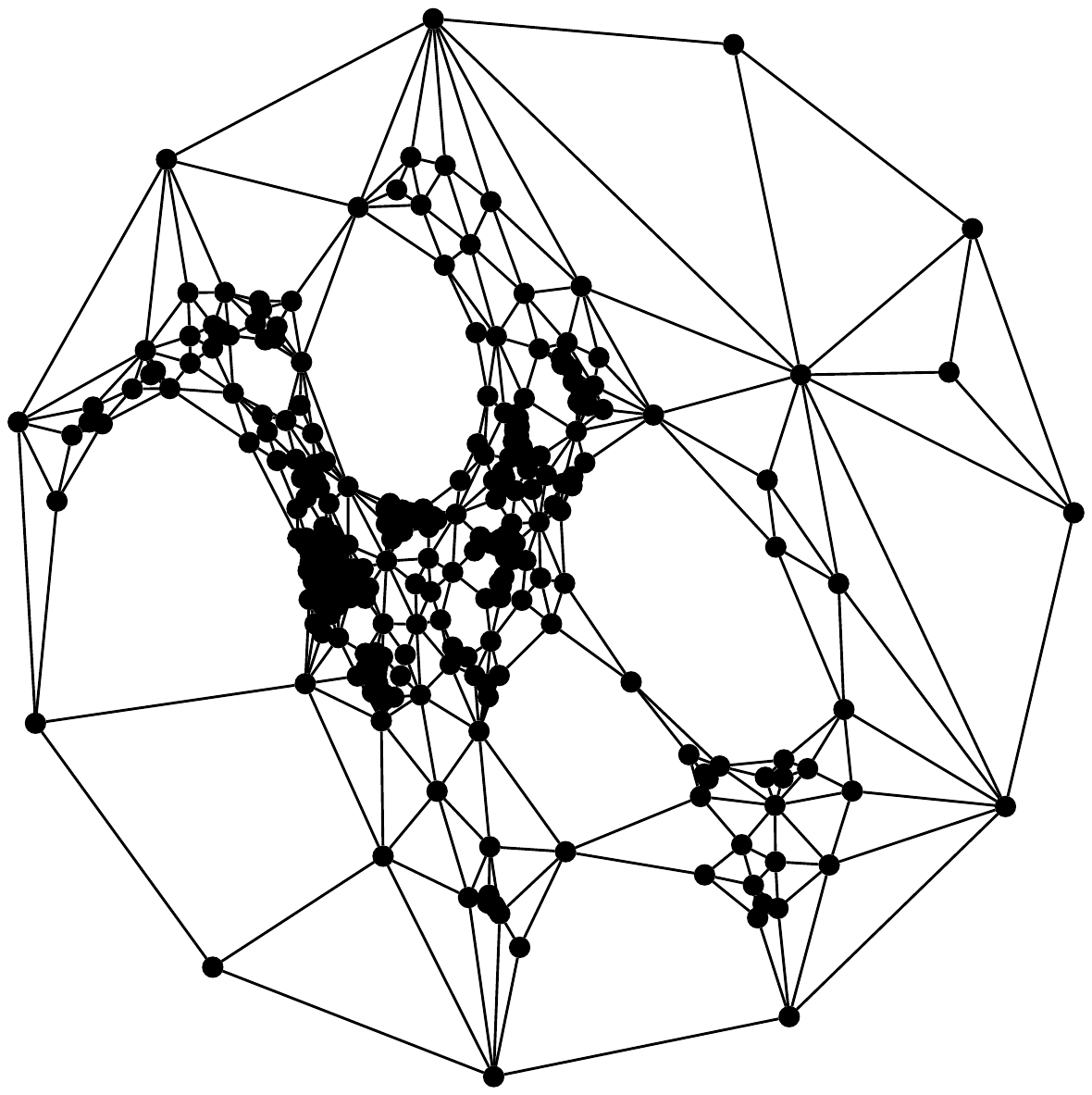} & 
        \includegraphics[width=.175\linewidth]{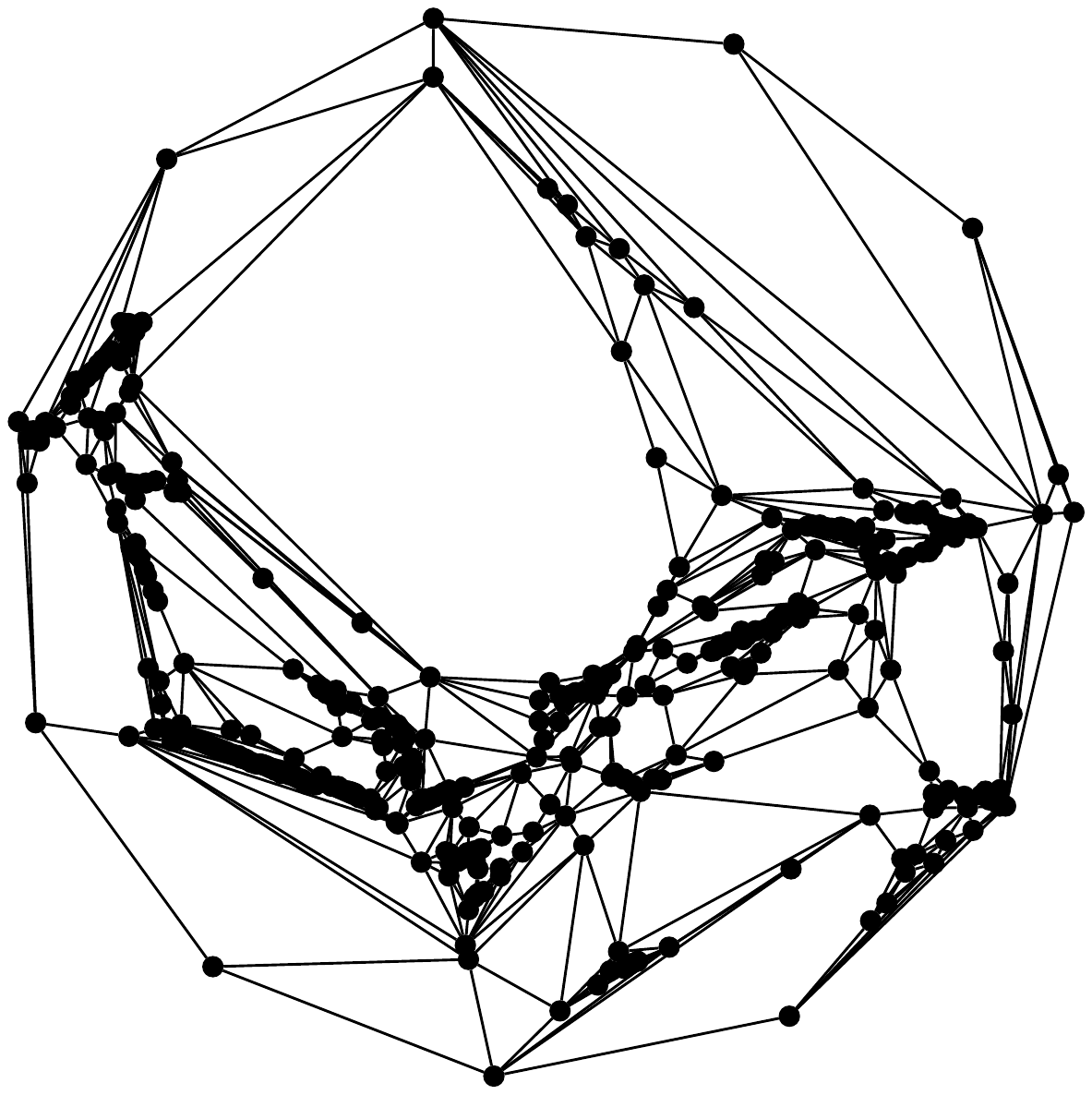} &  
        \includegraphics[width=.175\linewidth]{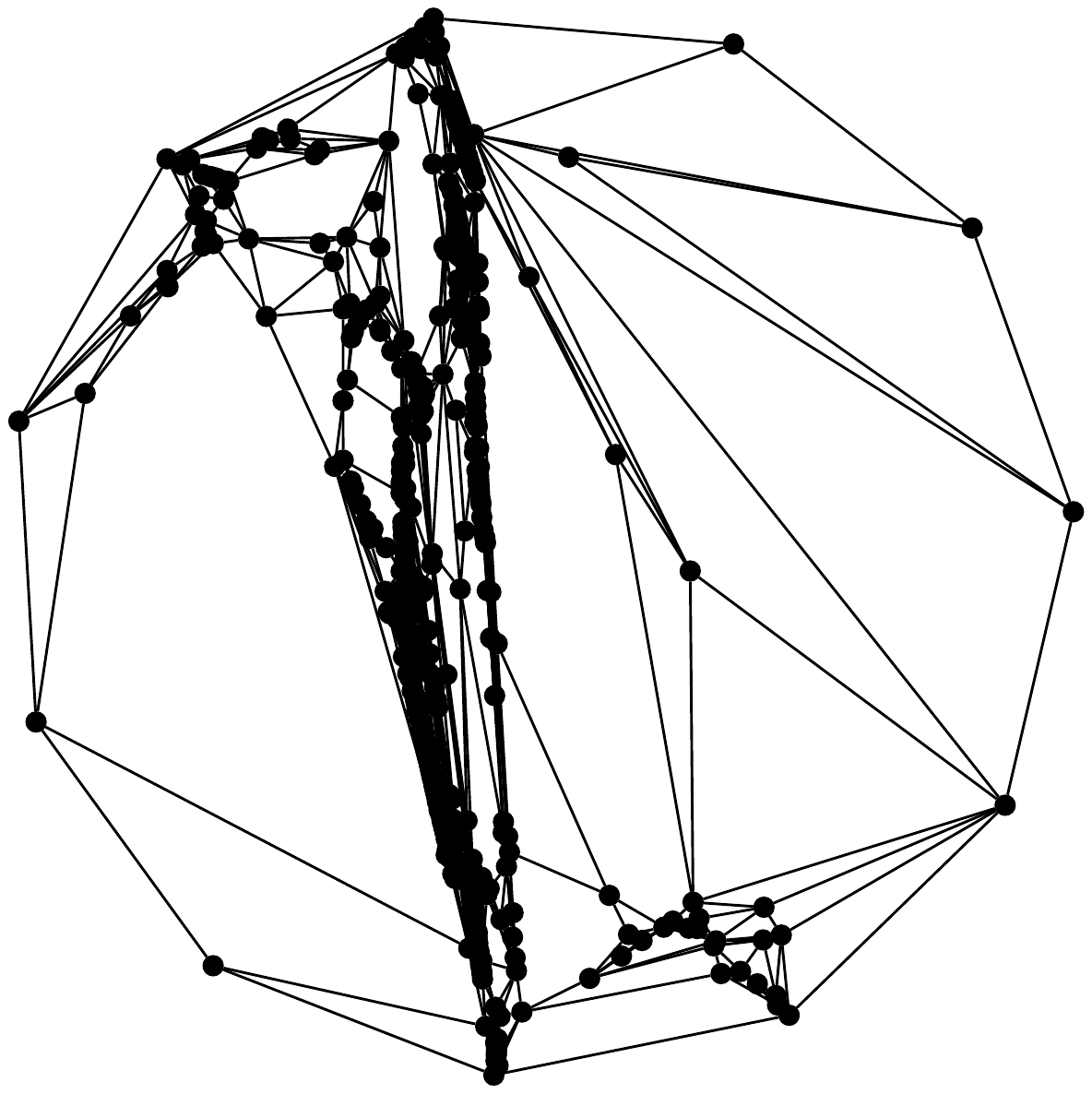} & 
        \includegraphics[width=.175\linewidth]{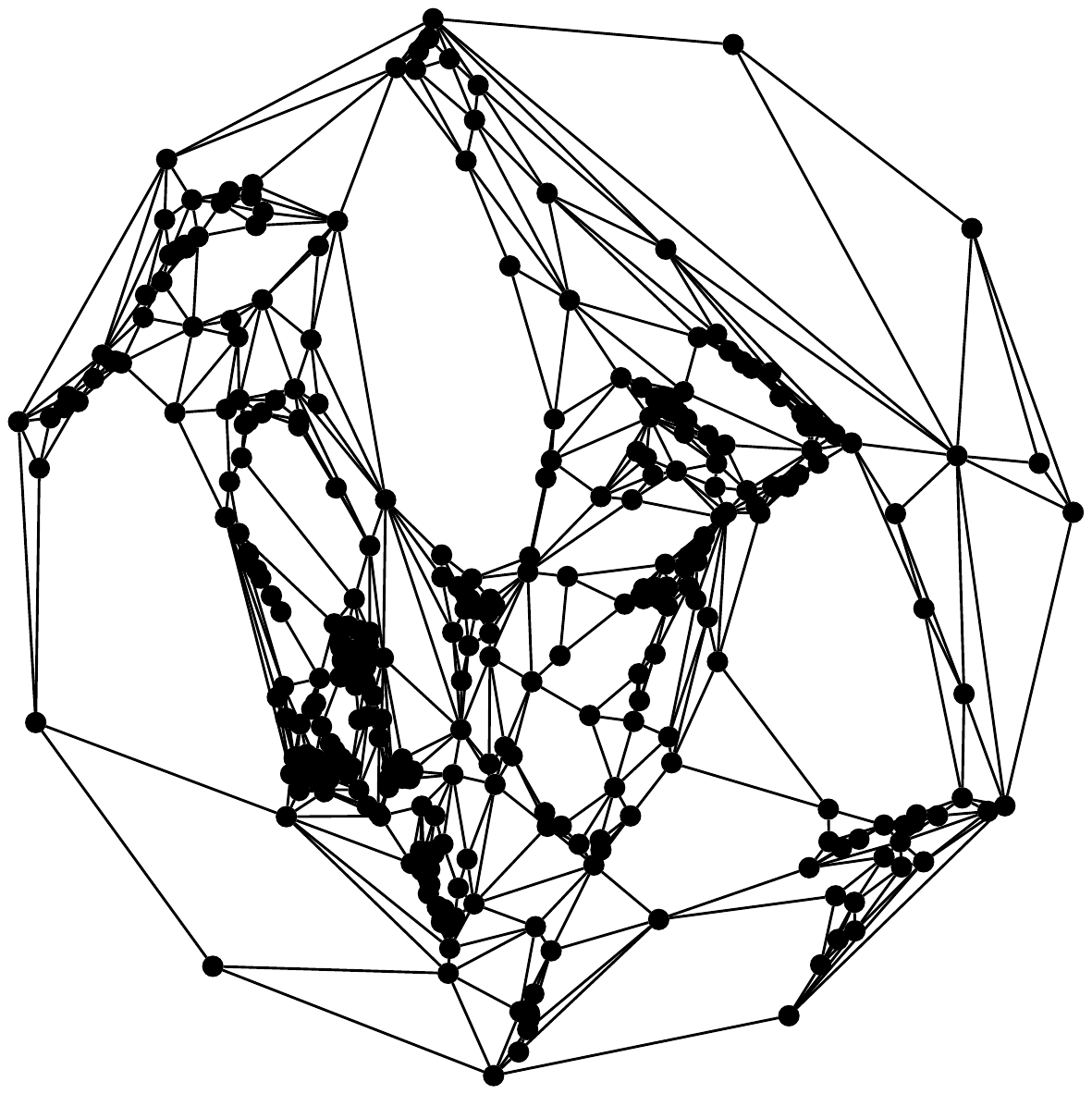} & 
        \includegraphics[width=.175\linewidth]{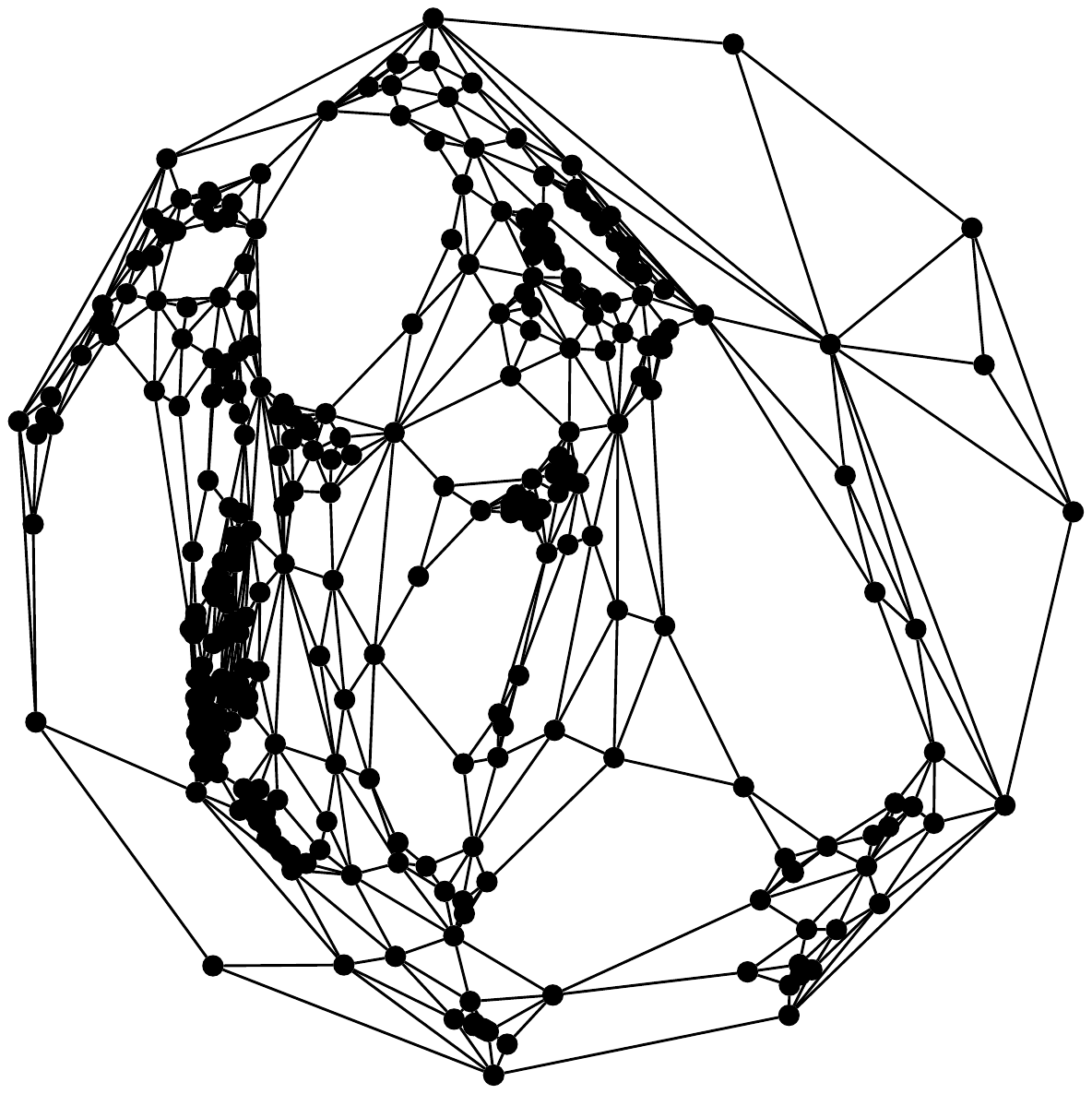} \\
        \hline
        & $\rho(\Gamma) =56162$ & $\rho(\Gamma) =538$ & $\rho(\Gamma) =414$ & $\rho(\Gamma) =5054$ & $\rho(\Gamma) = 9092, r=2$ \\
        \hline
        
  \end{tabular}
\end{table}

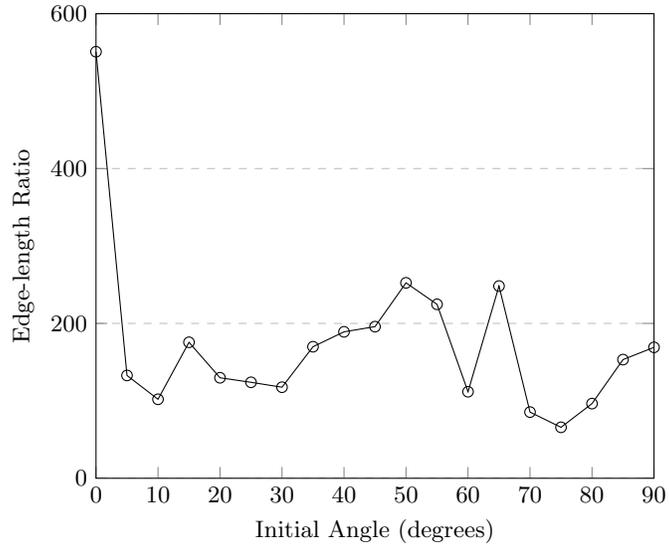
\begin{figure}[h]
\begin{center}
\begin{tikzpicture}
    \begin{axis}[
        xlabel={Initial Angle (degrees)},
        ylabel={Edge-length Ratio},
        xmin=0, xmax=90,
        ymin=0, ymax=600,
        xtick={0,10,20,30,40,50,60,70,80,90},
        ytick={0,200,400,600},
        legend pos=north east,
        ymajorgrids=true,
        grid style=dashed,
    ]
        
    \addplot[
    color=black,
    mark=o,
    ]
    coordinates {
    (0.000000, 550.875079)(5.000000, 132.786926)(10.000000, 102.001785)(15.000000, 175.628853)(20.000000, 129.793622)(25.000000, 123.938641)(30.000000, 117.591254)(35.000000, 169.879275)(40.000000, 189.127692)(45.000000, 195.824308)(50.000000, 252.412950)(55.000000, 224.674119)(60.000000, 111.753705)(65.000000, 248.236382)(70.000000, 85.455080)(75.000000, 65.798750)(80.000000, 96.430330)(85.000000, 153.267069)(90.000000, 169.102253)
    };
    
    \end{axis}
\end{tikzpicture}
\end{center}

\caption{Edge-length ratios for kaleidoscope $xy$-morphs for $G(90,240)$, increments of 5 degrees.}
\label{fig:kaleidoscope1}
\end{figure}

We also experimented with a ``kaleidoscope'' drawing paradigm, where we
rotate the $x$- and $y$-axes by small angular increments and 
compute an $xy$-morph for each angle.
The edge-length ratios can vary dramatically in such drawings, so the 
minima offer good choices.
We show an example plot of edge-length ratios
in Fig.~\ref{fig:kaleidoscope1},
with its worst and best rotations in Table.~\ref{tbl:kal1}. 


\begin{table}[h]
\caption{Worst and best rotations for the graph of Fig.~\ref{fig:kaleidoscope1}.}
\label{tbl:kal1}
\begin{center}
      \begin{tabular}
            {ccccc} \hline & Tutte & $x$-spread & $y$-spread & $xy$-morph\\
            \hline 
        
            \multirow{-10}{*}{\rotatebox[origin=c]{90}{$\alpha = 0^\circ$}} &
            \includegraphics[width=.23\linewidth]{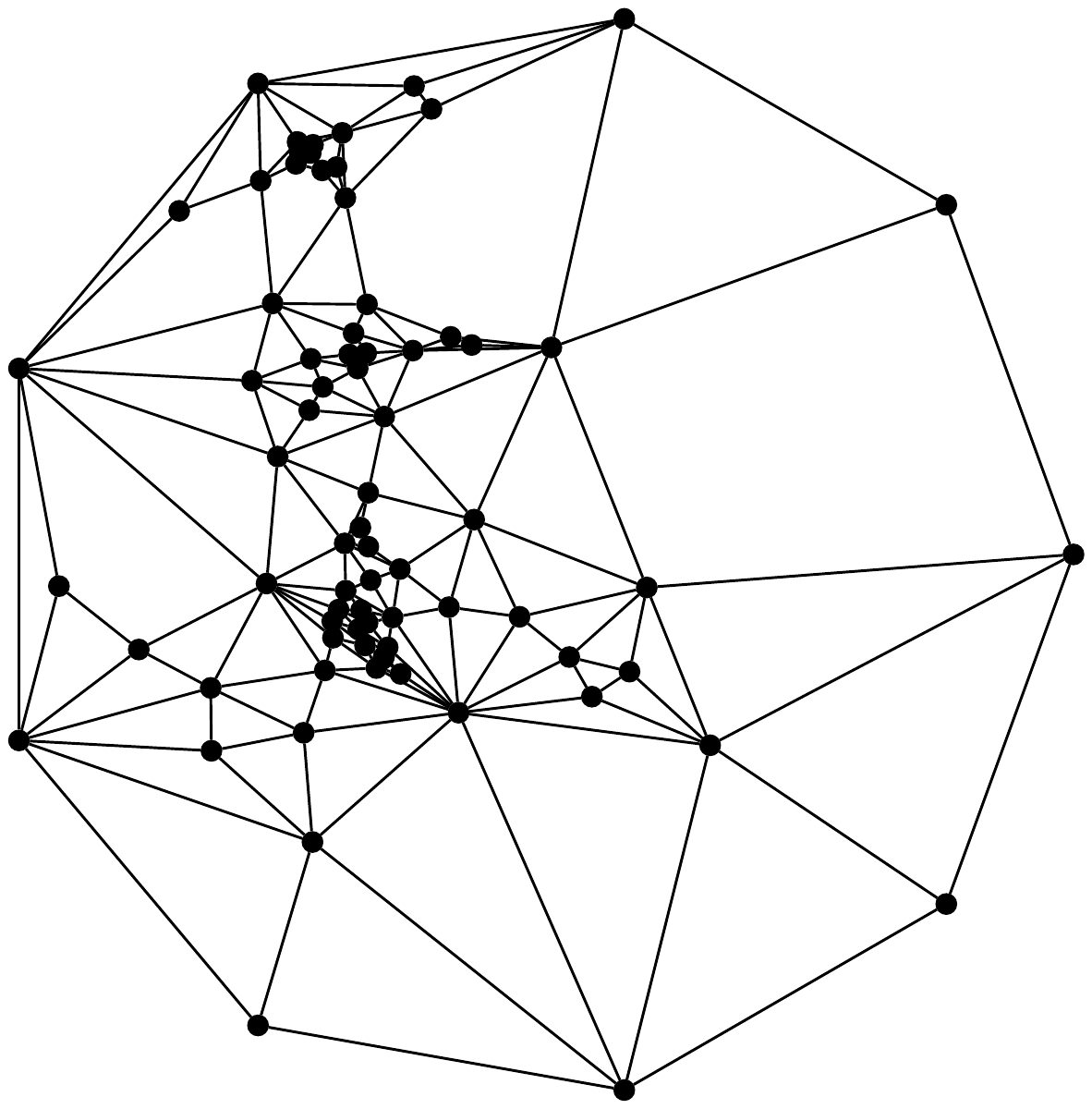} &
            \includegraphics[width=.23\linewidth]{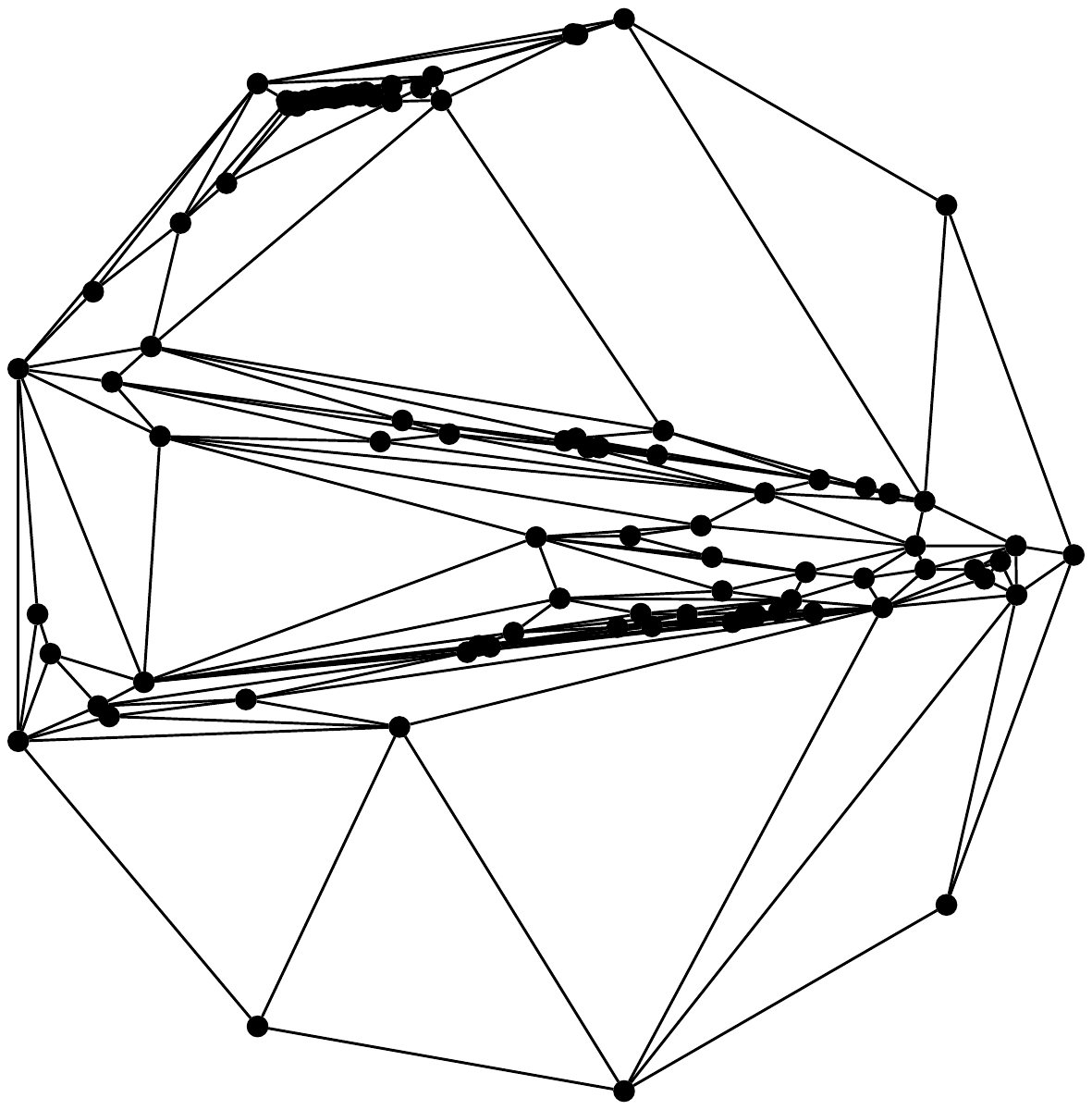} &  
            \includegraphics[width=.23\linewidth]{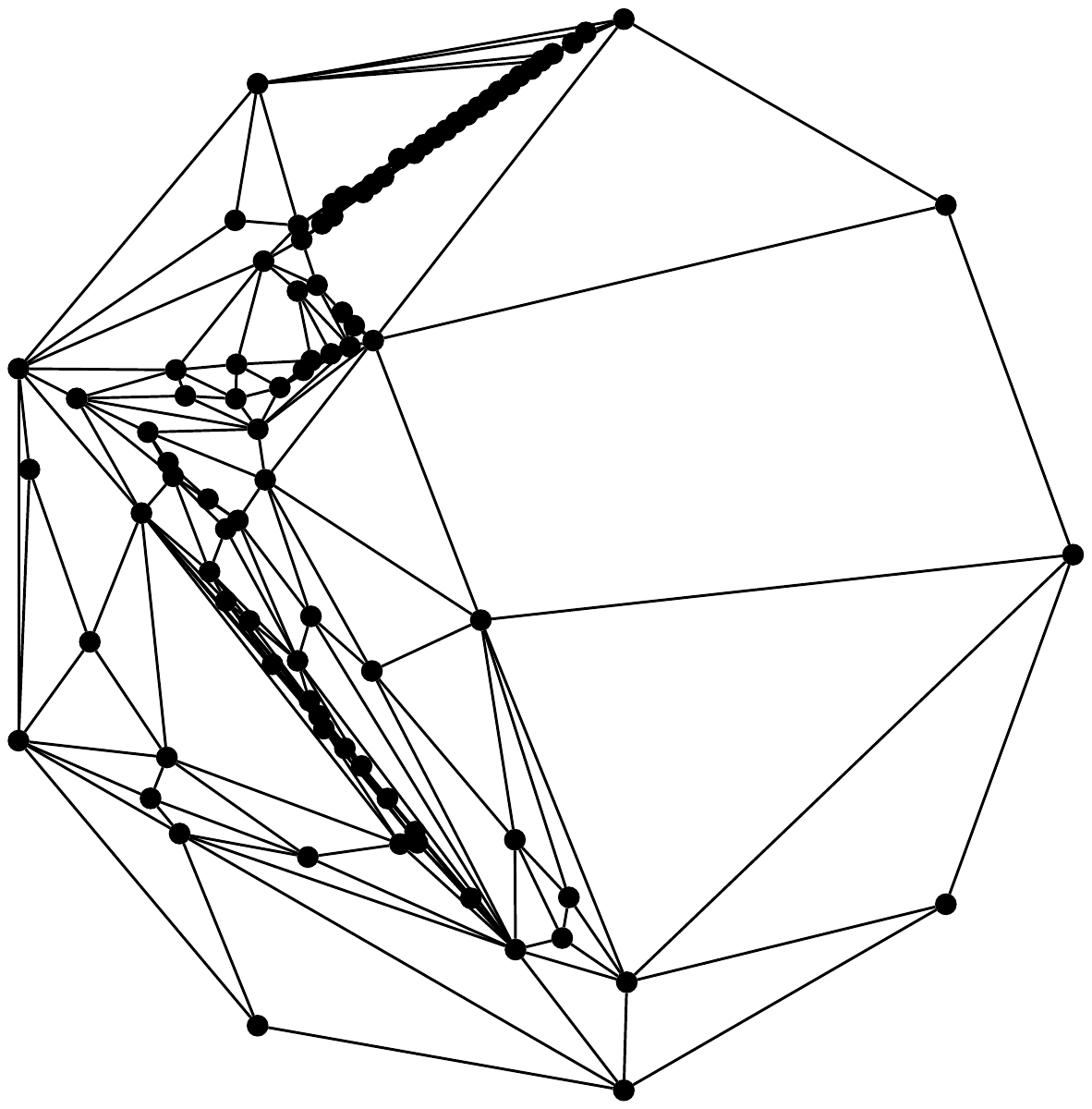} & 
            \includegraphics[width=.23\linewidth]{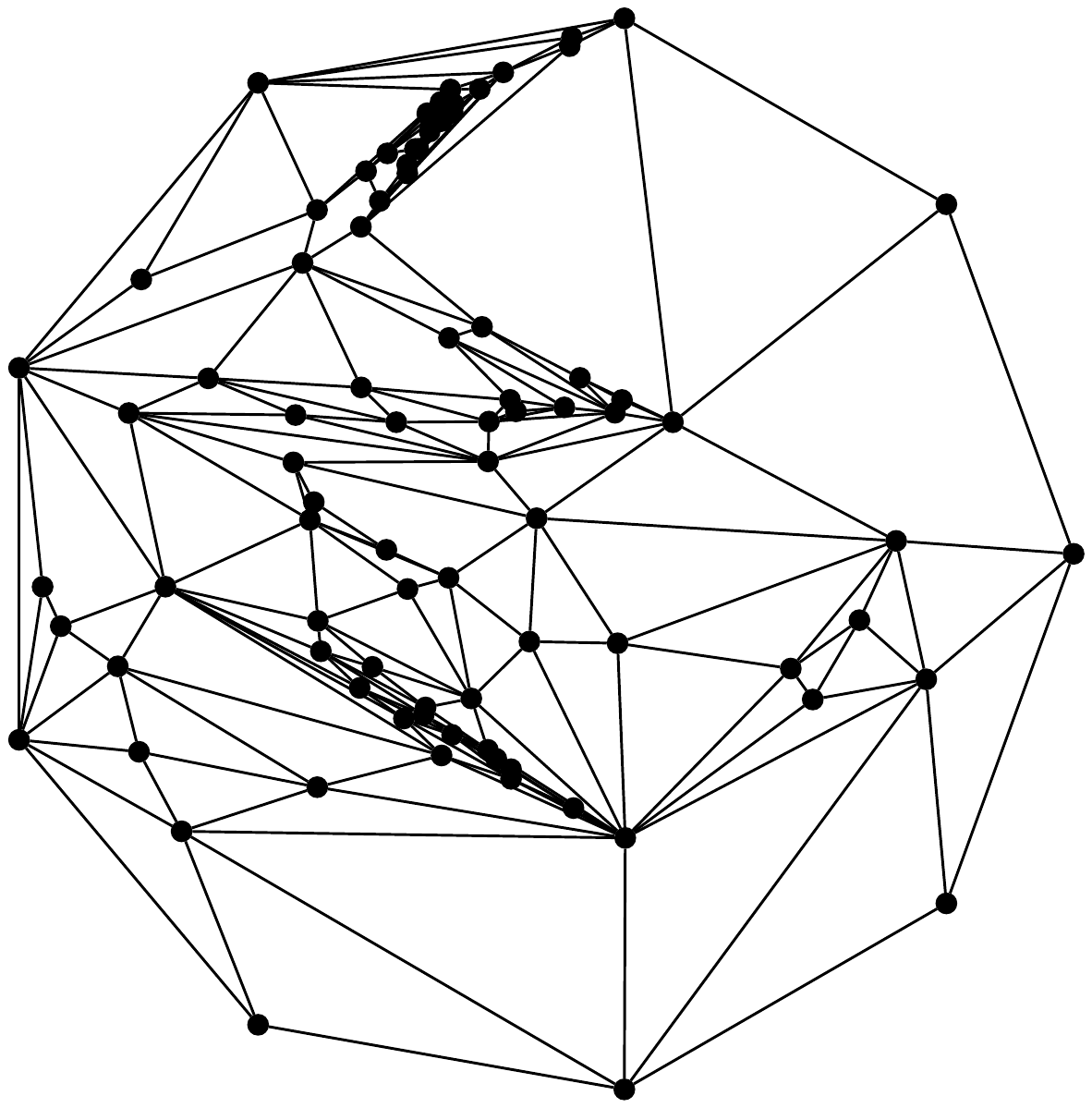} \\
            \hline
            \multirow{-10}{*}{\rotatebox[origin=c]{90}{$\alpha = 75^\circ$}} &
            \includegraphics[width=.23\linewidth]{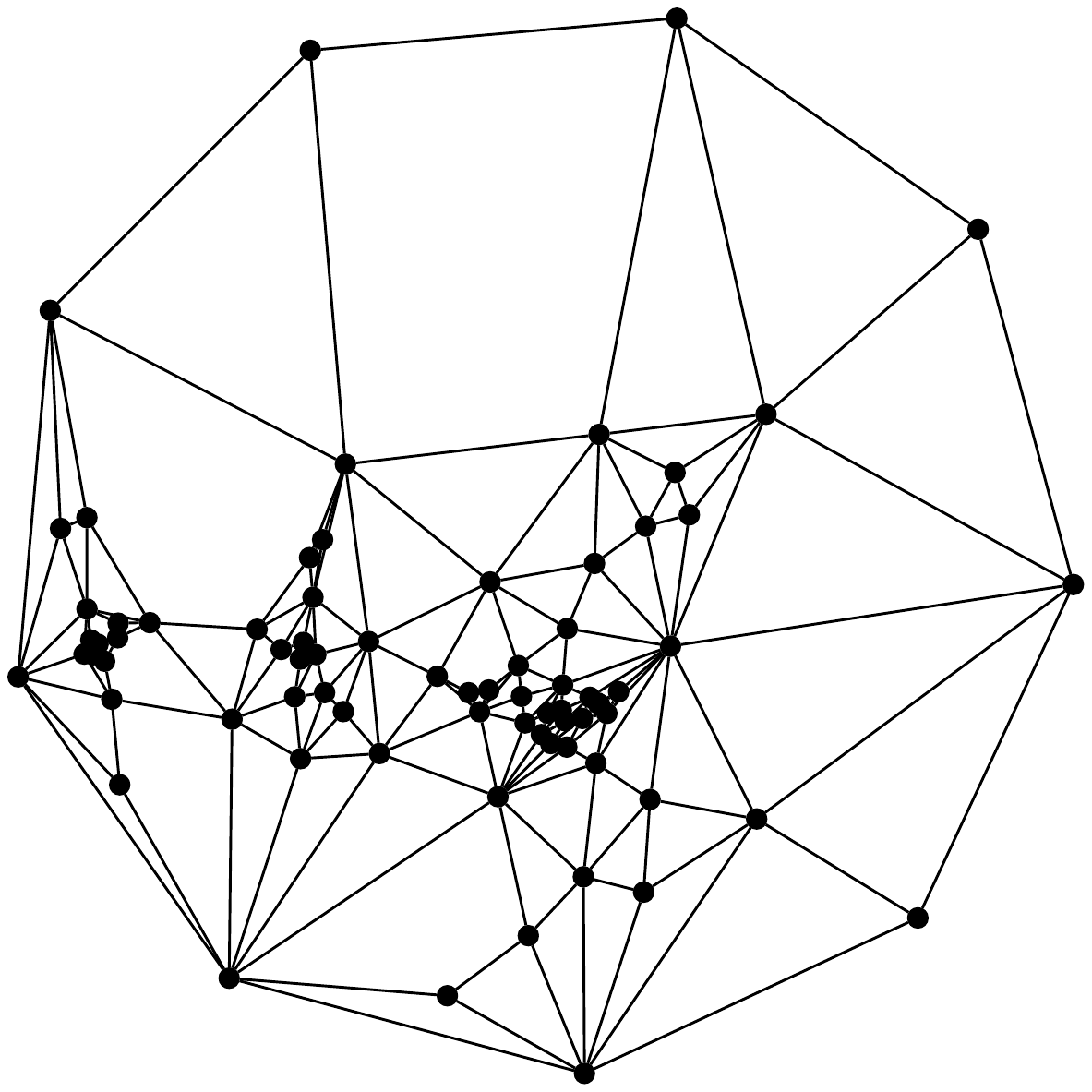} &
            \includegraphics[width=.23\linewidth]{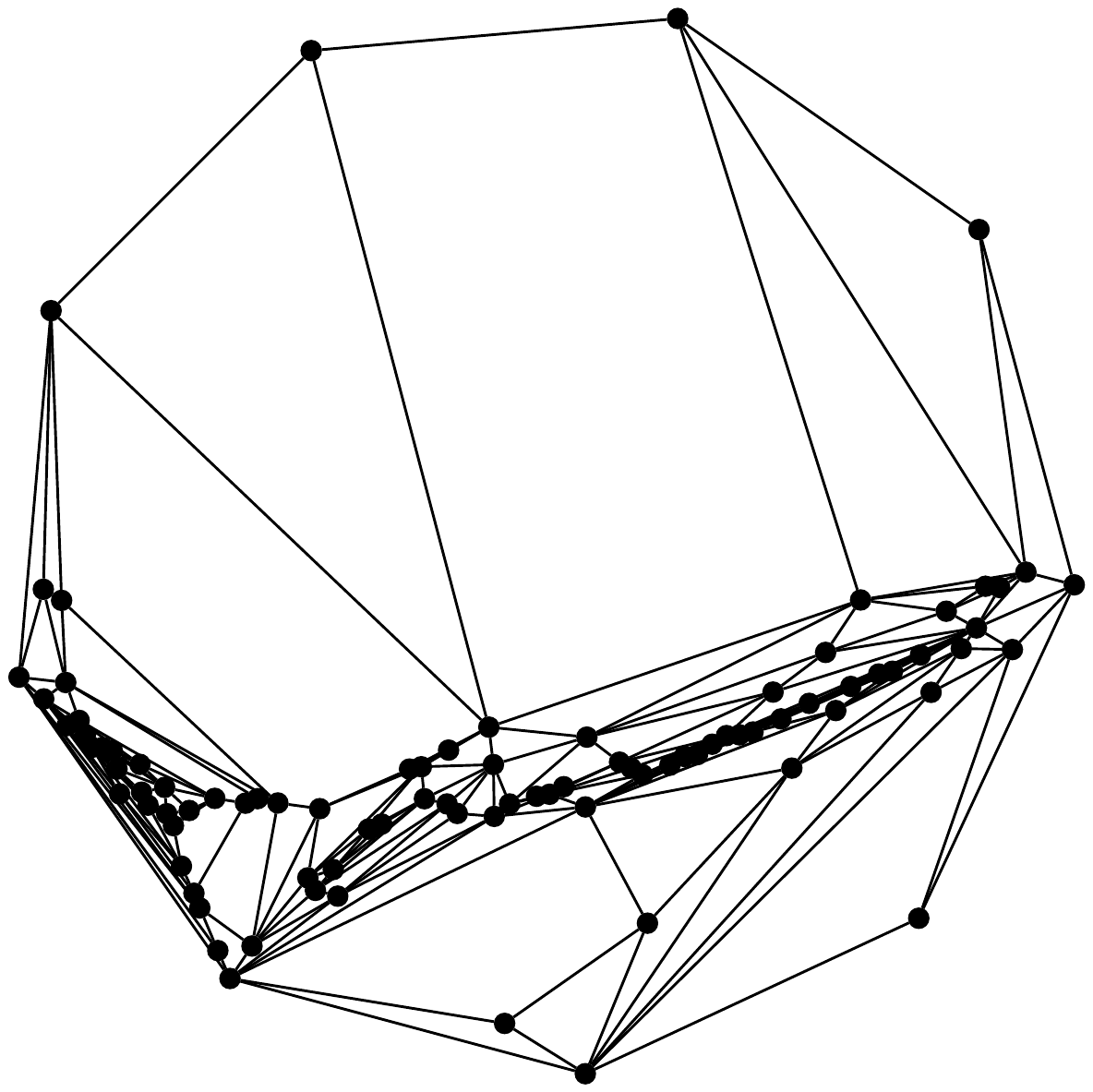} &  
            \includegraphics[width=.23\linewidth]{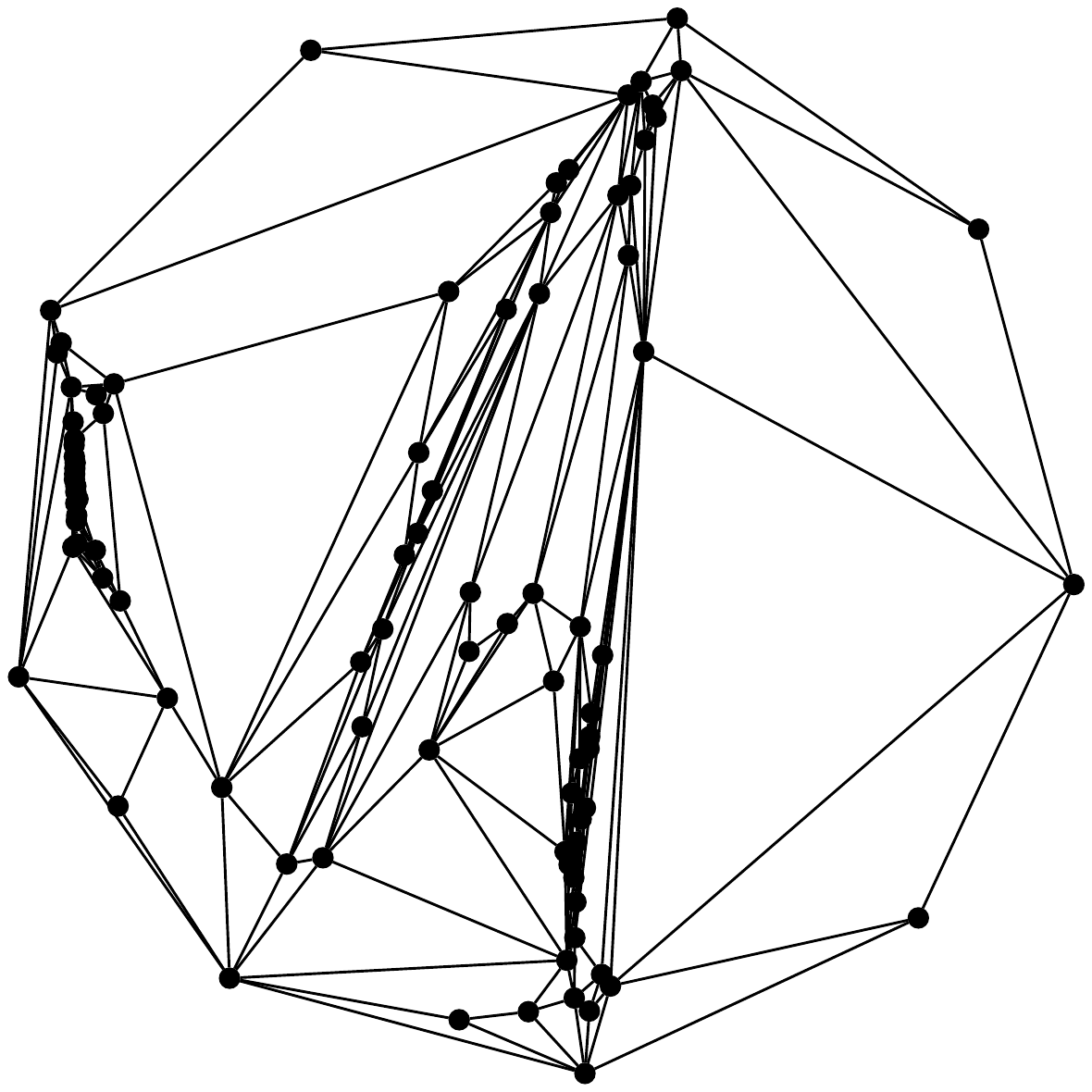} & 
            \includegraphics[width=.23\linewidth]{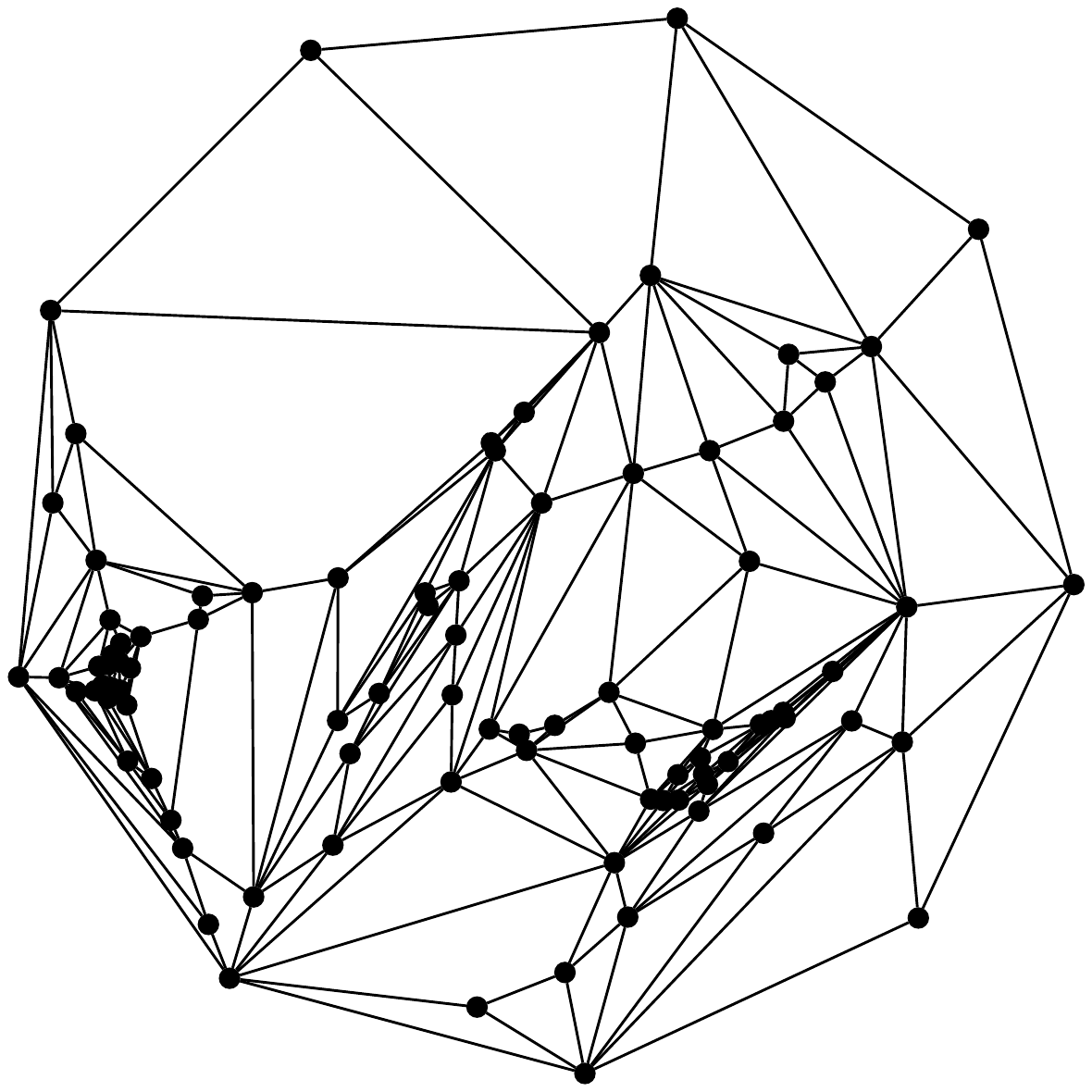} \\
      \end{tabular}
\end{center}
\end{table}

\clearpage

\ifAnon
\else
\section*{Acknowledgements}
This research was supported in part by NSF grant CCF-2212129.
\fi

    \bibliographystyle{splncs04}
    \bibliography{refs}

\clearpage
\begin{appendix}
\section{Appendix}
\begin{figure}[p]
    \begin{center}
    \begin{tikzpicture}
        \begin{axis}[
            xlabel={Initial Angle (degrees)},
            ylabel={Edge-length Ratio},
            xmin=0, xmax=90,
            ymin=400, ymax=4800,
            xtick={0,10,20,30,40,50,60,70,80,90},
            ytick={400,1500,2600,3700,4800},
            legend pos=north east,
            ymajorgrids=true,
            grid style=dashed,
        ]
            
        \addplot[
        color=black,
        mark=o,
        ]
        coordinates {
        (0.000000, 711.367417)(5.000000, 2587.758614)(10.000000, 788.010778)(15.000000, 1797.085945)(20.000000, 1136.497051)(25.000000, 624.293986)(30.000000, 1153.592017)(35.000000, 1379.172334)(40.000000, 956.469940)(45.000000, 937.318902)(50.000000, 595.023919)(55.000000, 2937.849837)(60.000000, 4602.726554)(65.000000, 3034.409528)(70.000000, 2140.730421)(75.000000, 704.205898)(80.000000, 1841.614585)(85.000000, 497.289193)(90.000000, 647.831340)
        };
        
        \end{axis}
    \end{tikzpicture}
    \end{center}

    \caption{Edge-length ratios for kaleidoscope $xy$-morphs for $G(300,800)$, increments of 5 degrees.}
    \label{fig:kaleidoscope2}
    
\end{figure}
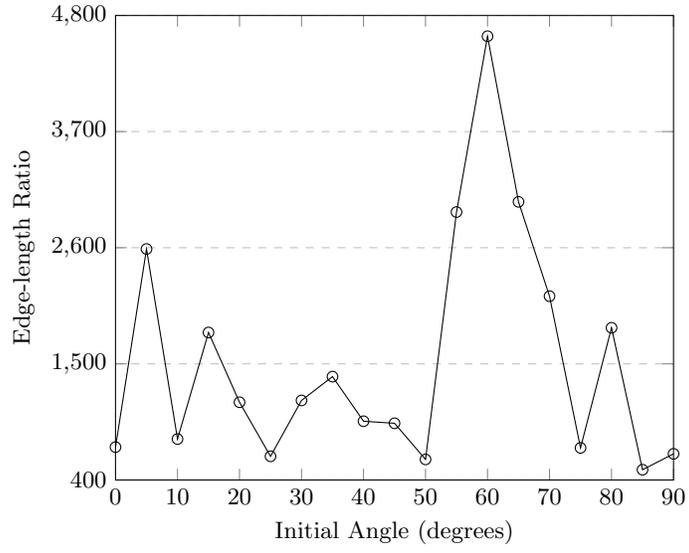

In this appendix, we provide additional drawings in 
Table~\ref{tab:gallery2} and Table~\ref{tab:gallery3}, as well as another kaleidoscope drawing in Figure~\ref{fig:kaleidoscope2} and Table ~\ref{tbl:kal2}.

\begin{table}[htp]
  \caption{Drawing Gallery. $\rho(\Gamma)$ is the edge-length ratio, $r$ is the scaling parameter.} 
\label{tab:gallery2}
  \begin{tabular}
        {cccccc} \hline & Tutte & $x$-spread & $y$-spread & $xy$-morph & BFS-spread \\
        \hline 
    
        \multirow{-10}{*}{\rotatebox[origin=c]{90}{$G(100,294)$}} &
        \includegraphics[width=.175\linewidth]{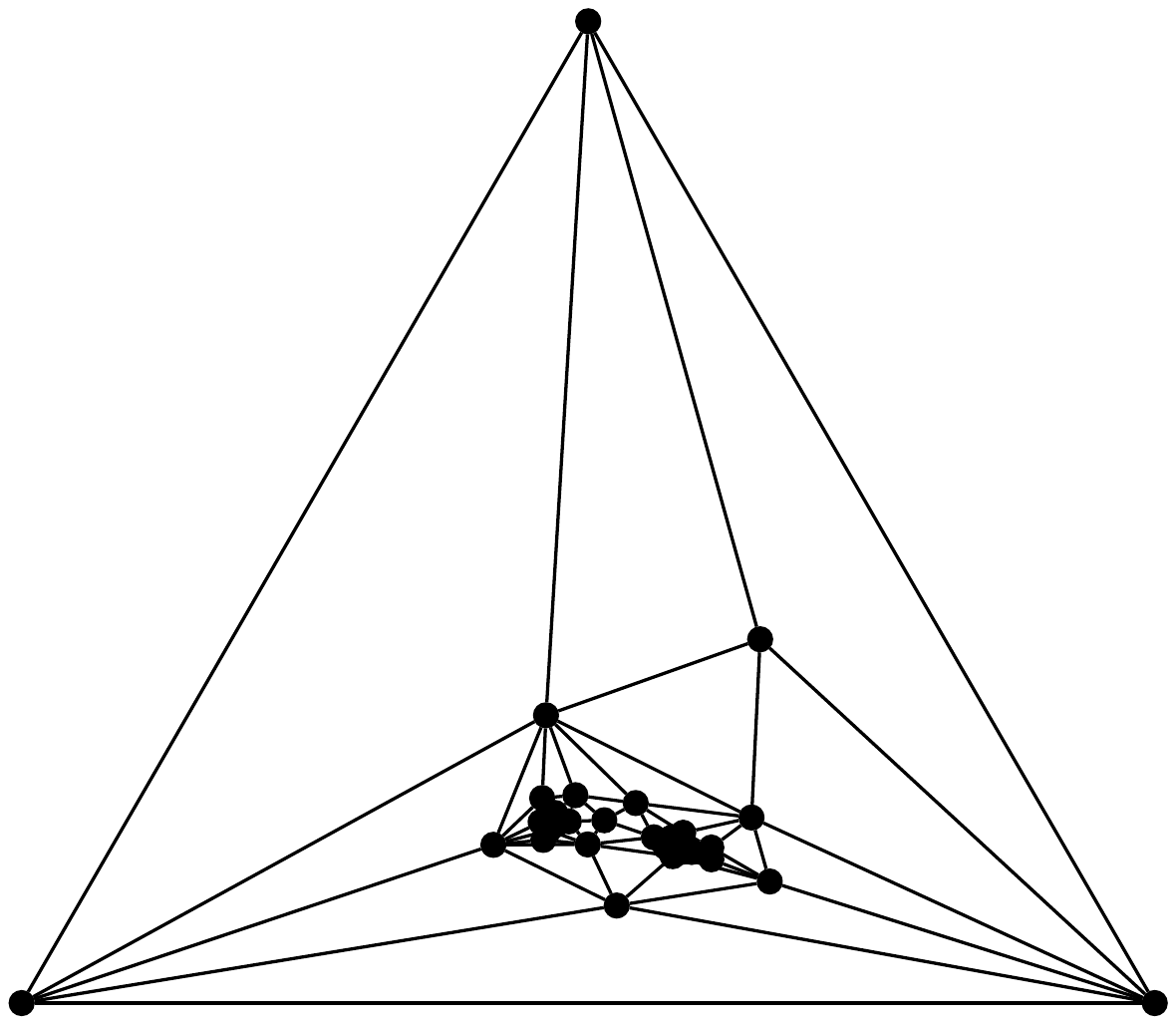} &
        \includegraphics[width=.175\linewidth]{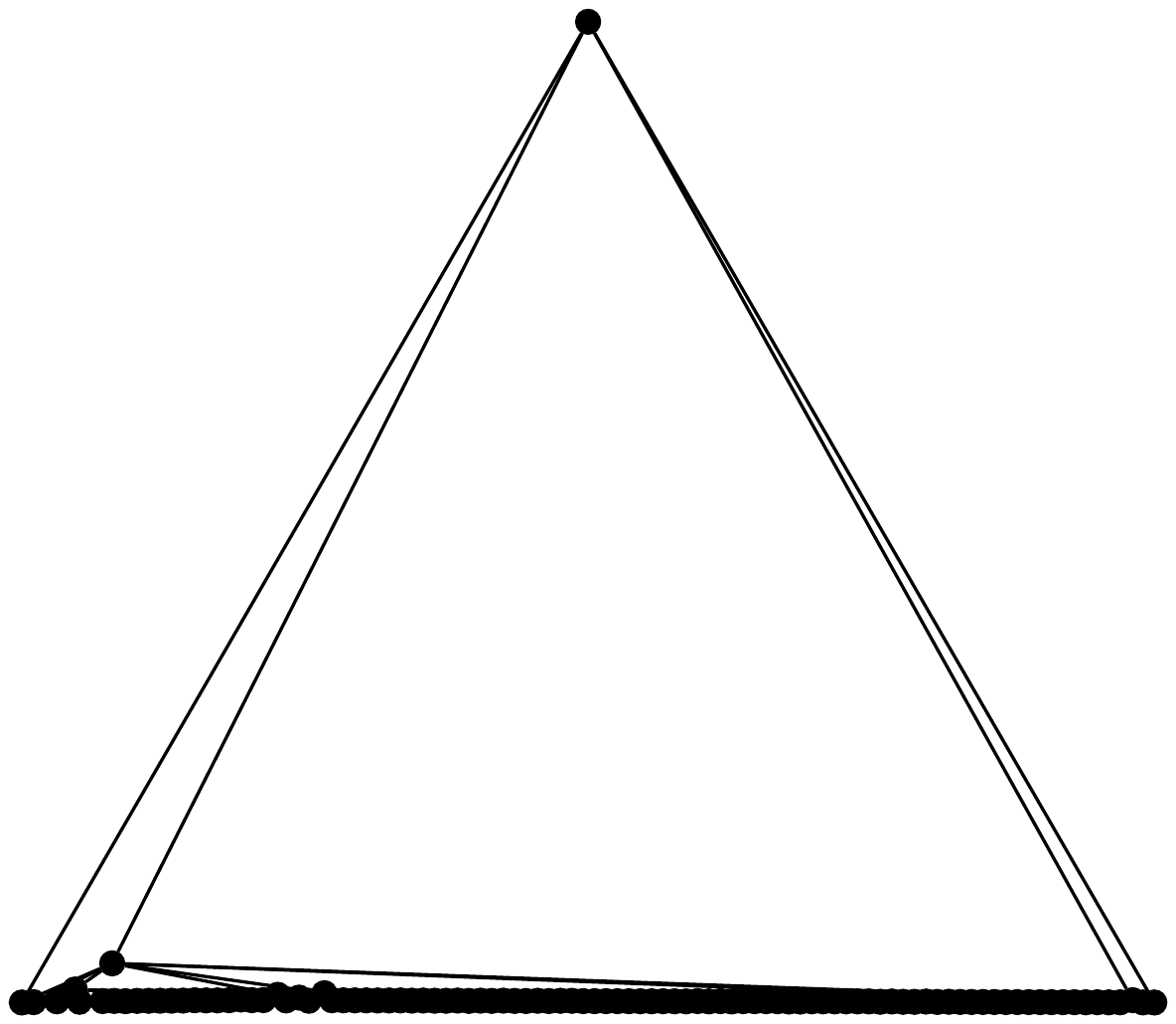} &  
        \includegraphics[width=.175\linewidth]{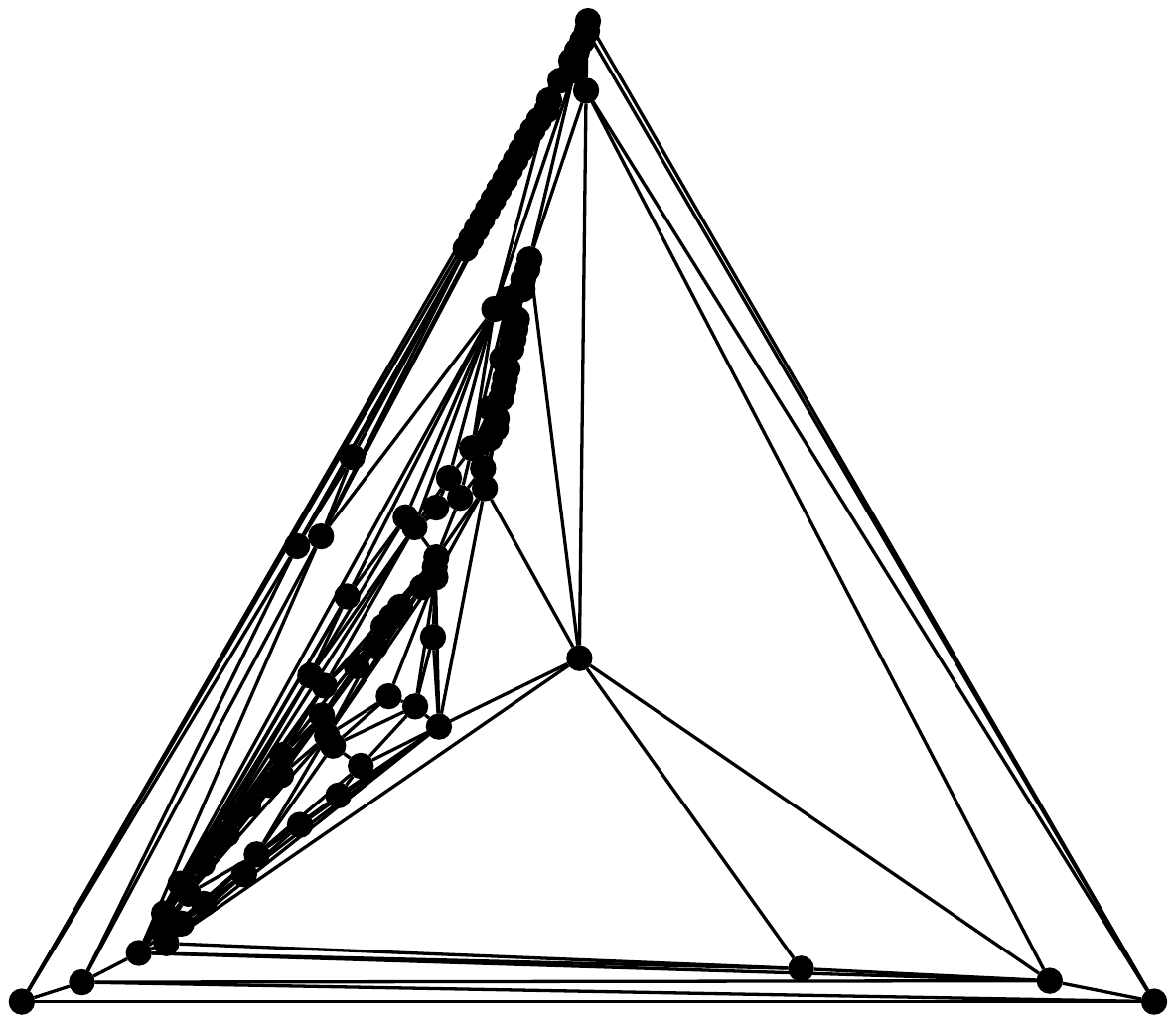} & 
        \includegraphics[width=.175\linewidth]{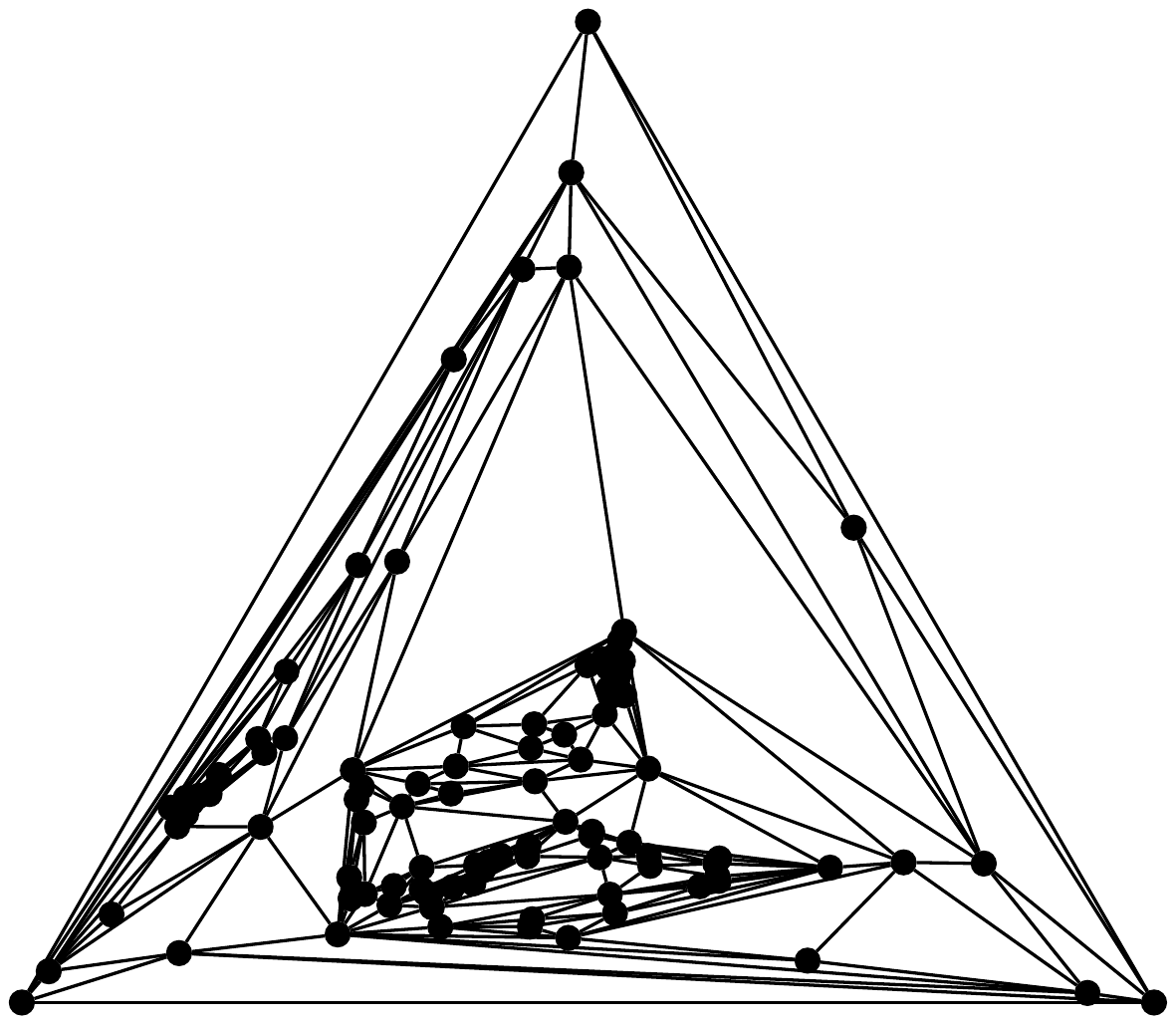} & 
        \includegraphics[width=.175\linewidth]{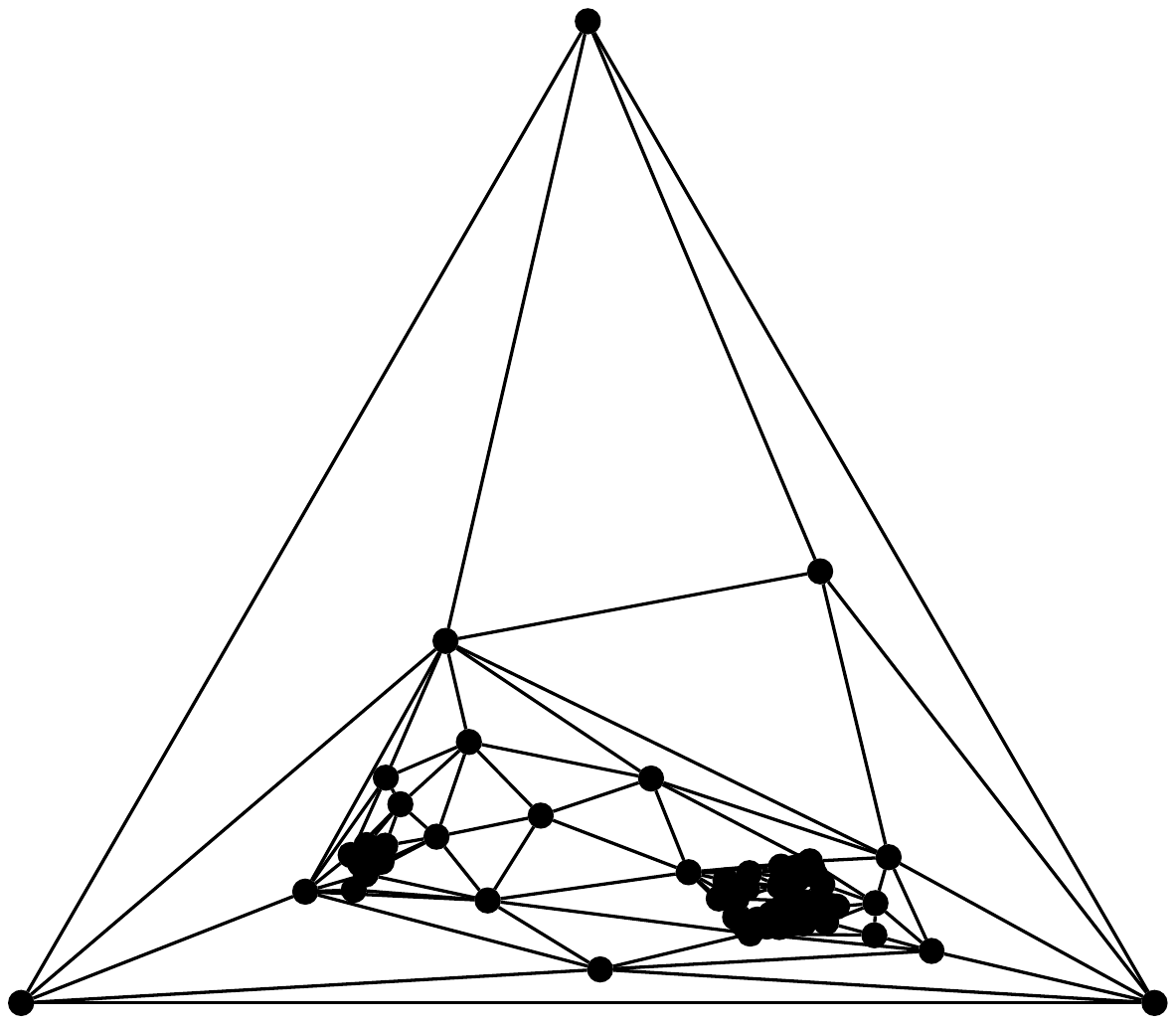} \\
        \hline
        & $\rho(\Gamma) =3016$ & $\rho(\Gamma) =110$ & $\rho(\Gamma) =114$ & $\rho(\Gamma) =285$ & $\rho(\Gamma) = 1397, r=3$ \\
        \hline
        \multirow{-10}{*}{\rotatebox[origin=c]{90}{$G(80,232)$}} &
        \includegraphics[width=.175\linewidth]{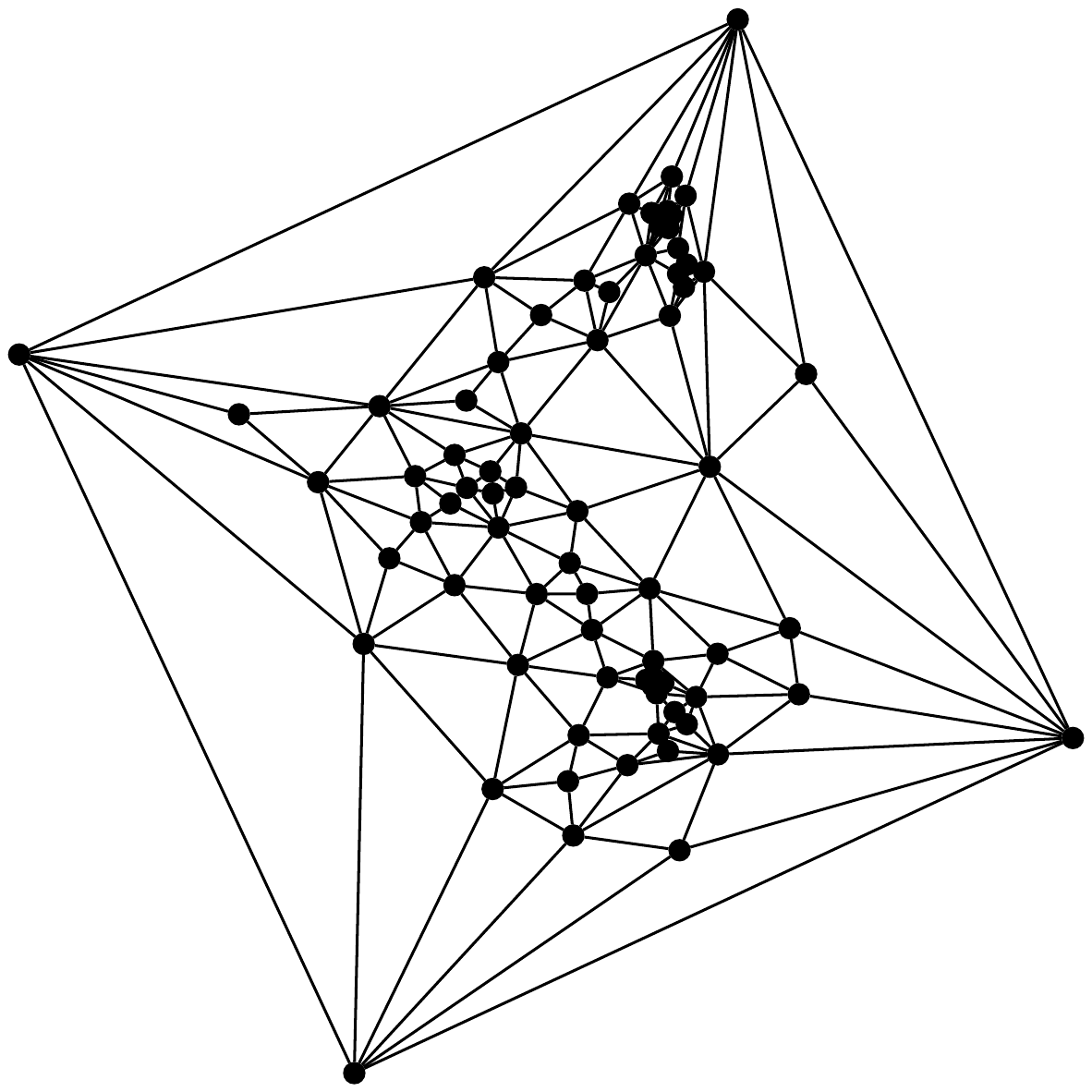} & 
        \includegraphics[width=.175\linewidth]{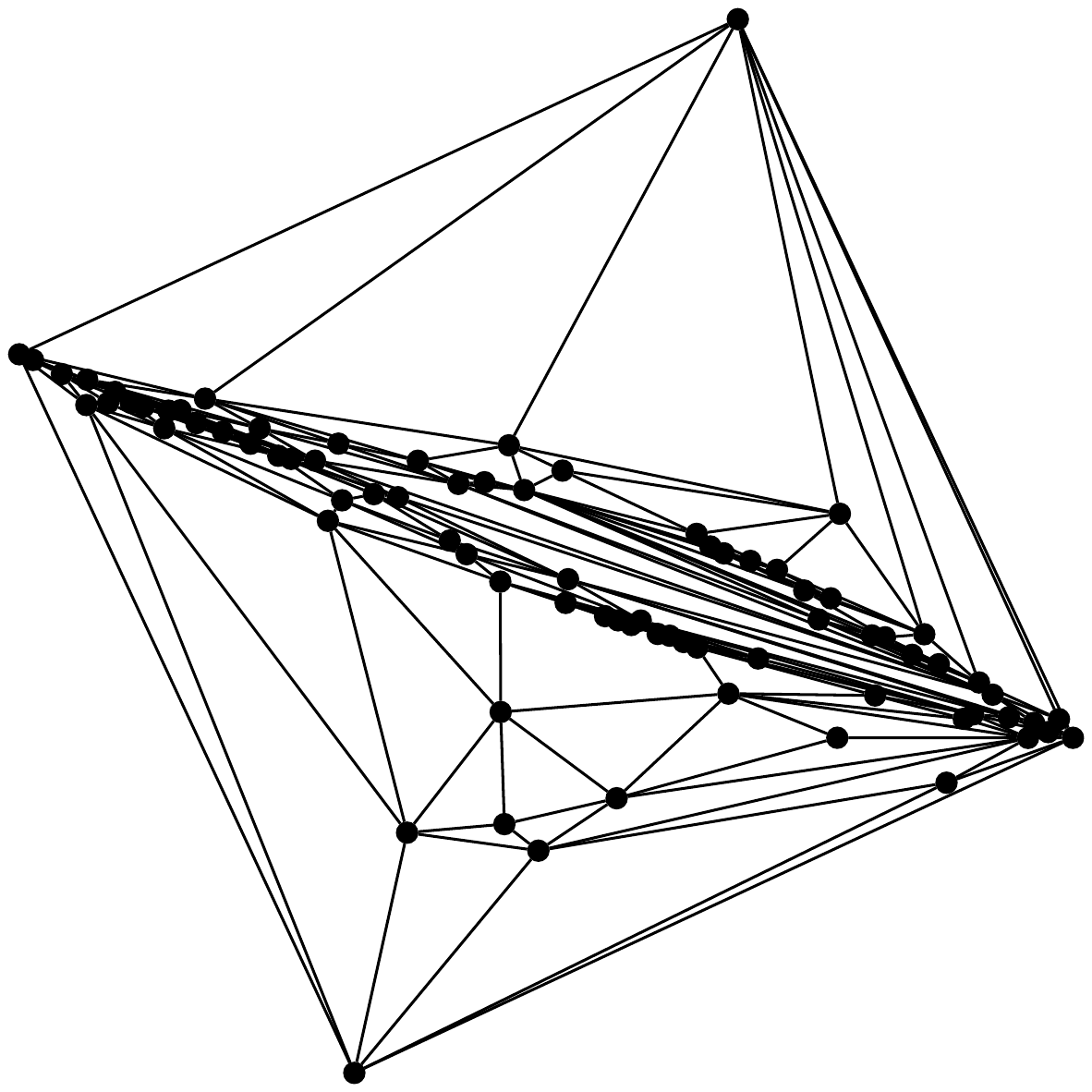} &  
        \includegraphics[width=.175\linewidth]{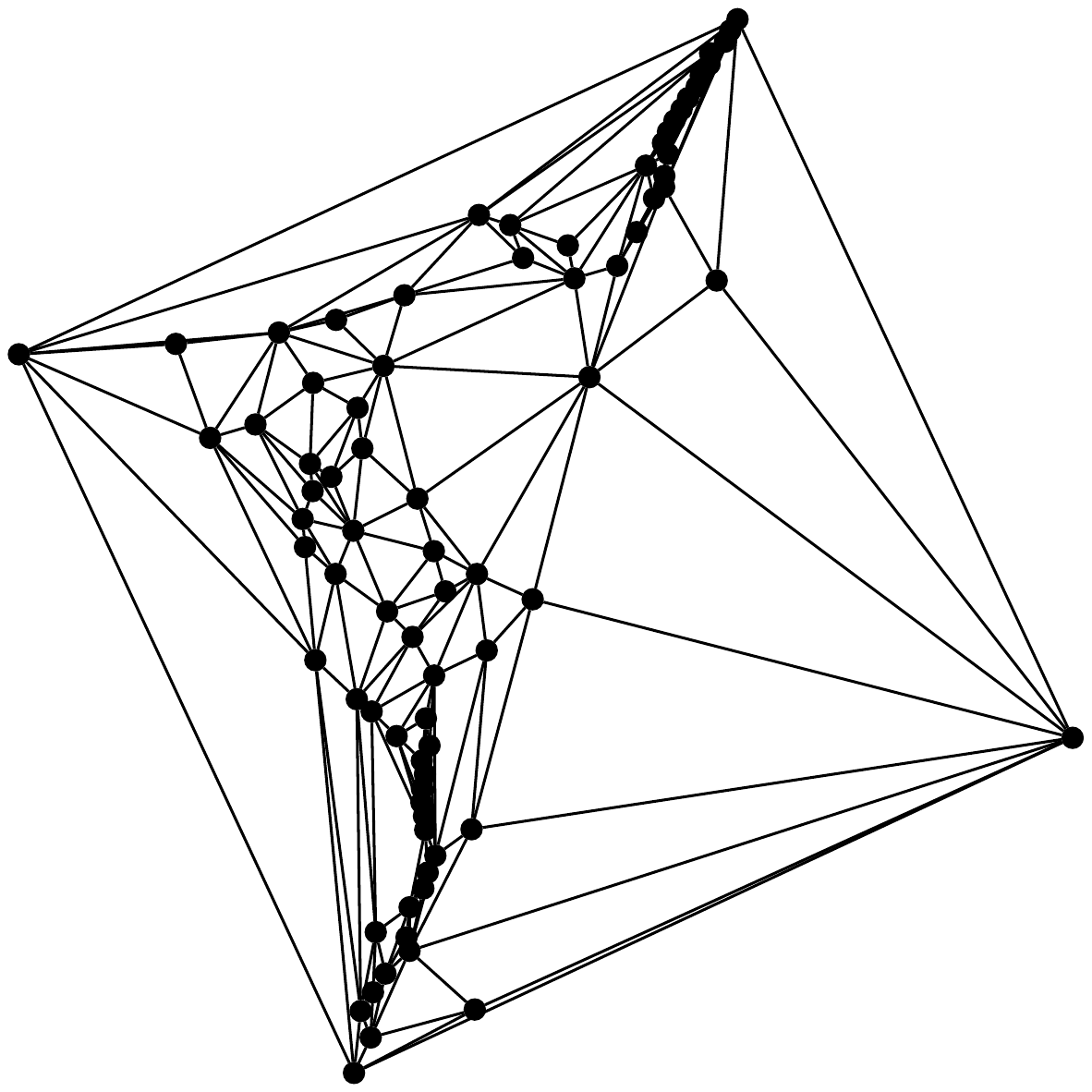} & 
        \includegraphics[width=.175\linewidth]{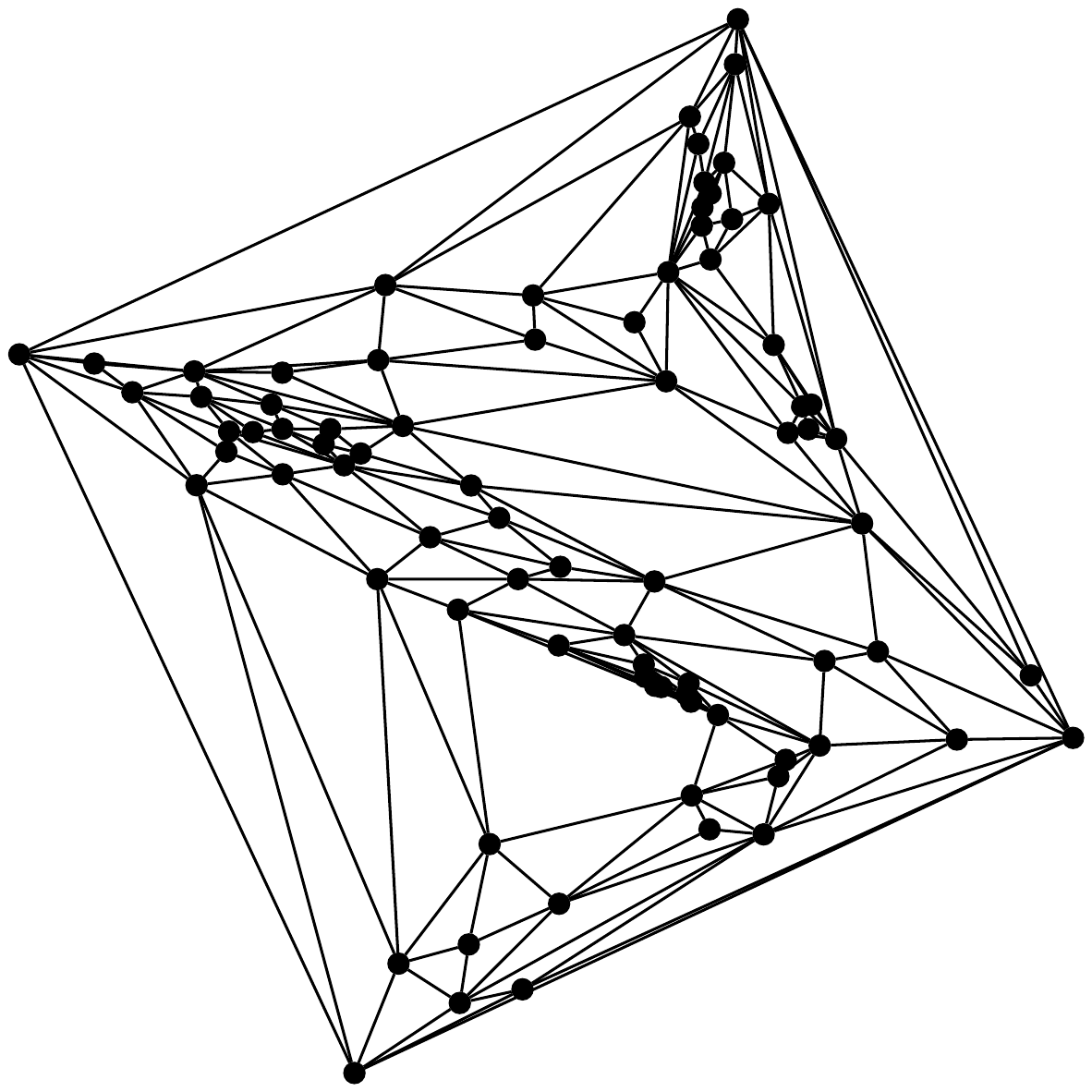} & 
        \includegraphics[width=.175\linewidth]{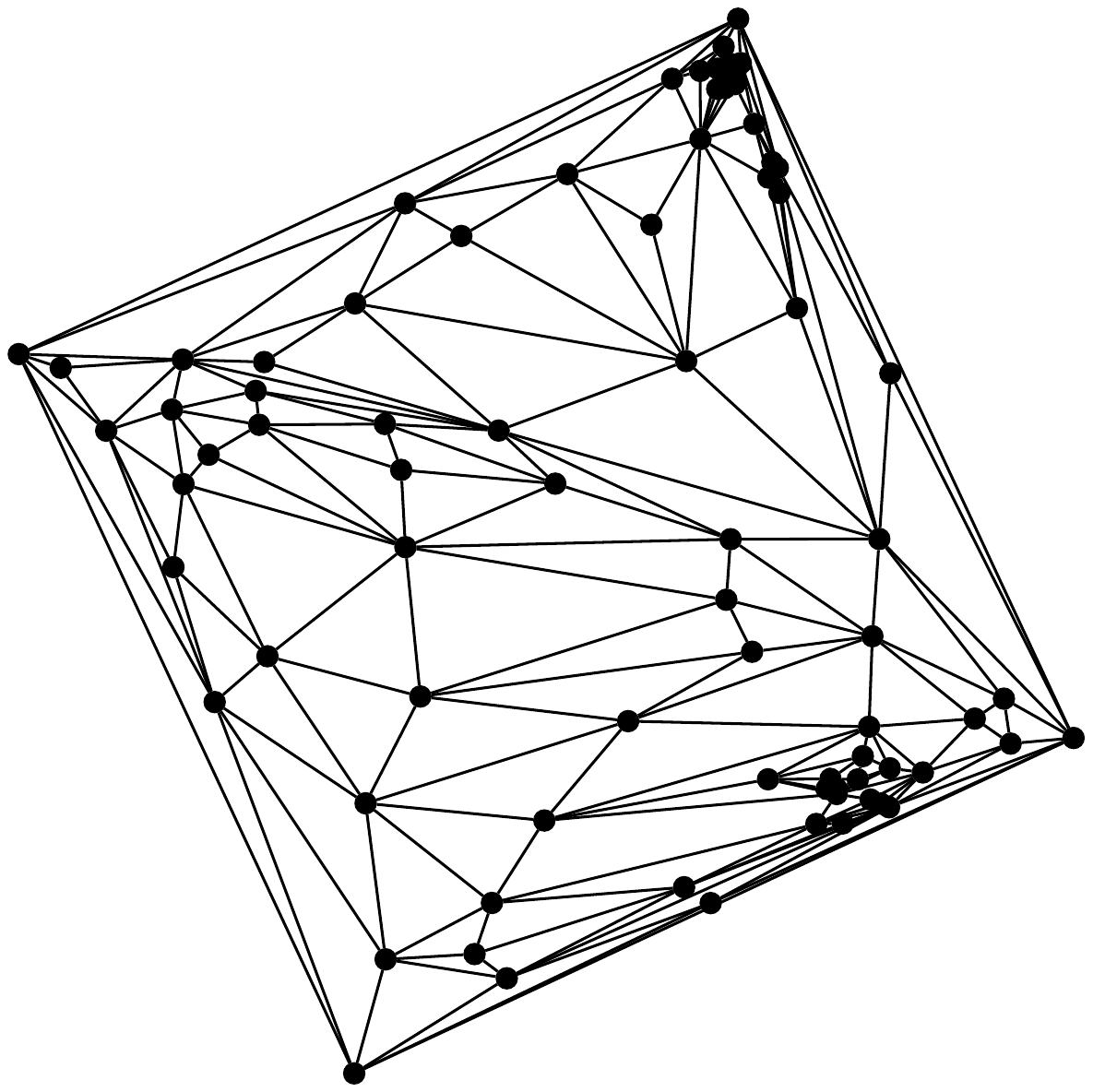} \\
        \hline
        & $\rho(\Gamma) =1013$ & $\rho(\Gamma) =77$ & $\rho(\Gamma) =72$ & $\rho(\Gamma) =126$ & $\rho(\Gamma) = 160, r=4$ \\
        \hline
        \multirow{-10}{*}{\rotatebox[origin=c]{90}{$G(58,136)$}} &
        \includegraphics[width=.175\linewidth]{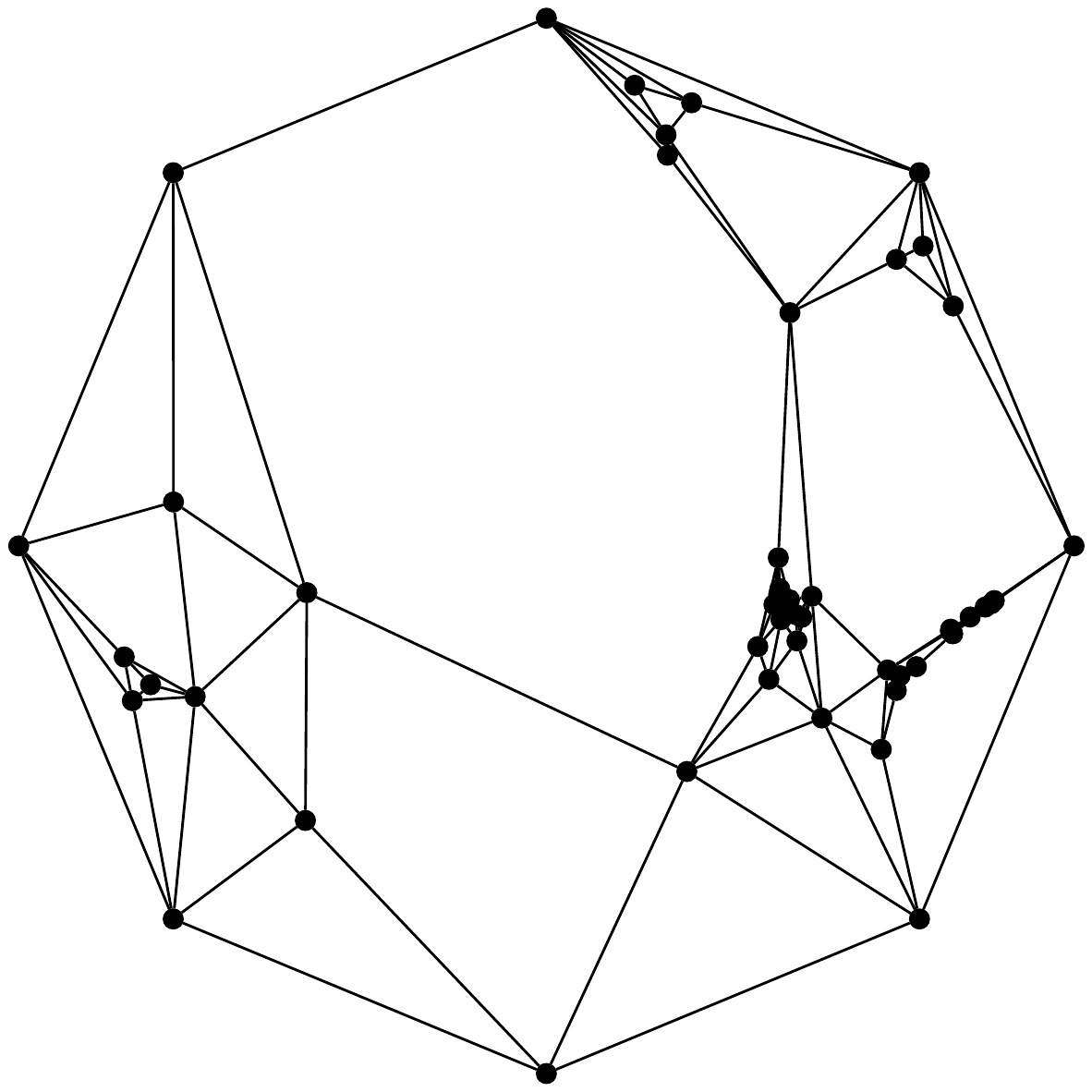} & 
        \includegraphics[width=.175\linewidth]{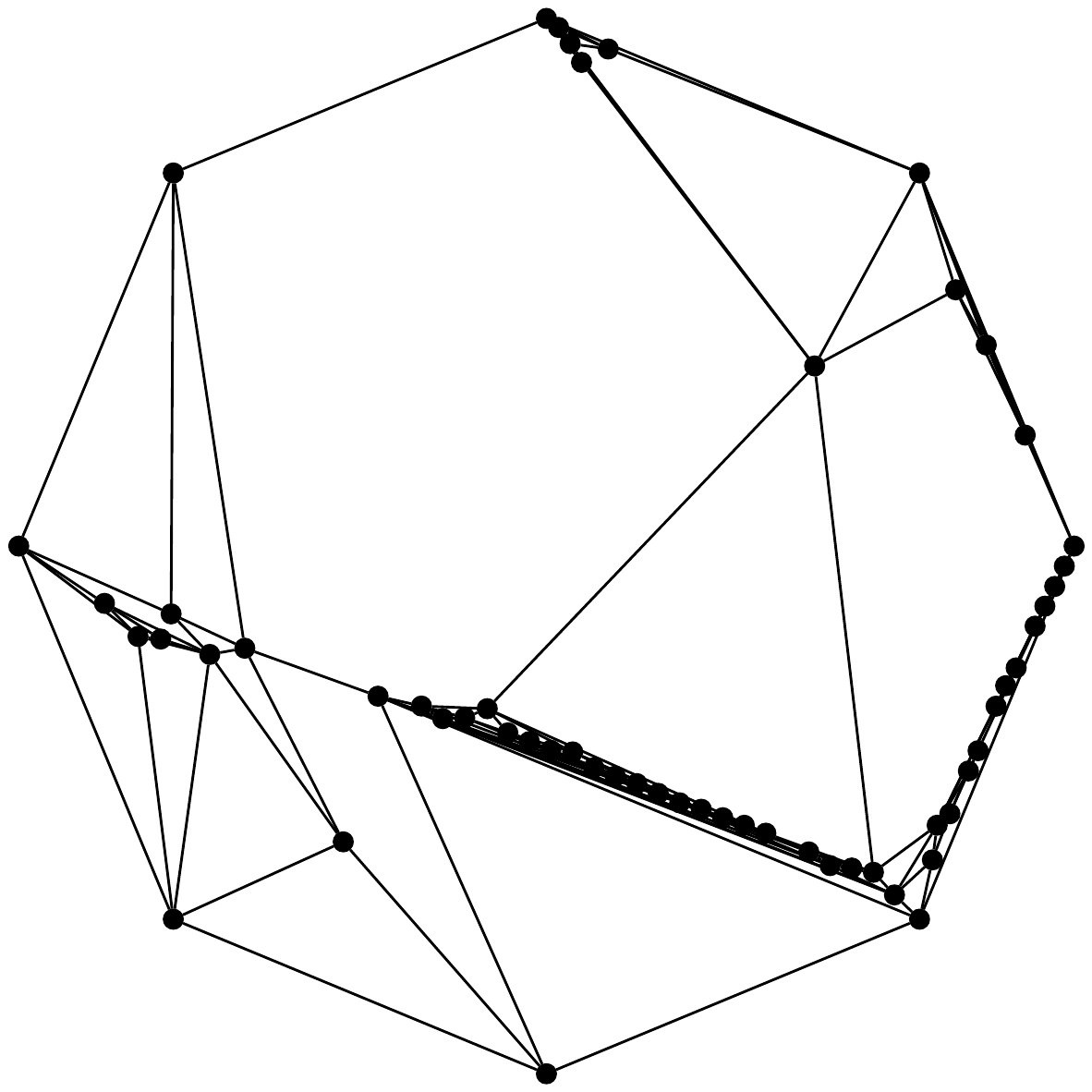} &  
        \includegraphics[width=.175\linewidth]{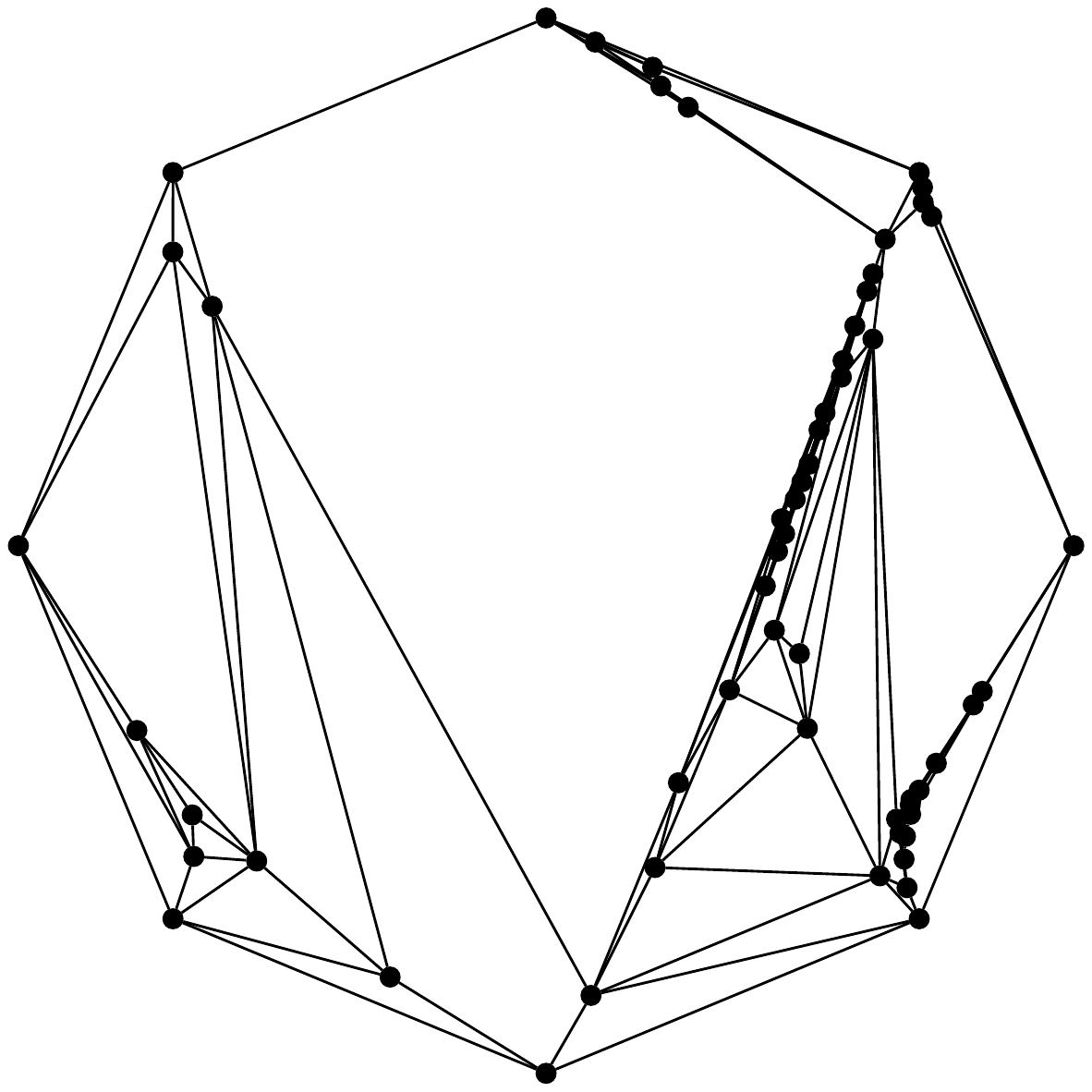} & 
        \includegraphics[width=.175\linewidth]{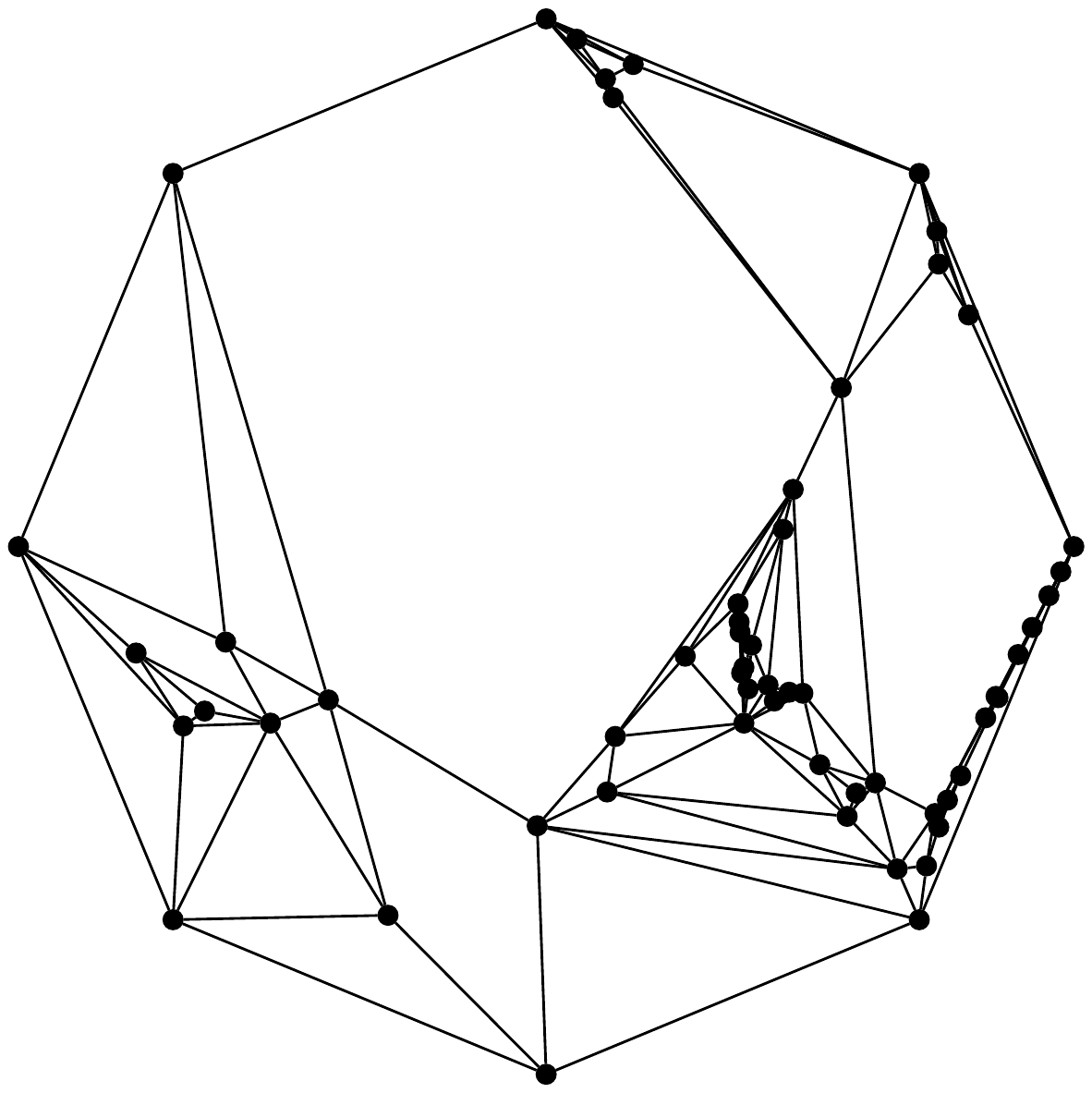} & 
        \includegraphics[width=.175\linewidth]{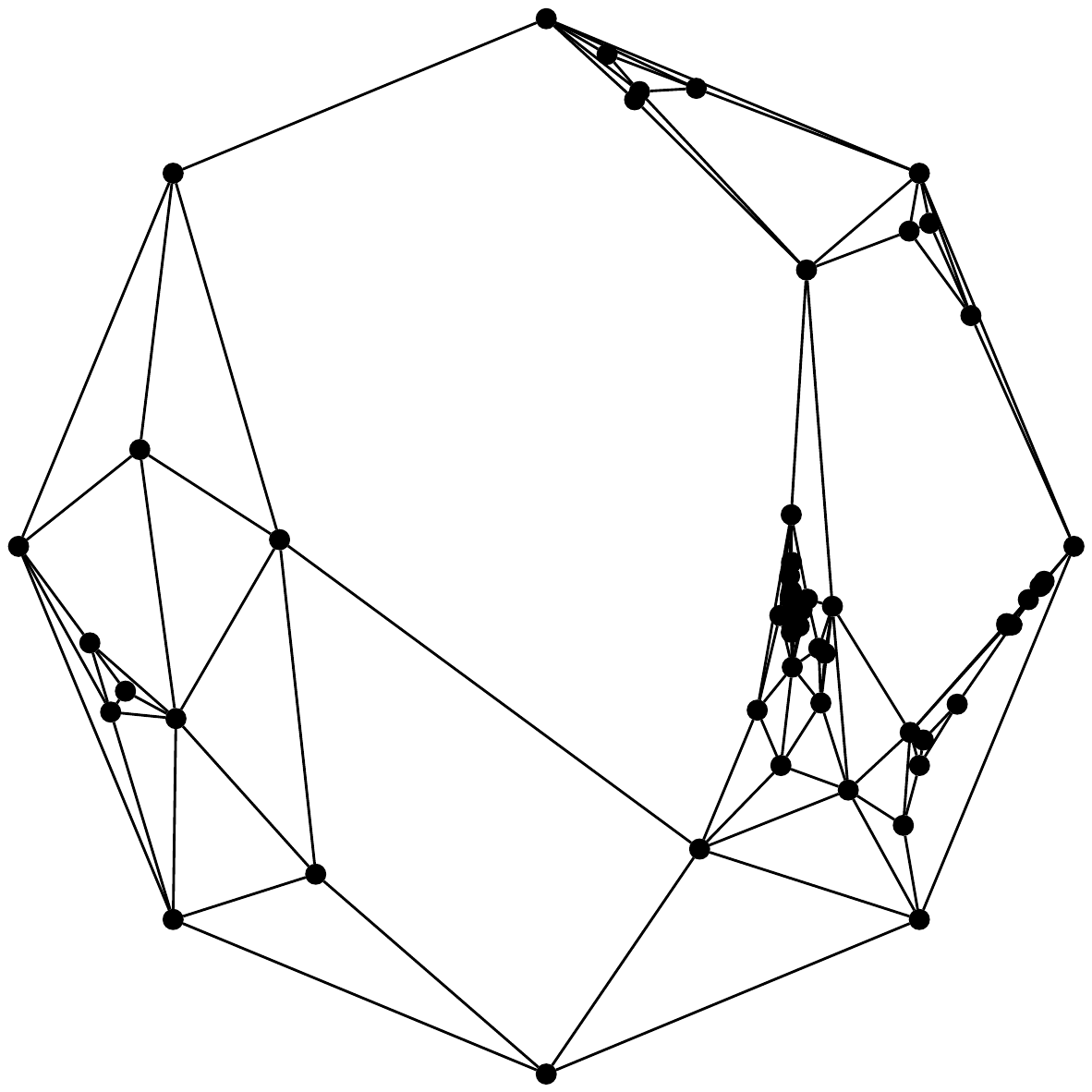} \\
        \hline
        & $\rho(\Gamma) =648$ & $\rho(\Gamma) =39$ & $\rho(\Gamma) =122$ & $\rho(\Gamma) =53$ & $\rho(\Gamma) = 236, r=2$ \\
        \hline
        \multirow{-10}{*}{\rotatebox[origin=c]{90}{$G(300,450)$}} &
        \includegraphics[width=.175\linewidth]{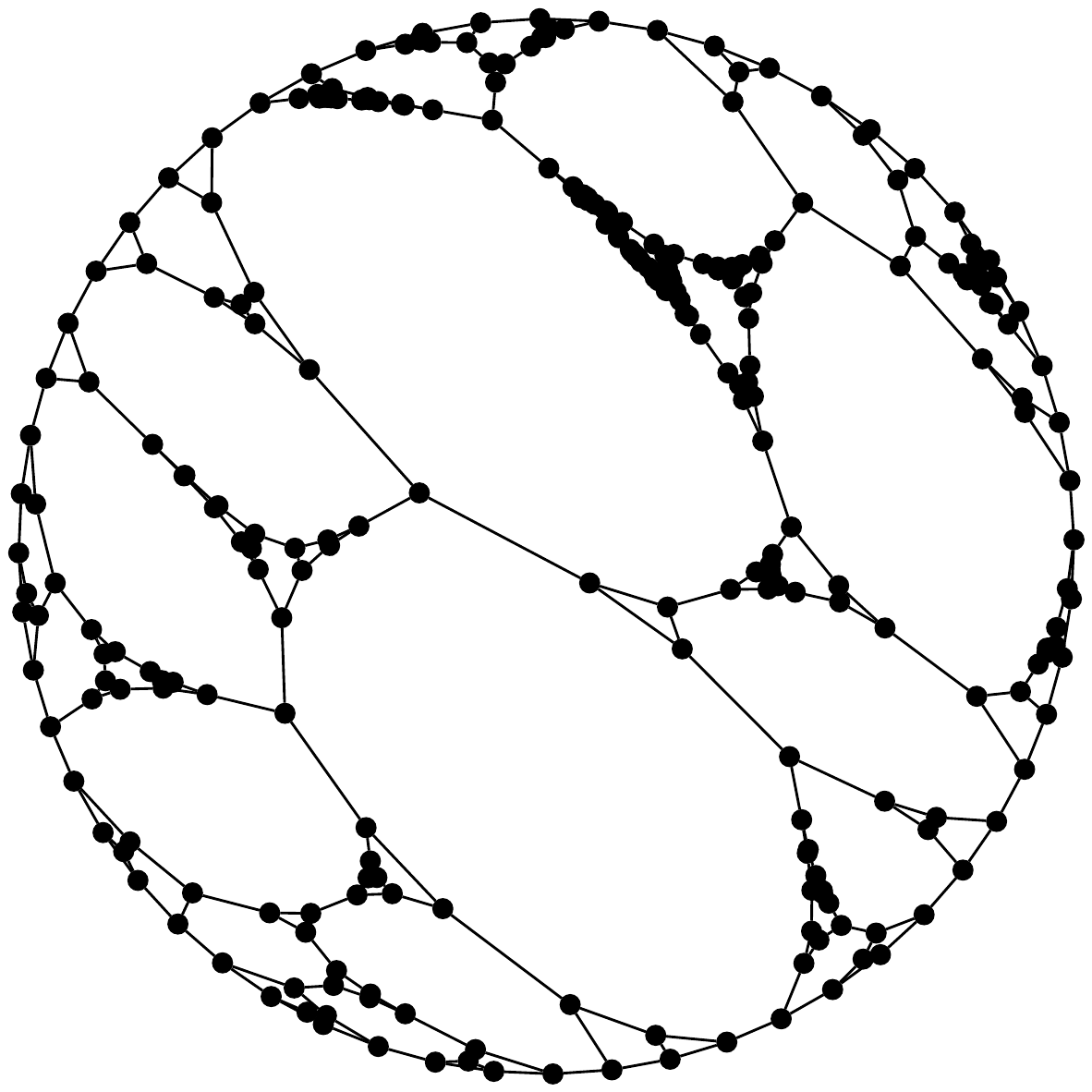} & 
        \includegraphics[width=.175\linewidth]{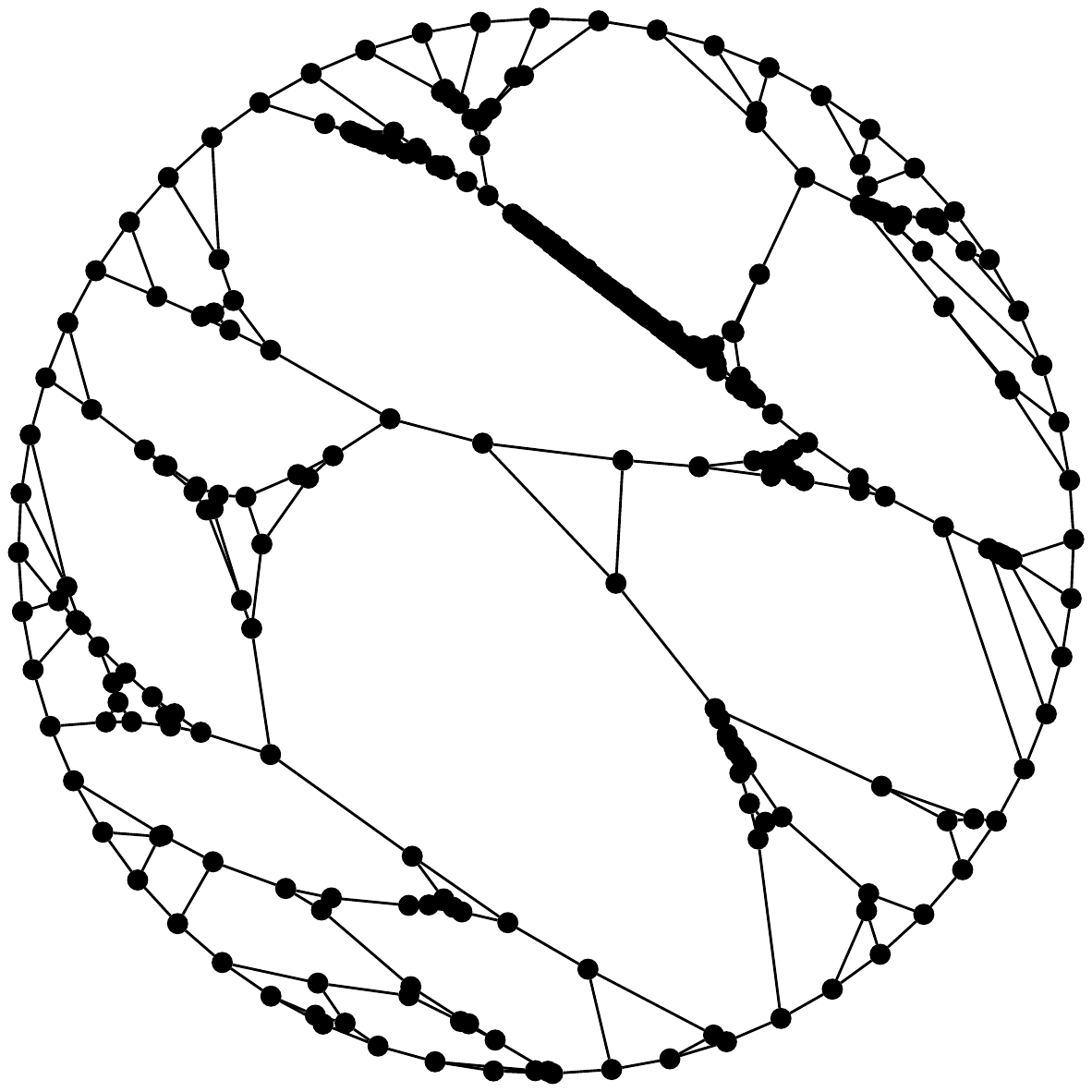} &  
        \includegraphics[width=.175\linewidth]{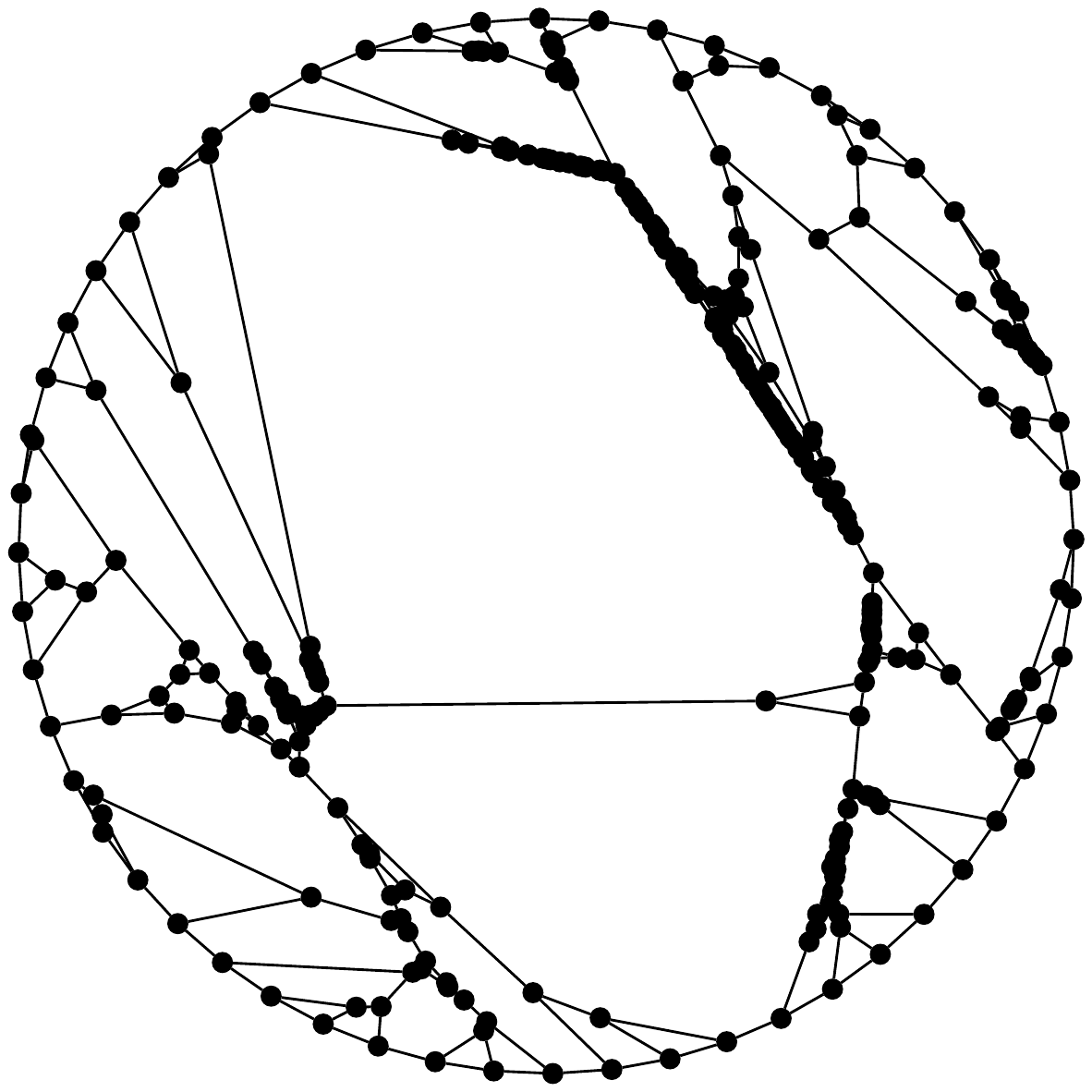} & 
        \includegraphics[width=.175\linewidth]{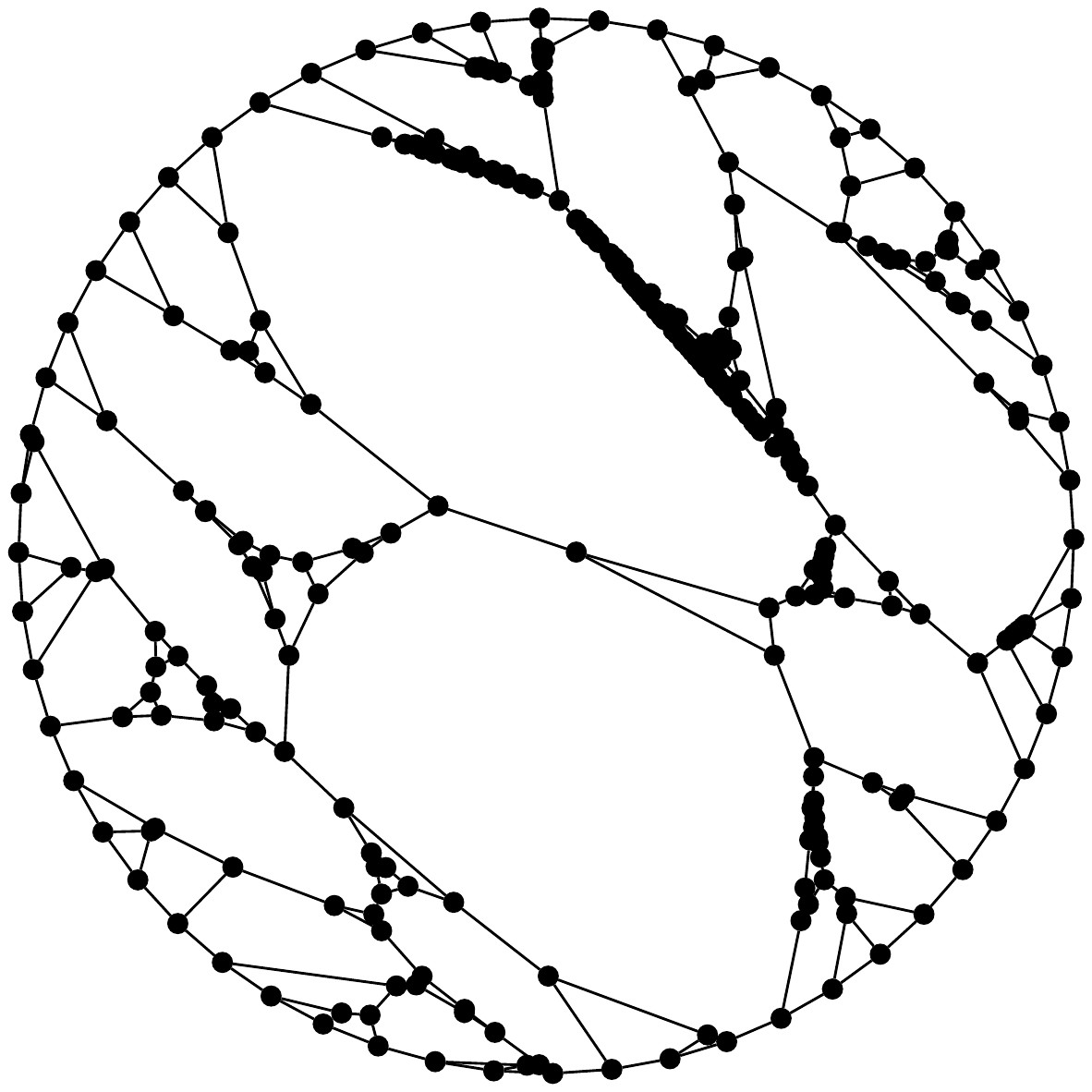} & 
        \includegraphics[width=.175\linewidth]{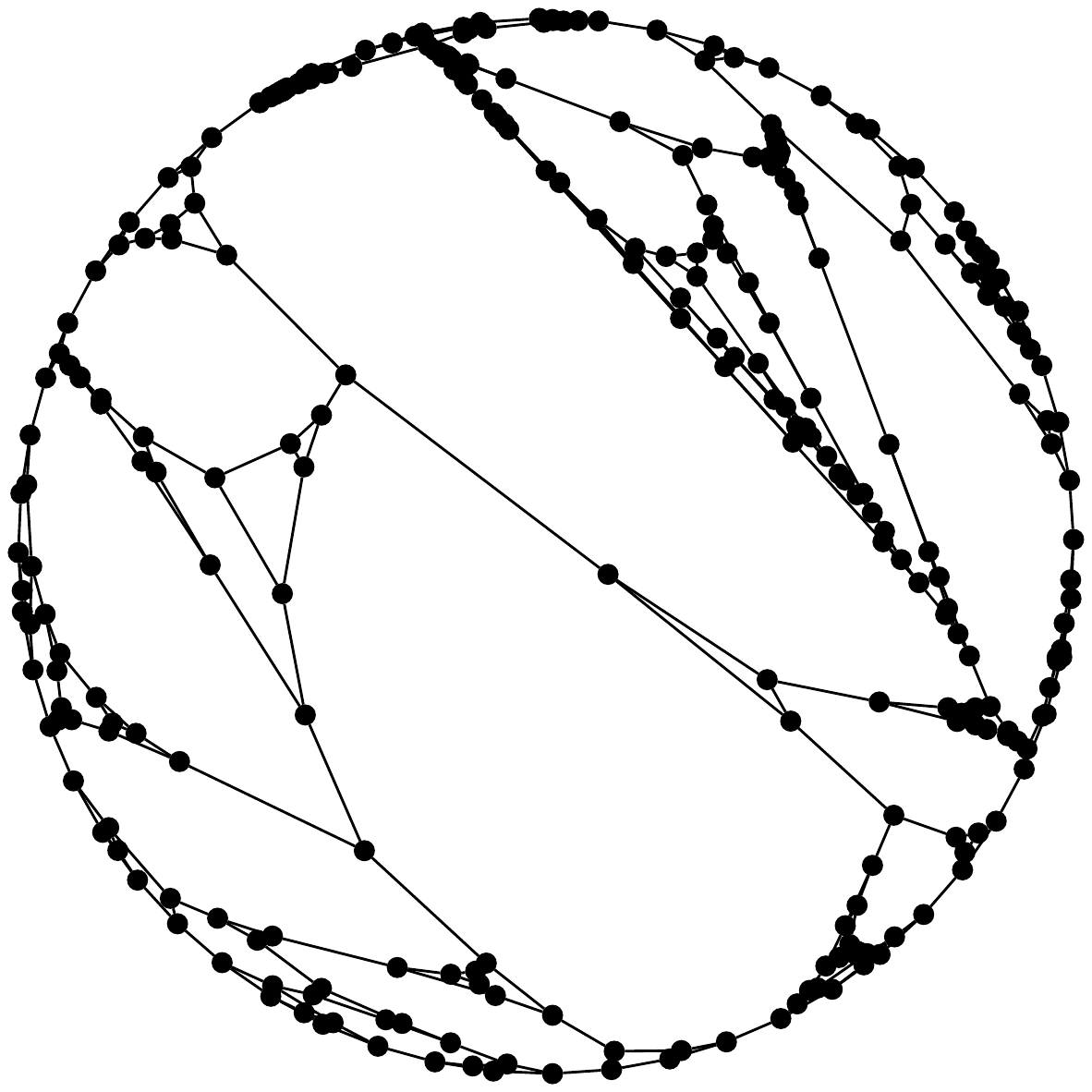} \\
        \hline
        & $\rho(\Gamma) =2599$ & $\rho(\Gamma) =244$  & $\rho(\Gamma) =1814$ & $\rho(\Gamma) =214$ & $\rho(\Gamma) = 1874, r=2$ \\
        \hline
        \multirow{-10}{*}{\rotatebox[origin=c]{90}{$G(444,1111)$}} &
        \includegraphics[width=.175\linewidth]{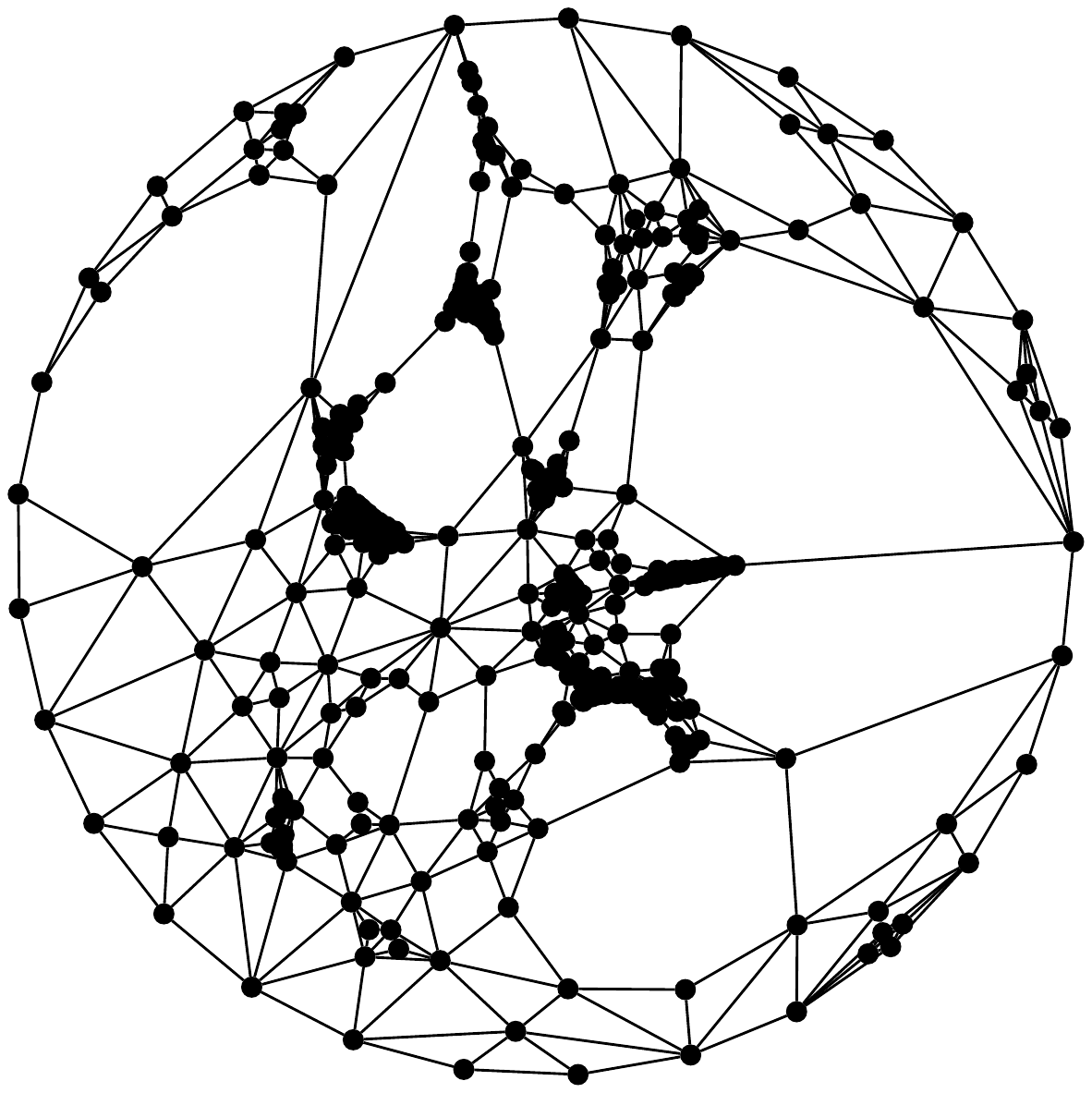} & 
        \includegraphics[width=.175\linewidth]{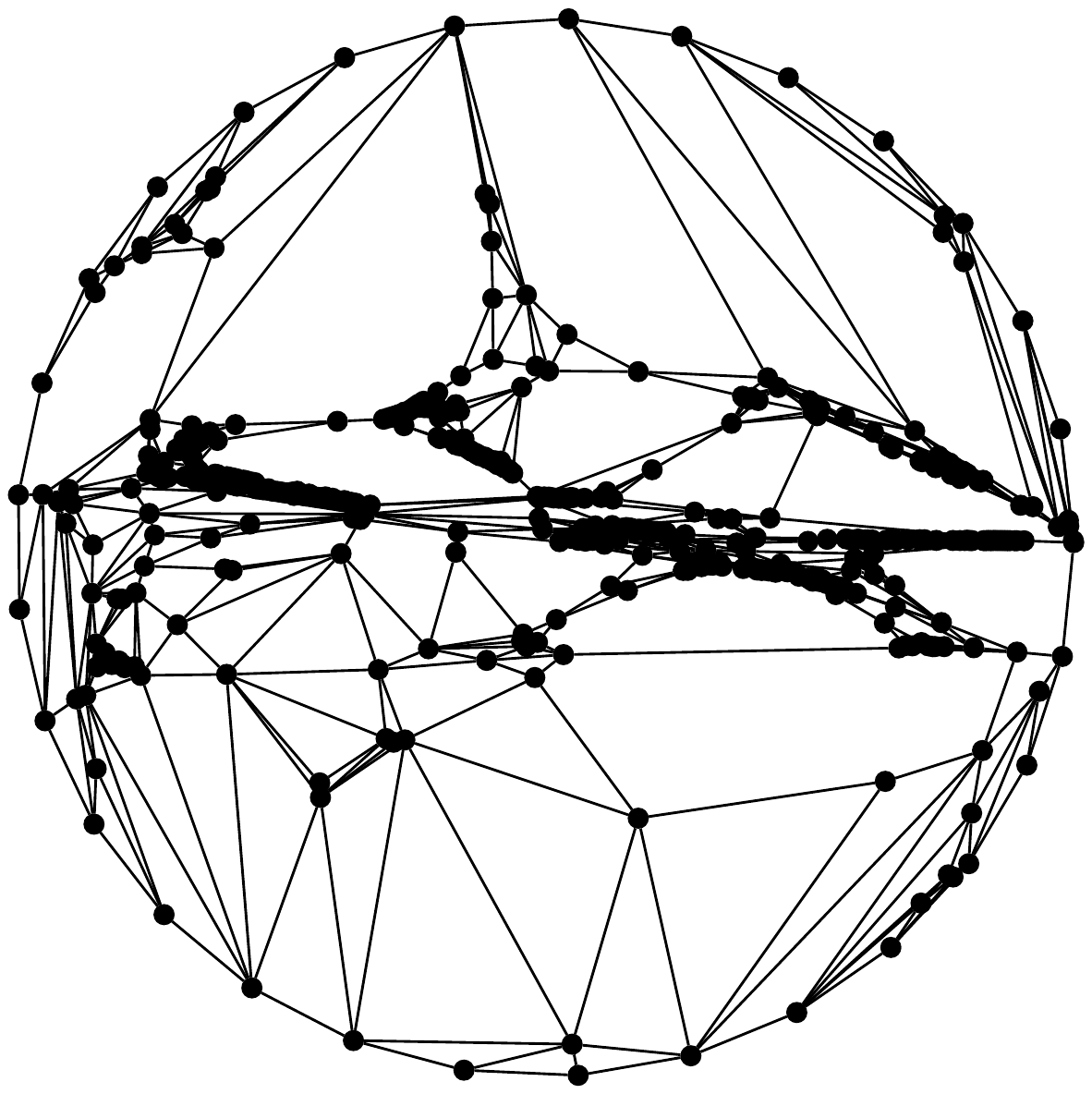} &  
        \includegraphics[width=.175\linewidth]{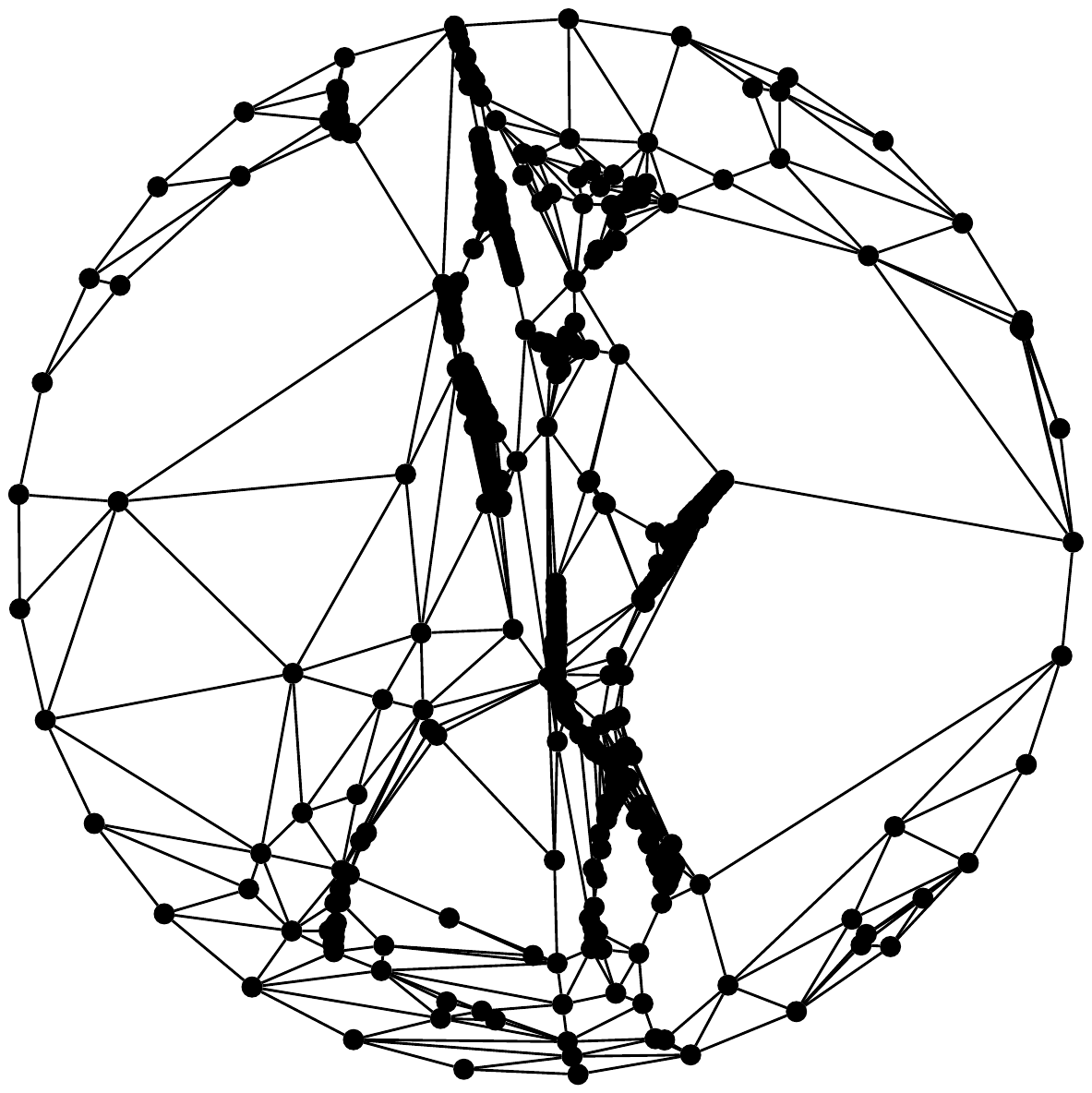} & 
        \includegraphics[width=.175\linewidth]{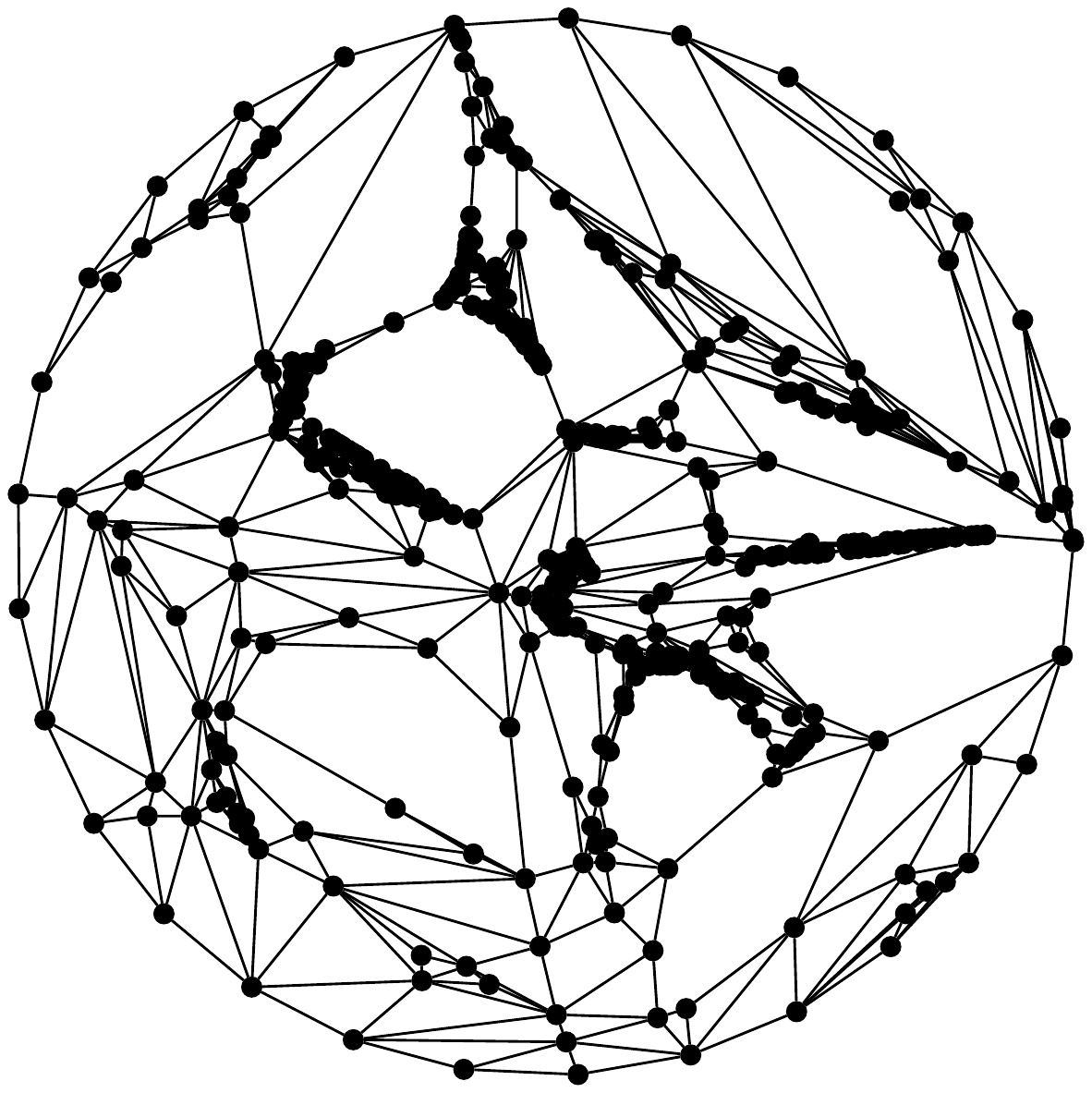} & 
        \includegraphics[width=.175\linewidth]{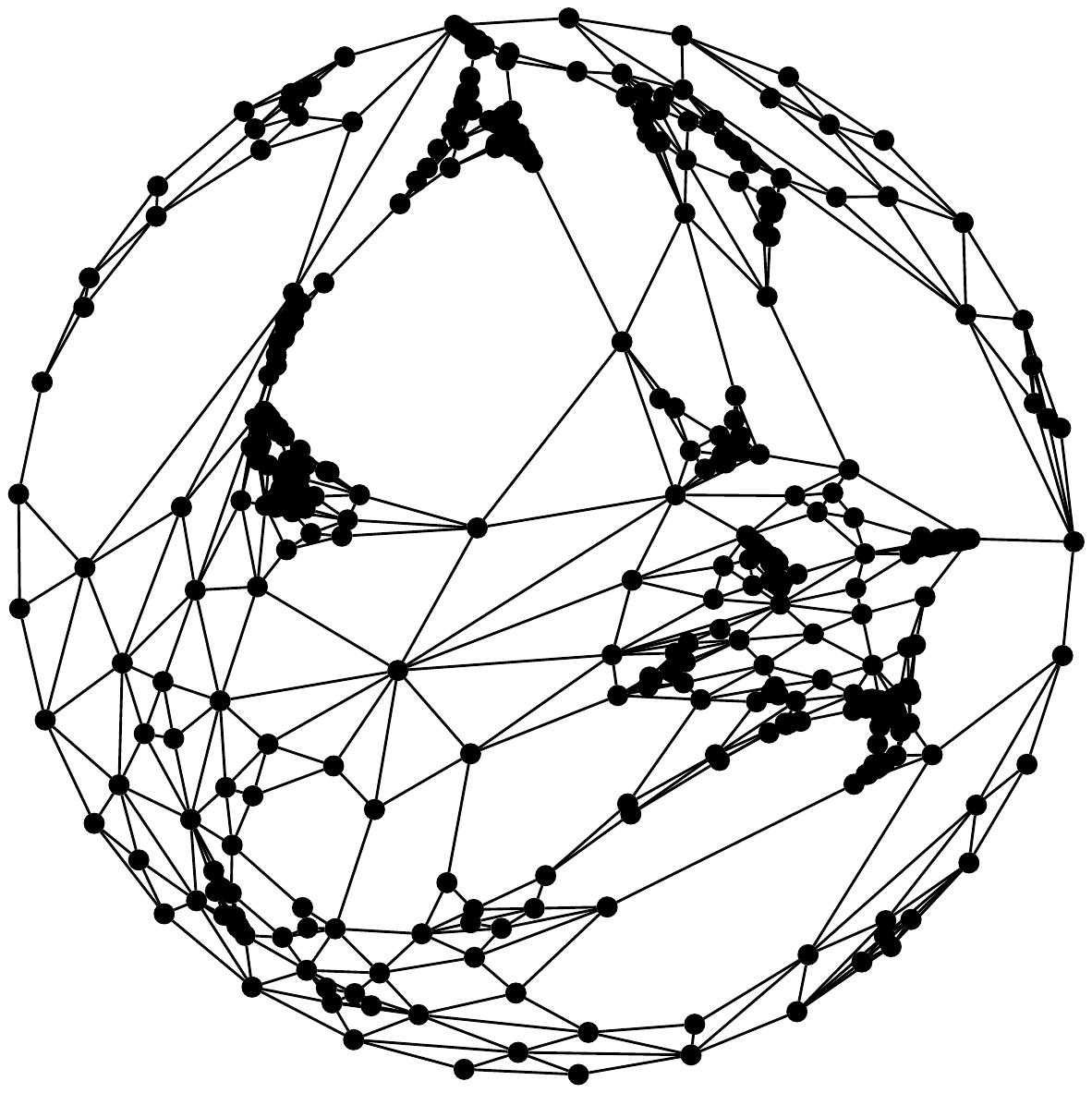} \\
        \hline
        & $\rho(\Gamma) =7774$ & $\rho(\Gamma) =280$ & $\rho(\Gamma) =623$ & $\rho(\Gamma) =542$ & $\rho(\Gamma) = 2557, r=2$ \\
        \hline
        
  \end{tabular}
\end{table}

\begin{table}[htp]
  \caption{Drawing Gallery. $\rho(\Gamma)$ is edge-length ratio, $r$ is the scaling parameter.} 
\label{tab:gallery3}
  \begin{tabular}
        {cccccc} \hline & Tutte & $x$-spread & $y$-spread & $xy$-morph & BFS-spread \\
        \hline 
        \multirow{-10}{*}{\rotatebox[origin=c]{90}{$G(390,1112)$}} &
        \includegraphics[width=.175\linewidth]{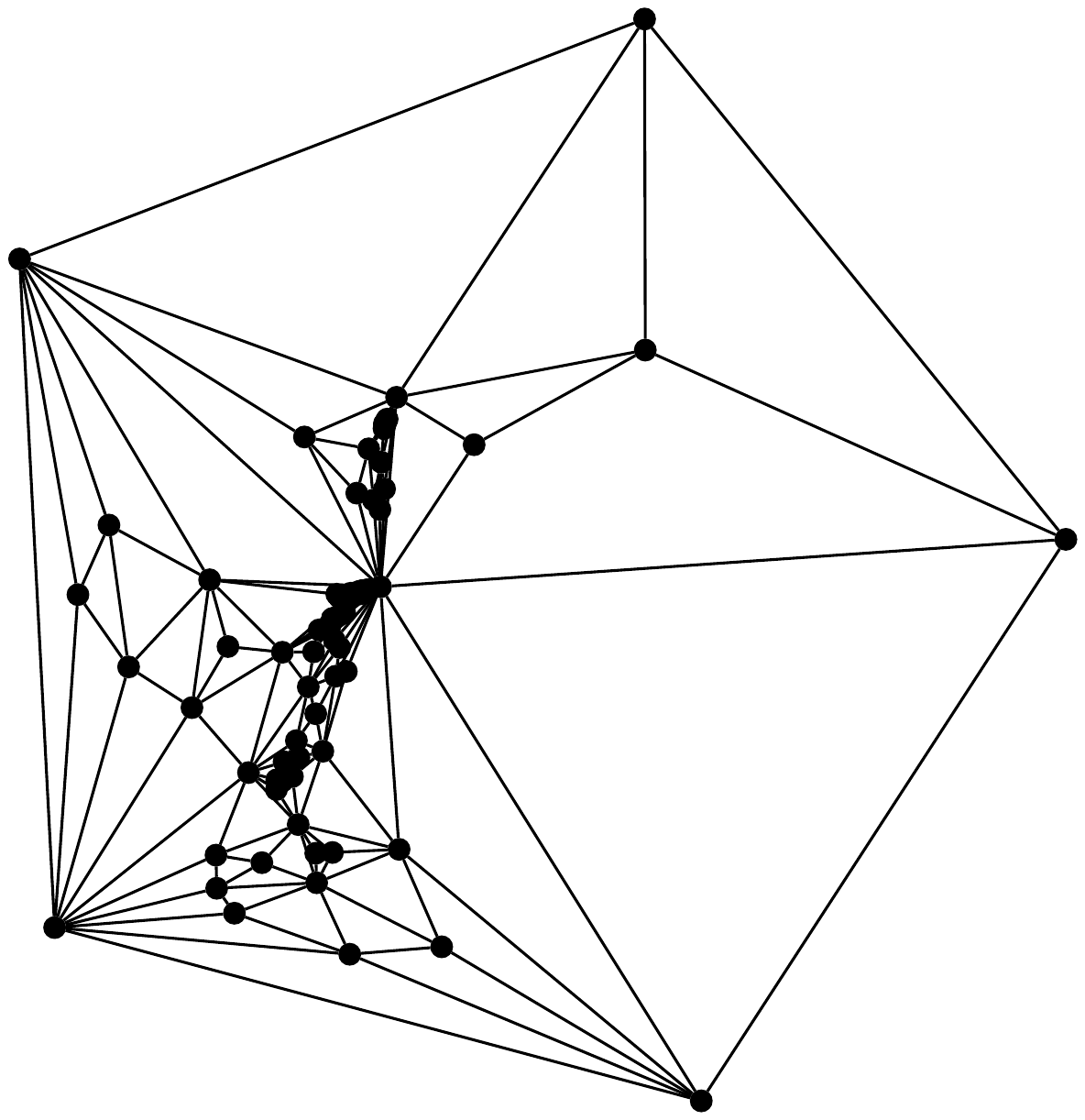} &
        \includegraphics[width=.175\linewidth]{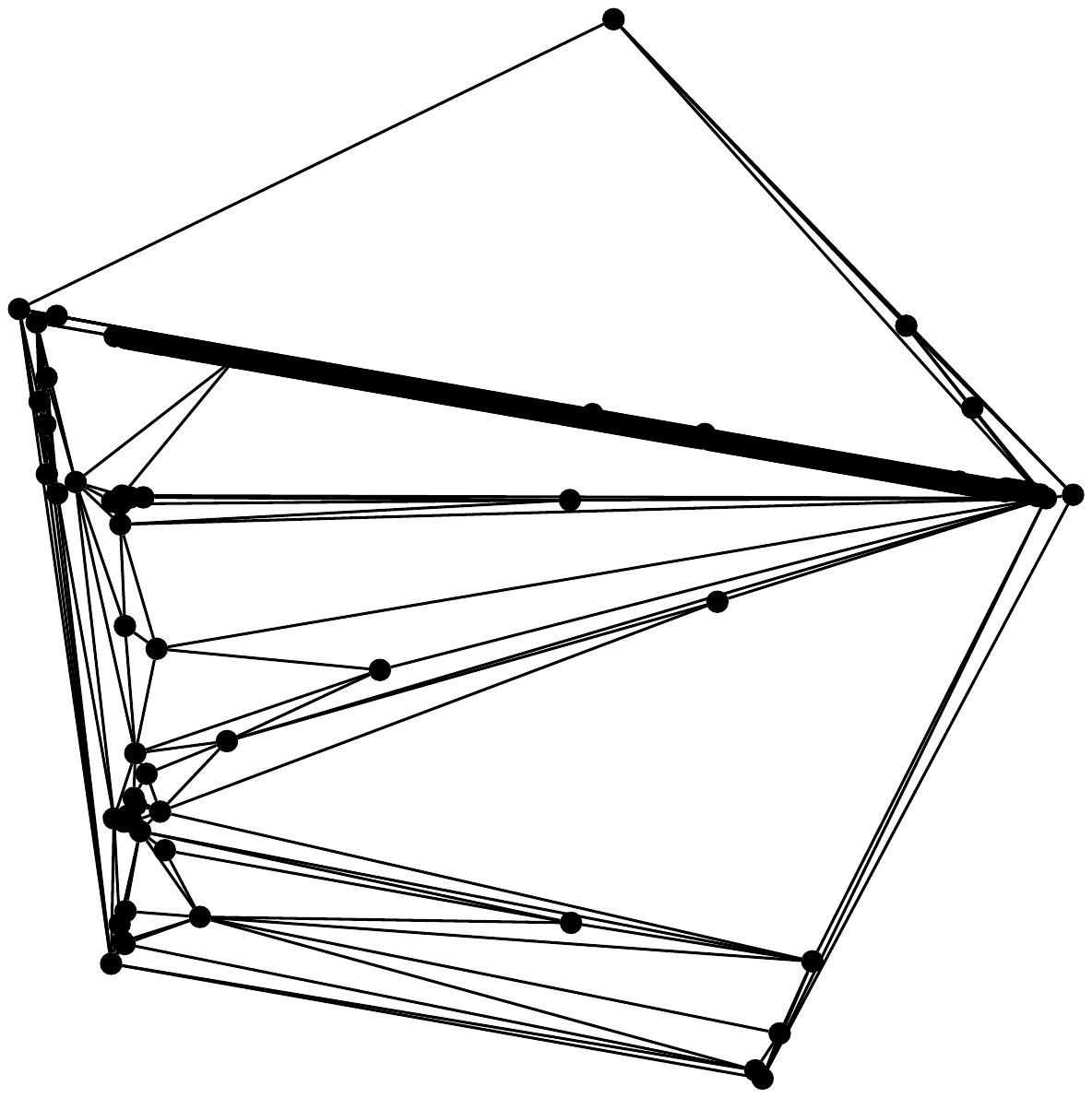} &  
        \includegraphics[width=.175\linewidth]{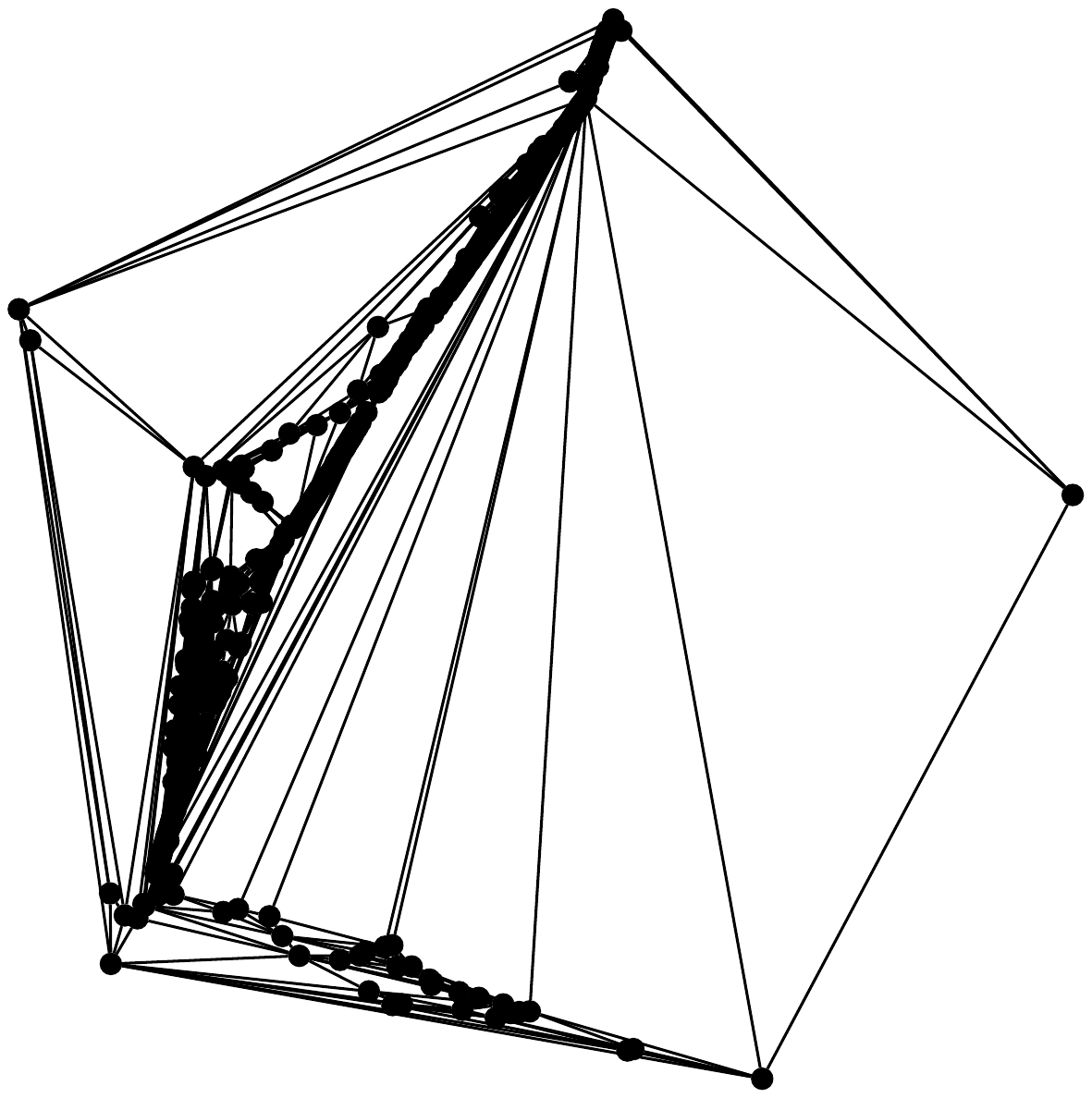} & 
        \includegraphics[width=.175\linewidth]{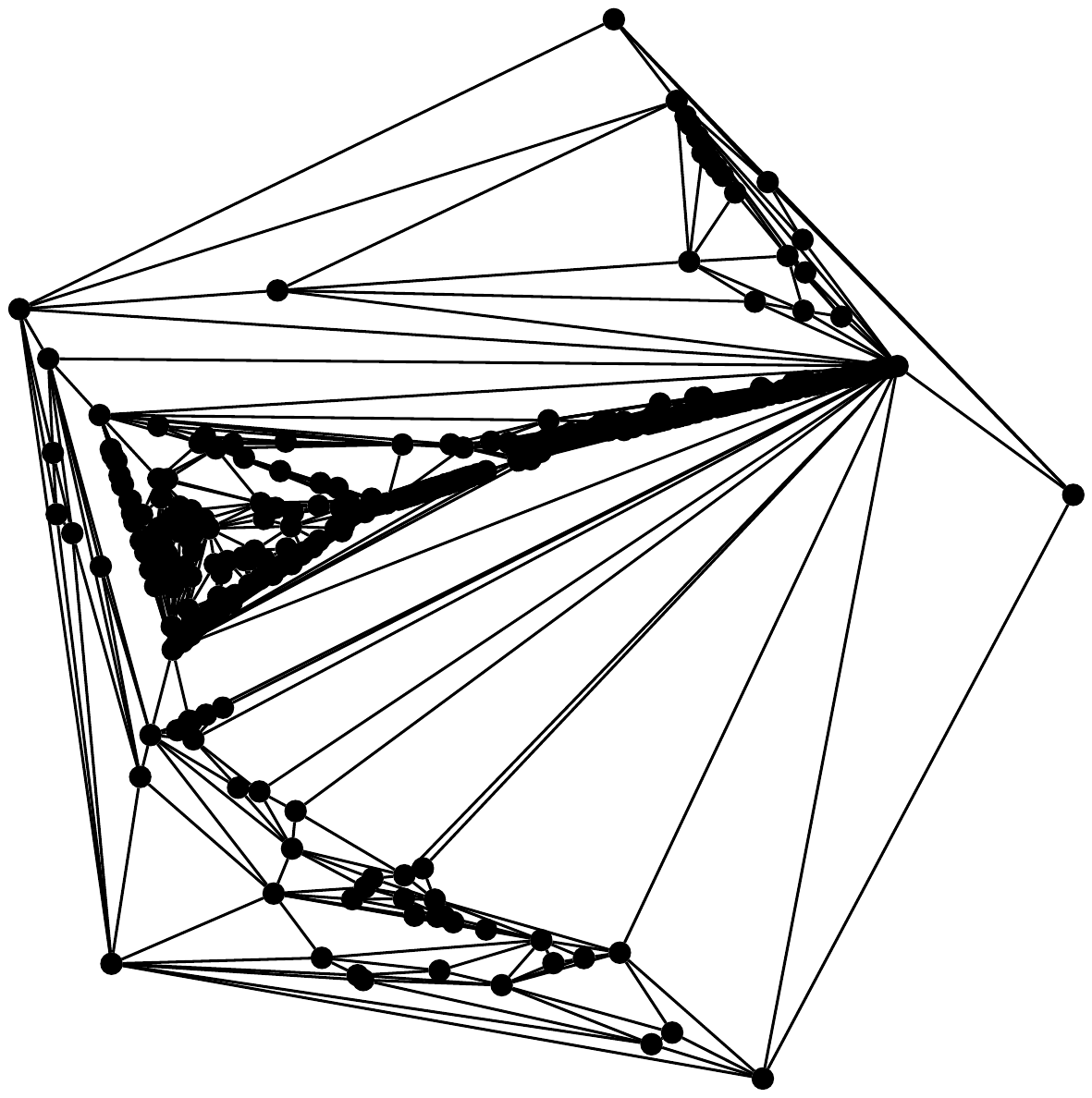} & 
        \includegraphics[width=.175\linewidth]{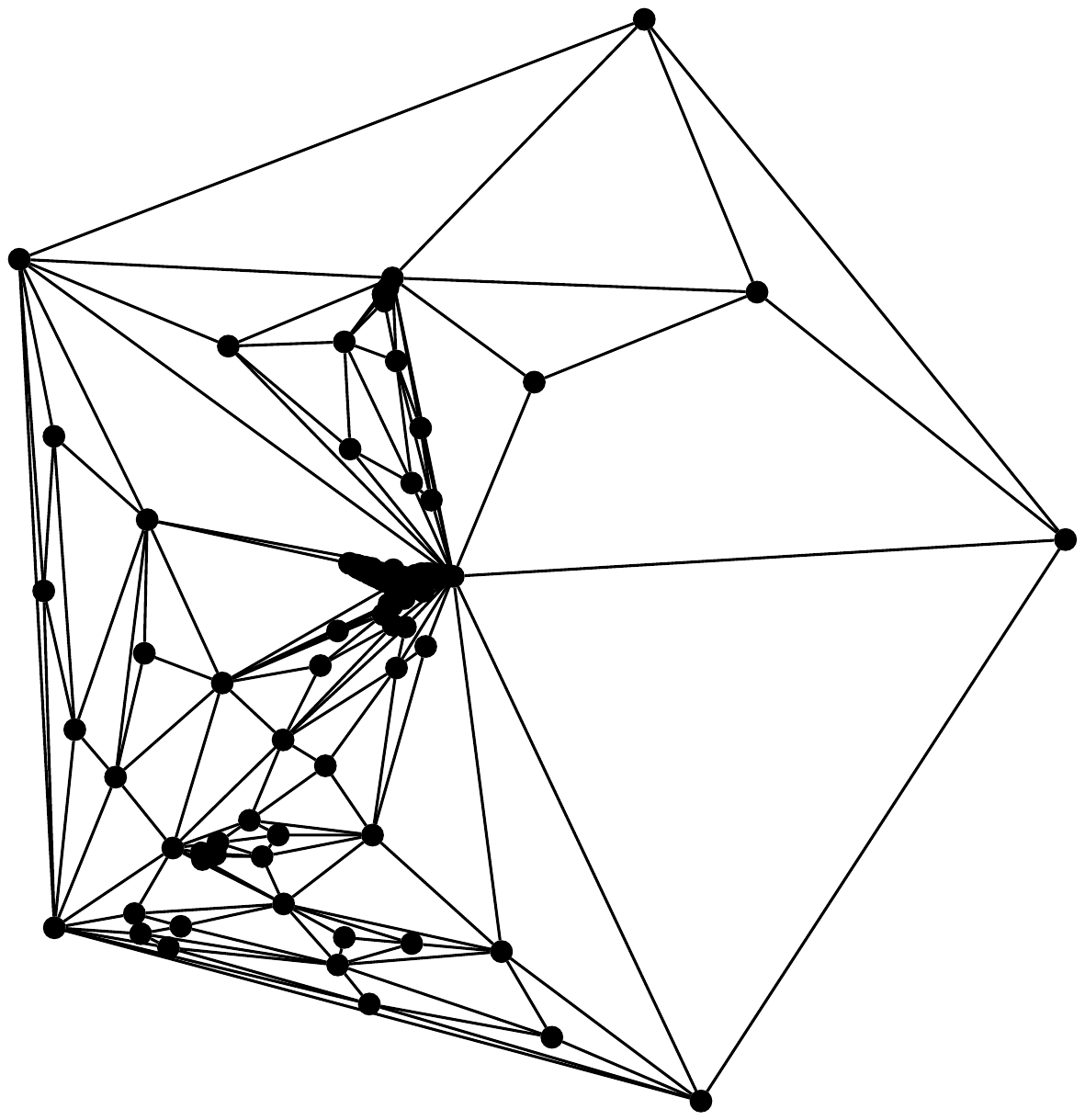} \\
        \hline
        & $\rho(\Gamma) =1043677$ & $\rho(\Gamma) =600$ & $\rho(\Gamma) =597$ & $\rho(\Gamma) =600$ & $\rho(\Gamma) = 172022, r=3$ \\
        \hline
        \multirow{-10}{*}{\rotatebox[origin=c]{90}{$G(66,147)$}} &
        \includegraphics[width=.175\linewidth]{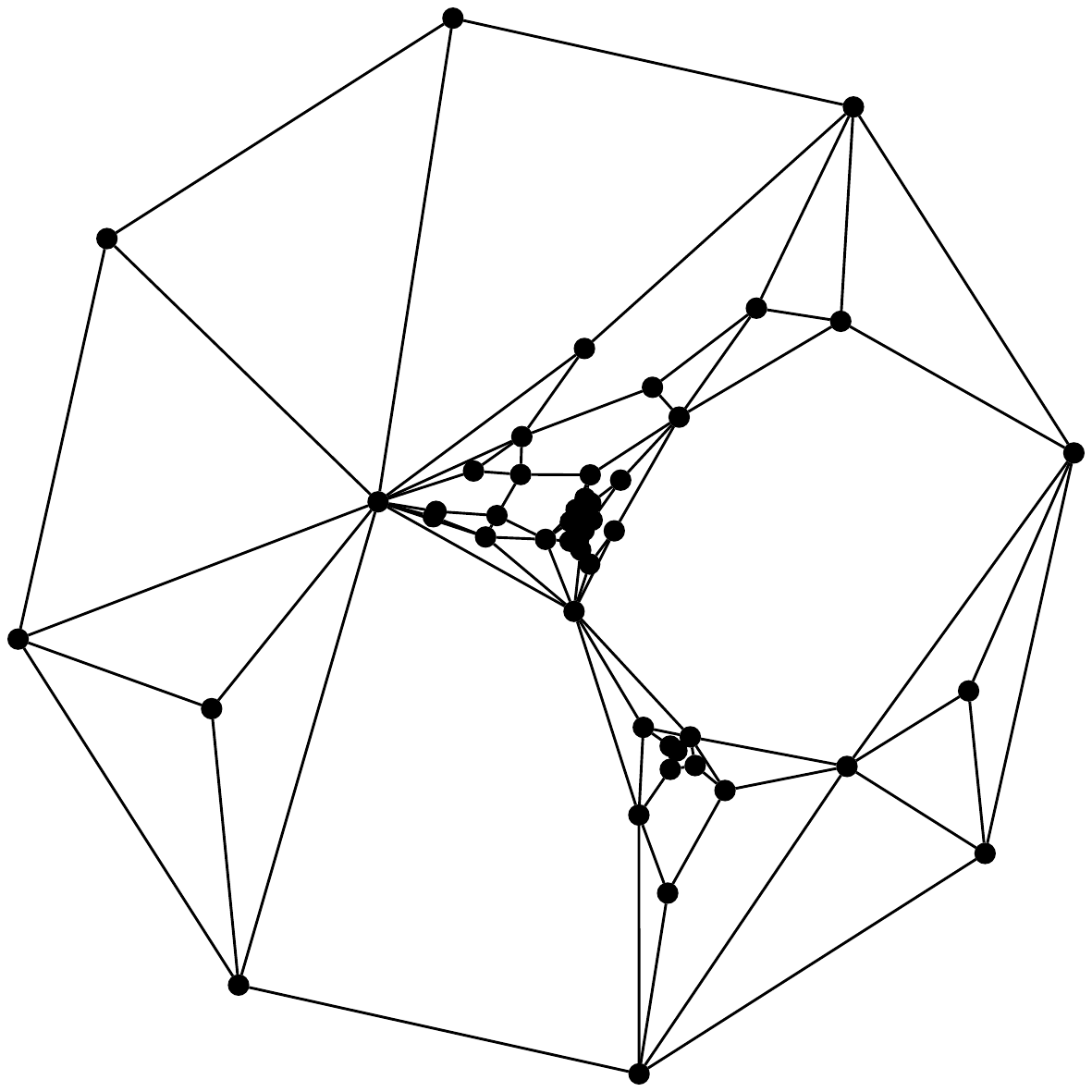} & 
        \includegraphics[width=.175\linewidth]{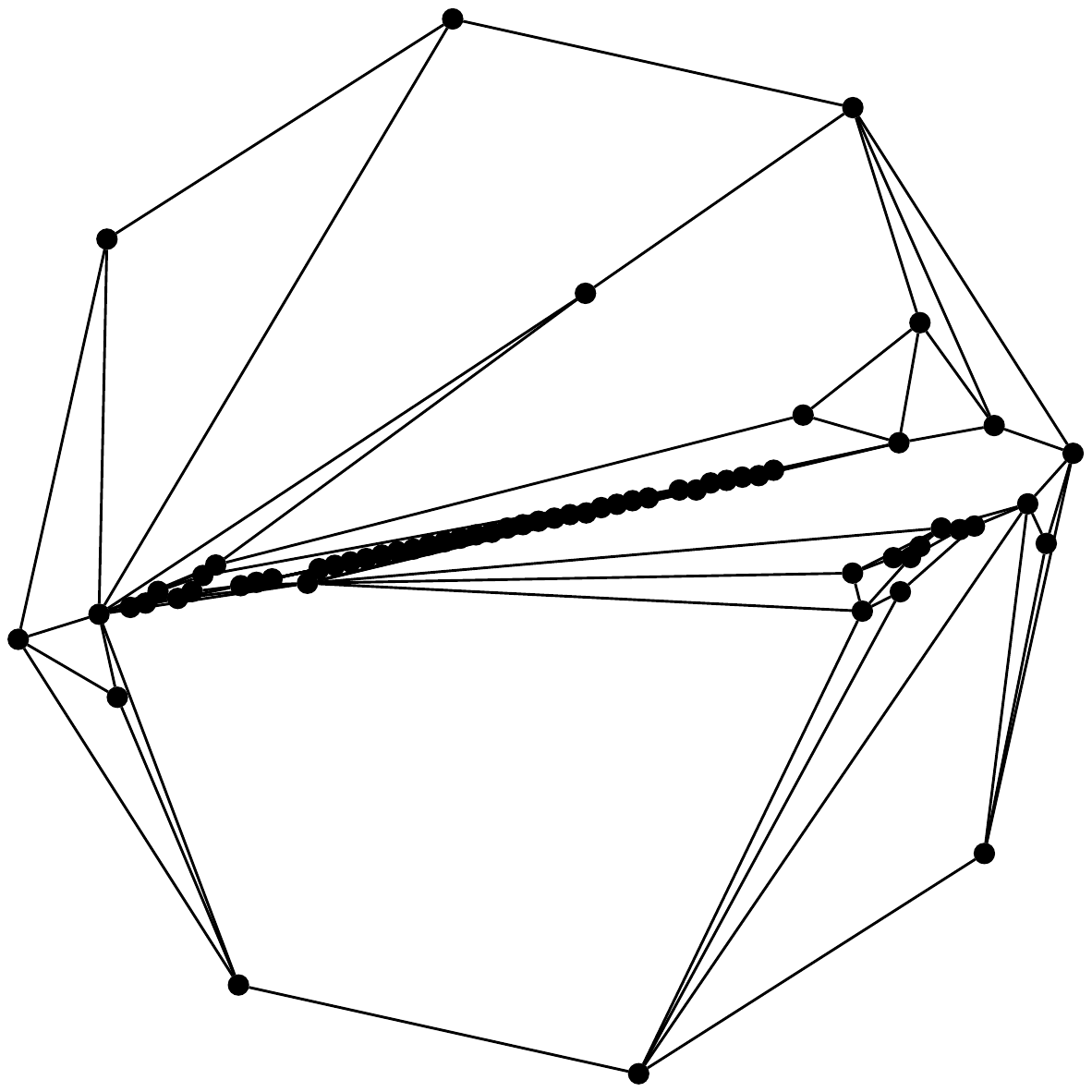} &  
        \includegraphics[width=.175\linewidth]{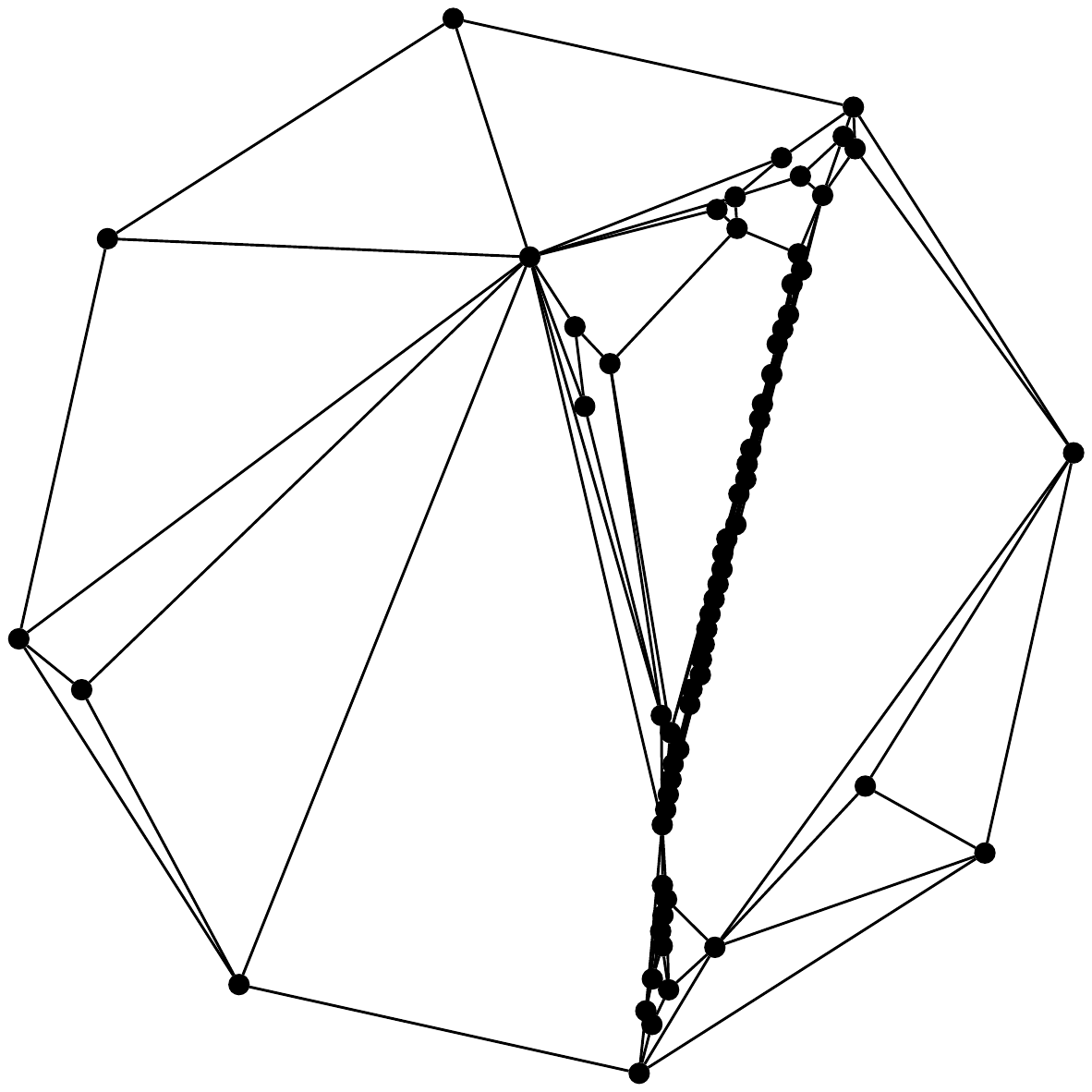} & 
        \includegraphics[width=.175\linewidth]{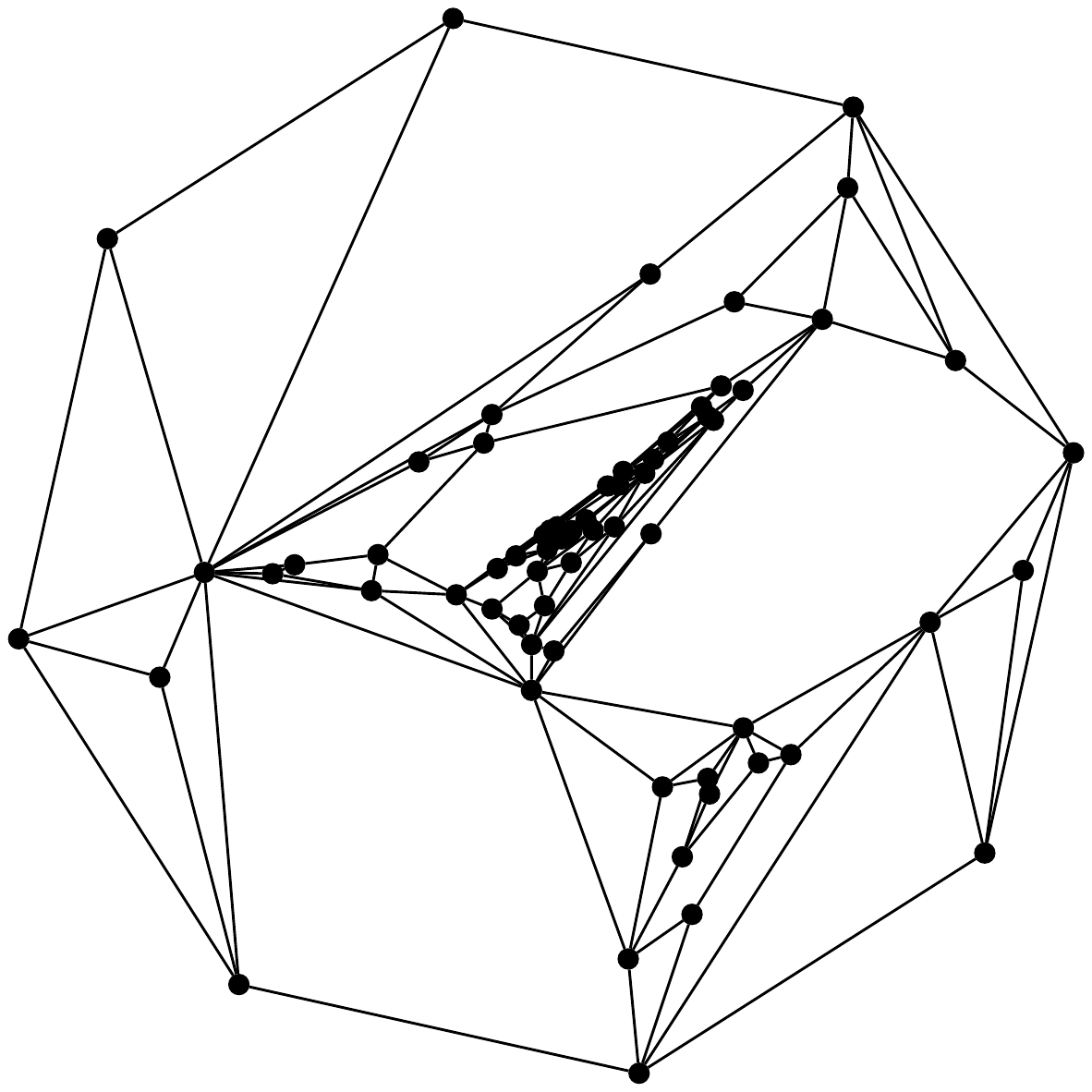} & 
        \includegraphics[width=.175\linewidth]{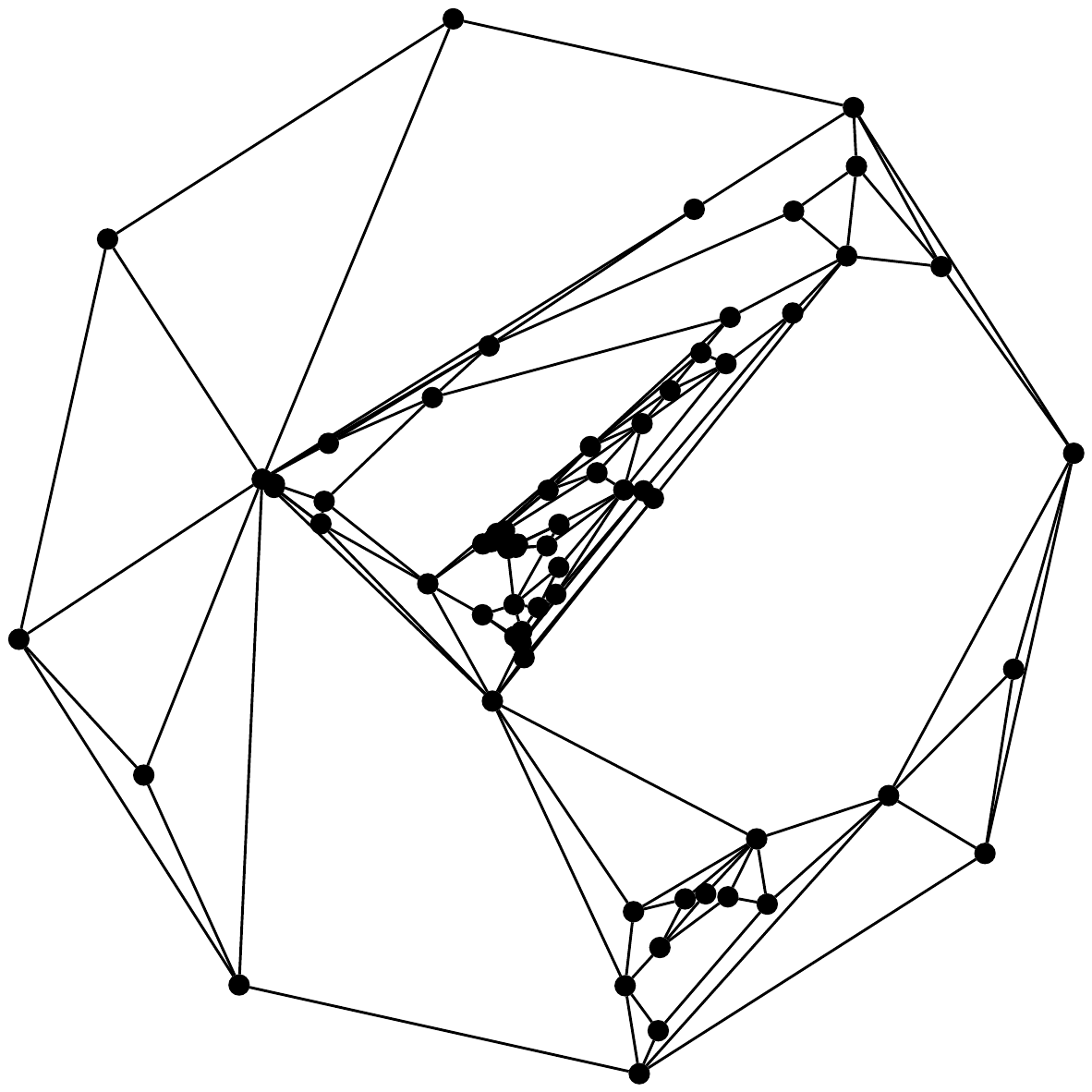} \\
        \hline
        & $\rho(\Gamma) =1341$ & $\rho(\Gamma) =48$ & $\rho(\Gamma) =53$ & $\rho(\Gamma) =77$ & $\rho(\Gamma) = 157, r=4$ \\
        \hline
        \multirow{-10}{*}{\rotatebox[origin=c]{90}{$G(74,212)$}} &
        \includegraphics[width=.175\linewidth]{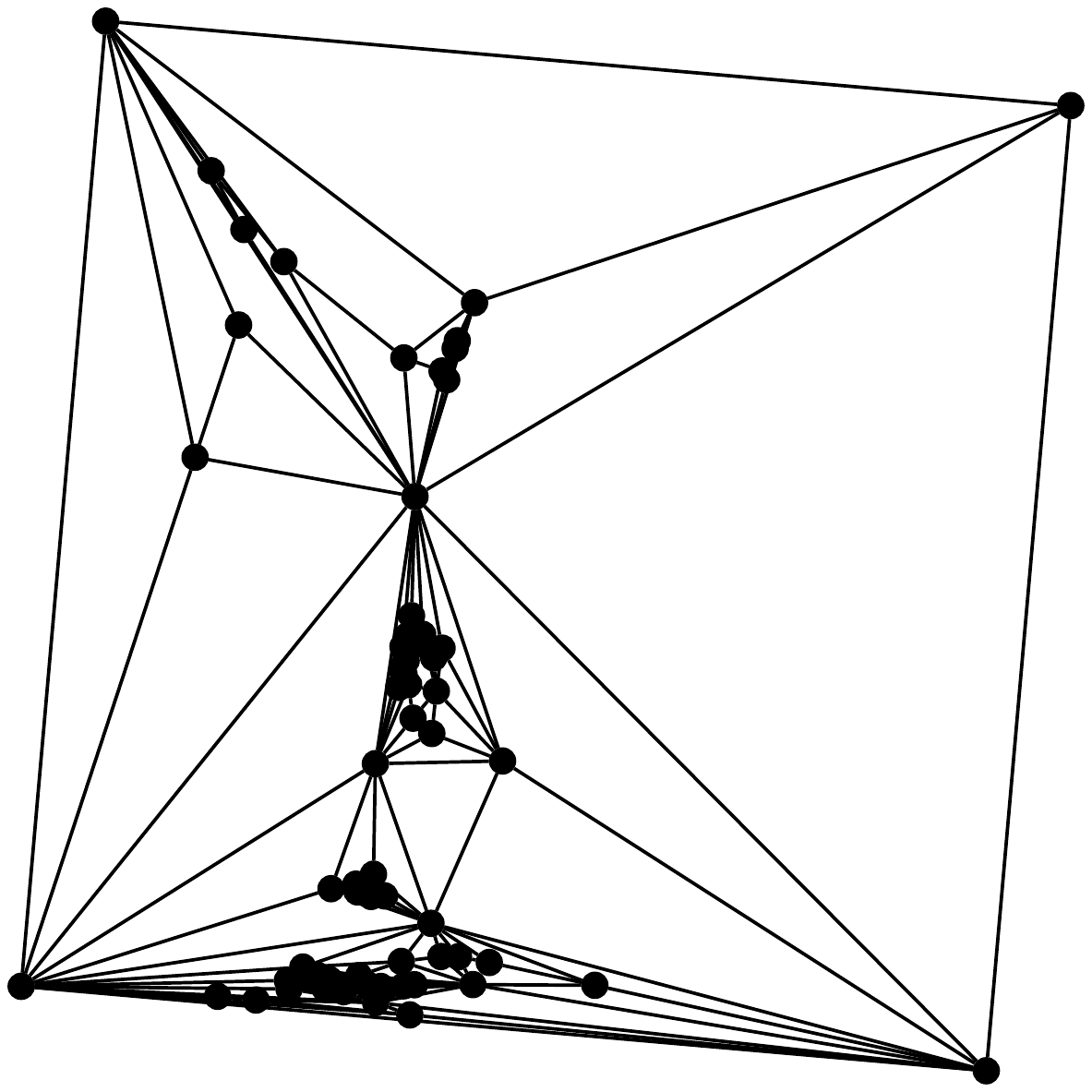} & 
        \includegraphics[width=.175\linewidth]{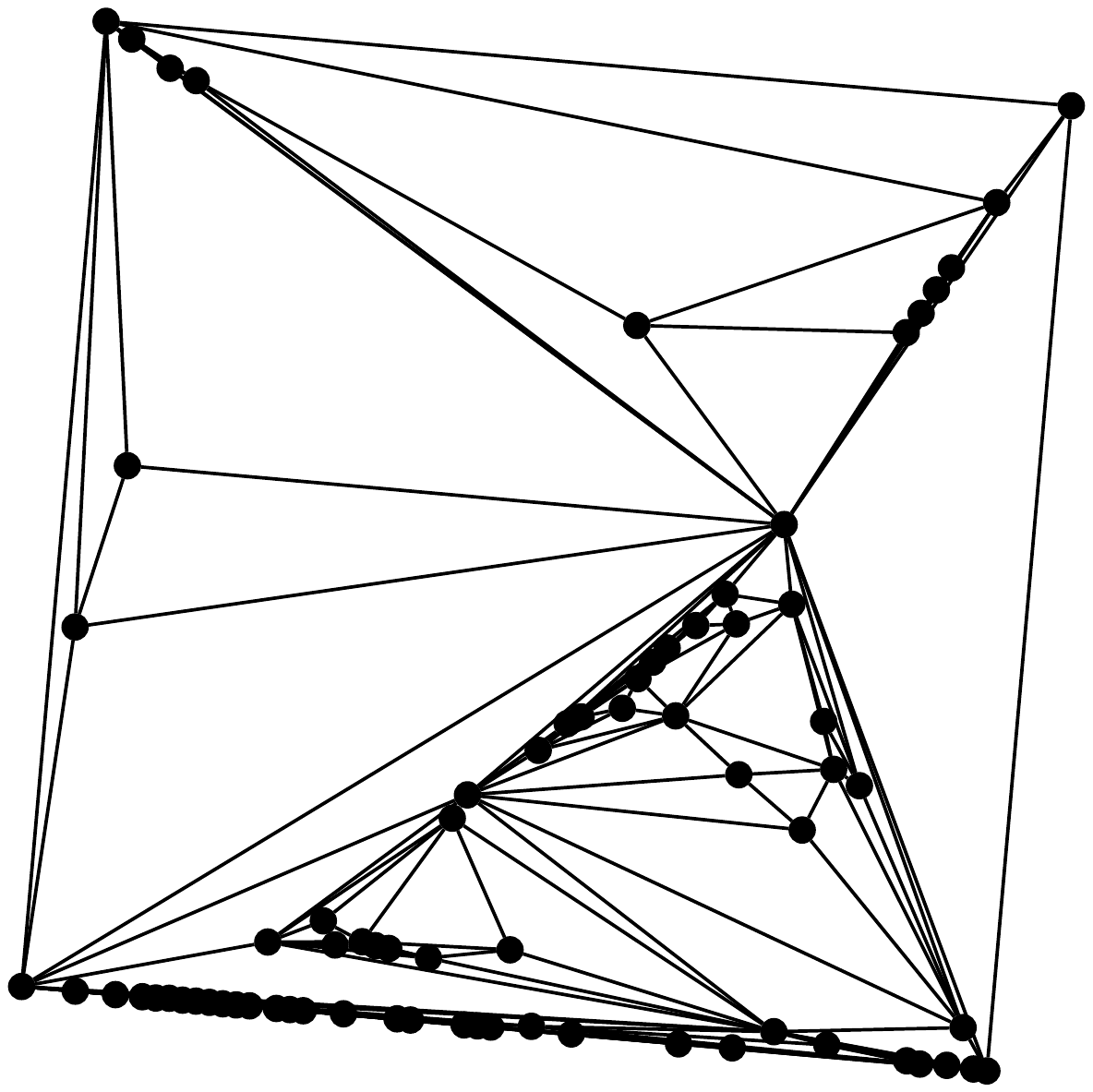} &  
        \includegraphics[width=.175\linewidth]{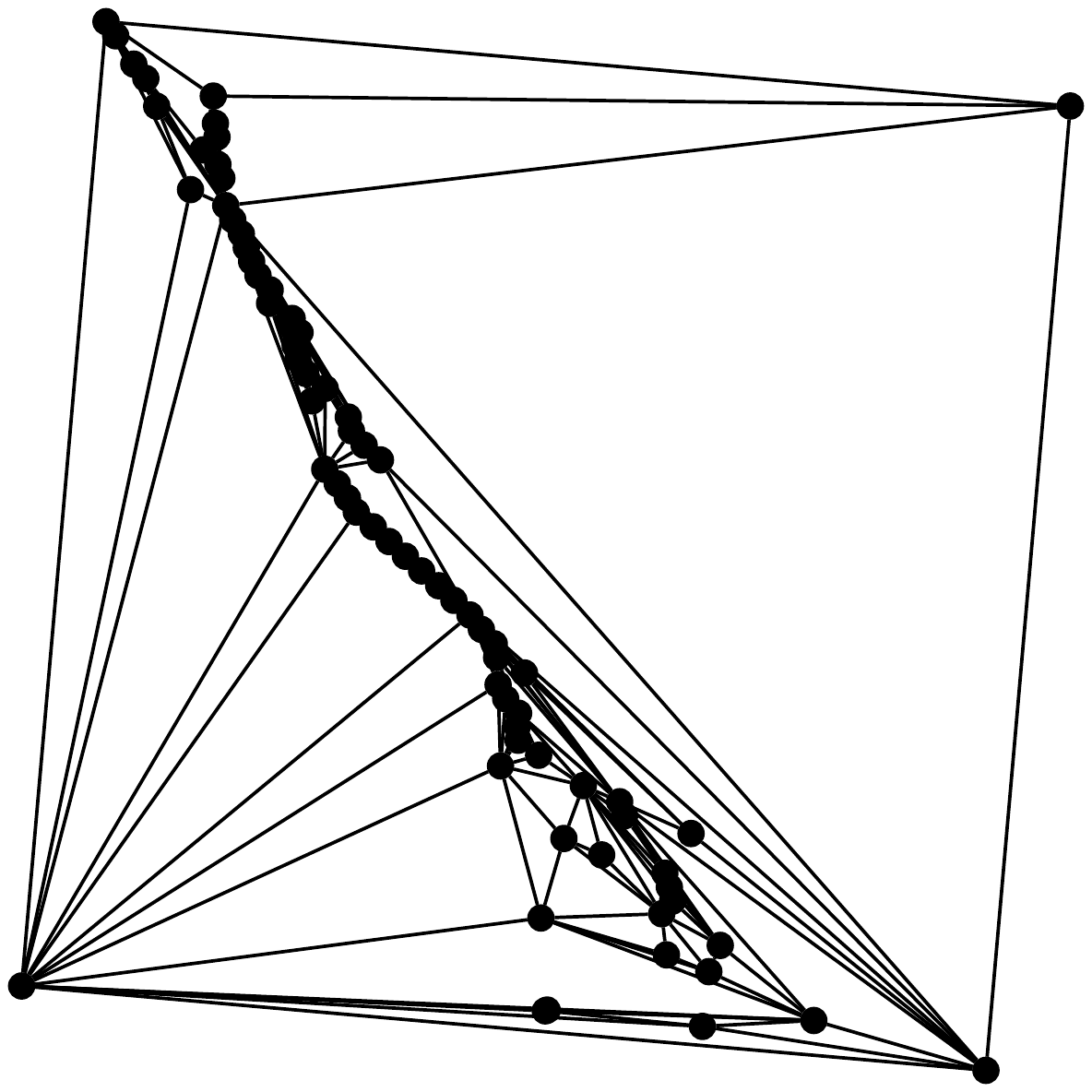} & 
        \includegraphics[width=.175\linewidth]{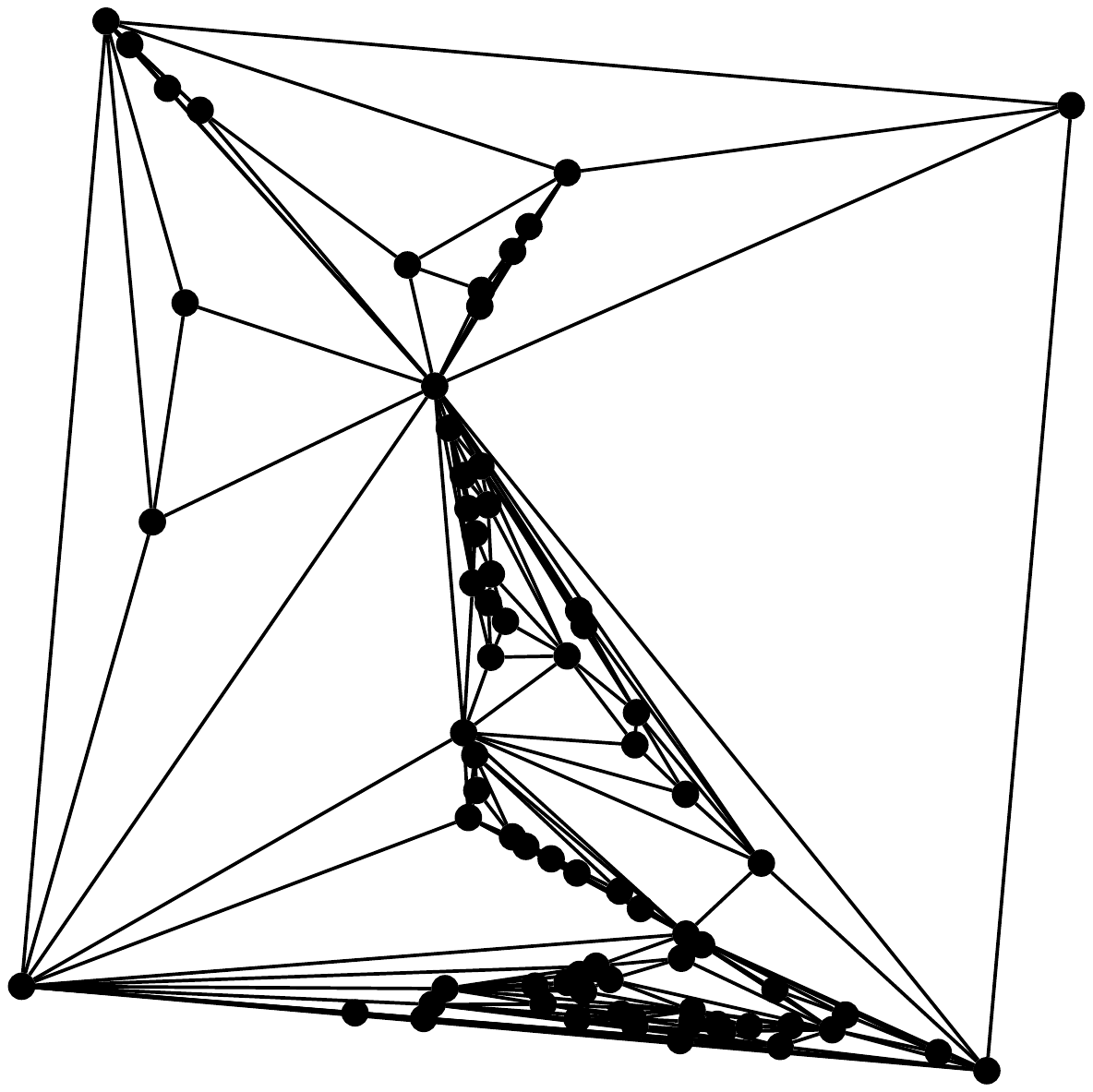} & 
        \includegraphics[width=.175\linewidth]{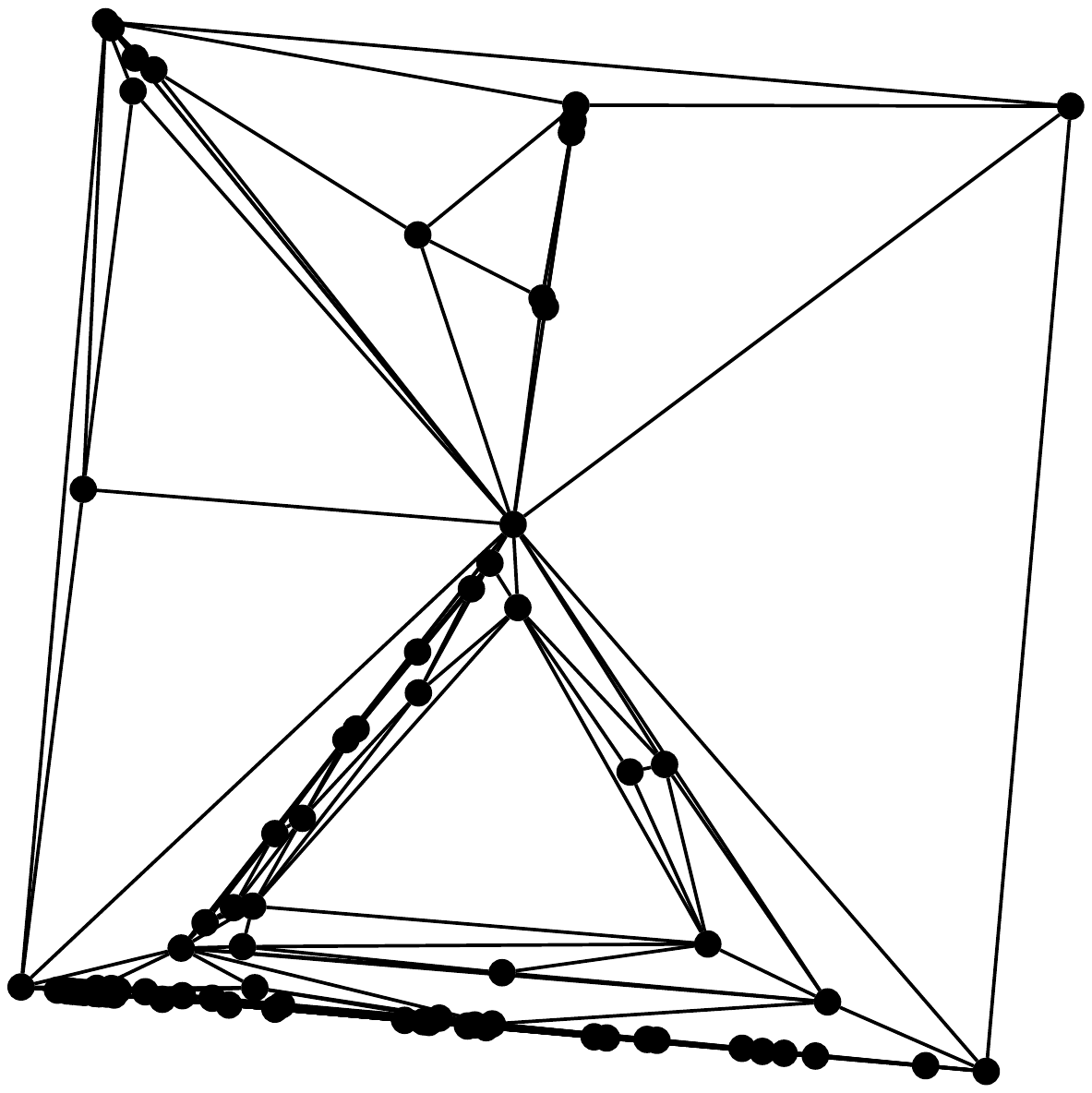} \\
        \hline
        & $\rho(\Gamma) =560$ & $\rho(\Gamma) =72$ & $\rho(\Gamma) =551$ & $\rho(\Gamma) =135$ & $\rho(\Gamma) = 182, r=12$ \\
        \hline
        \multirow{-10}{*}{\rotatebox[origin=c]{90}{$G(355,700)$}} &
        \includegraphics[width=.175\linewidth]{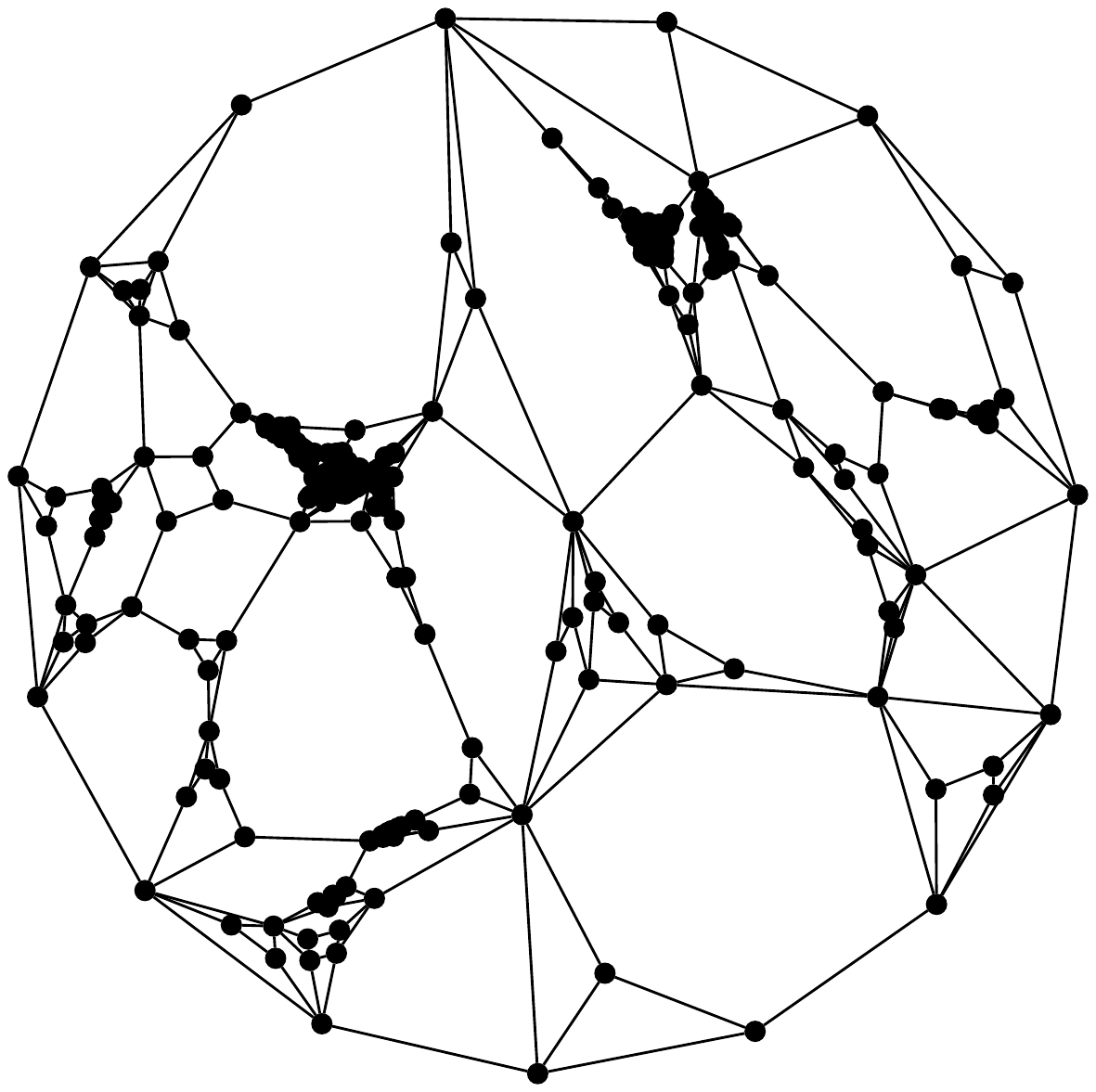} & 
        \includegraphics[width=.175\linewidth]{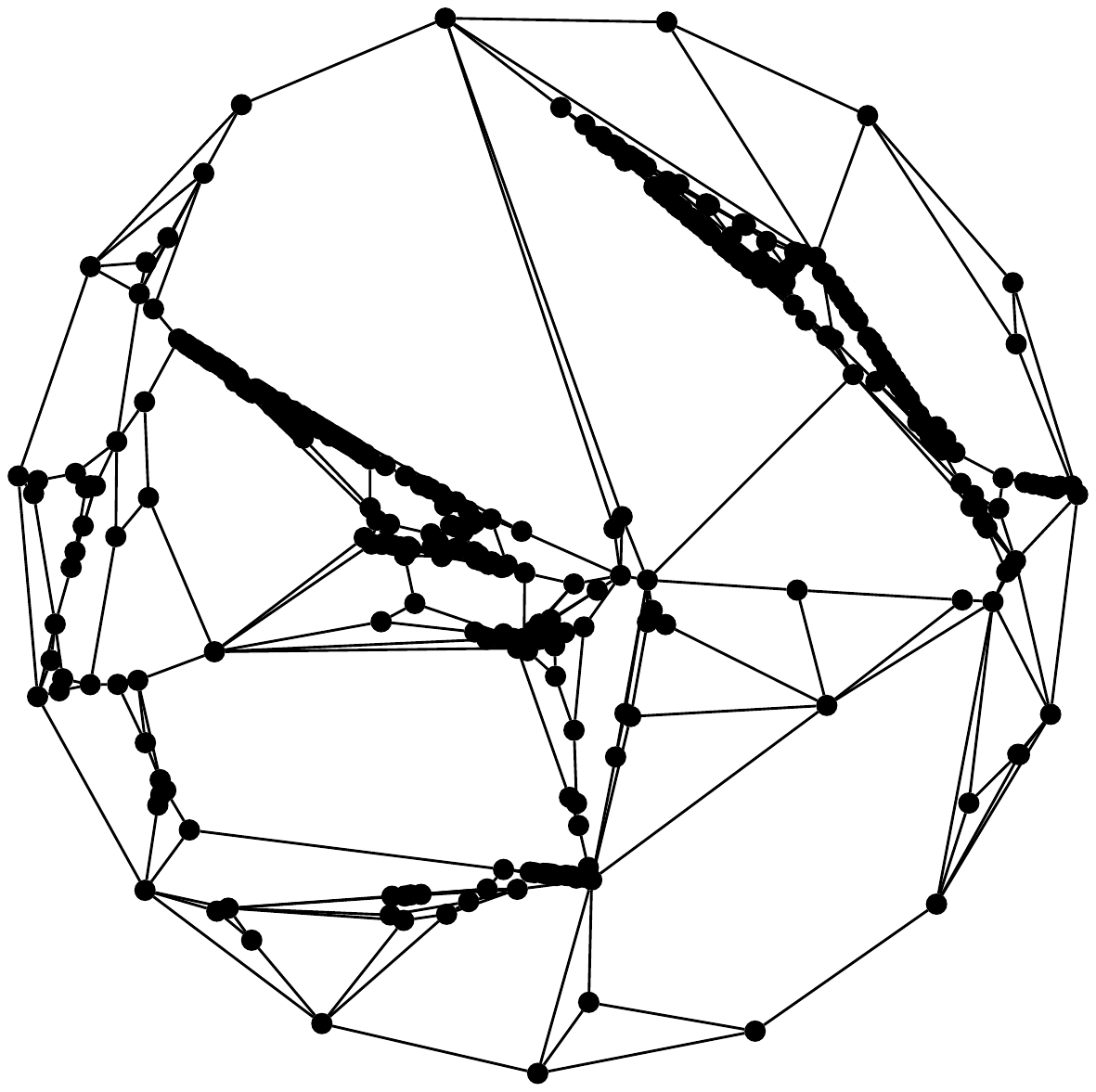} &  
        \includegraphics[width=.175\linewidth]{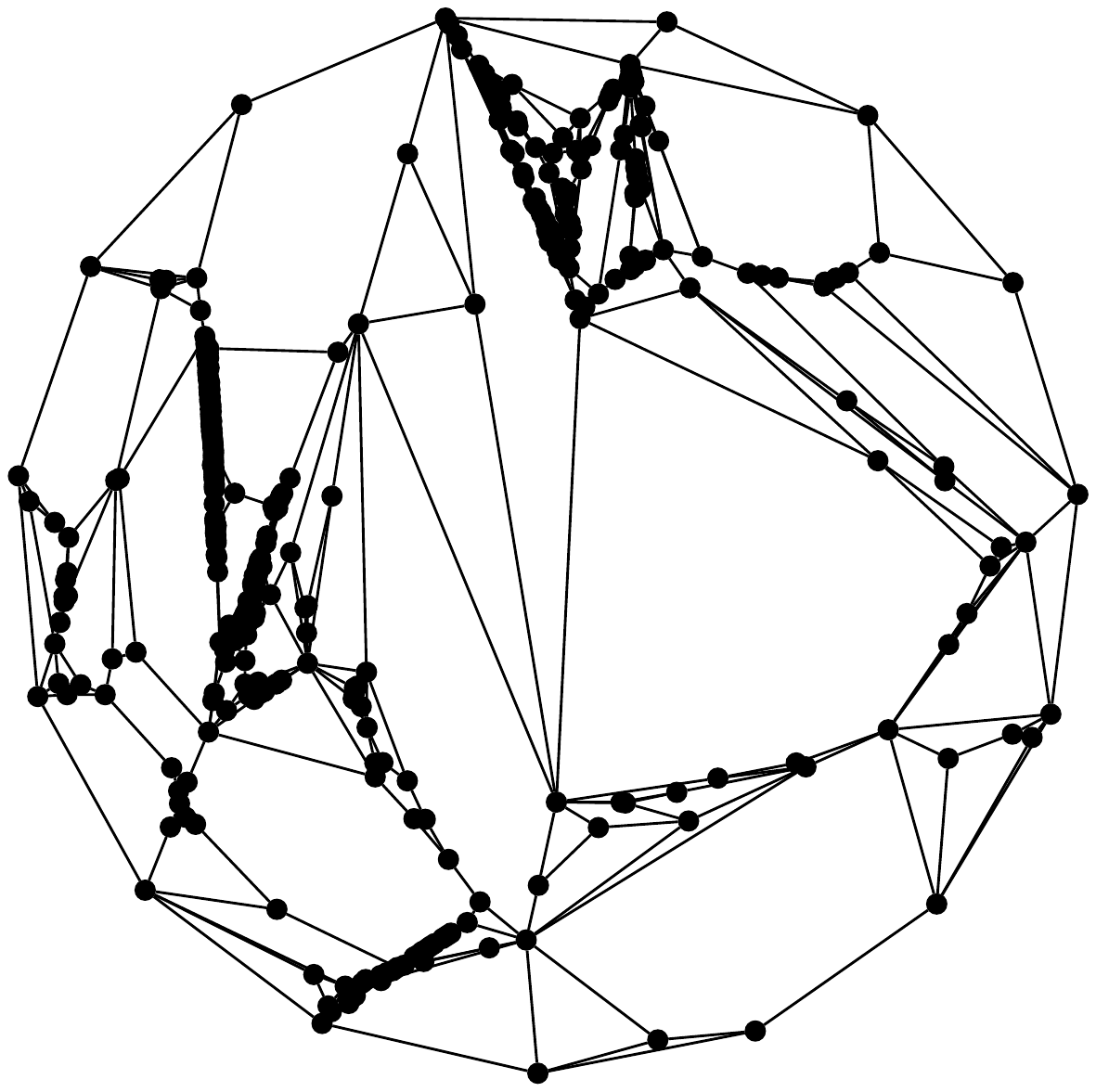} & 
        \includegraphics[width=.175\linewidth]{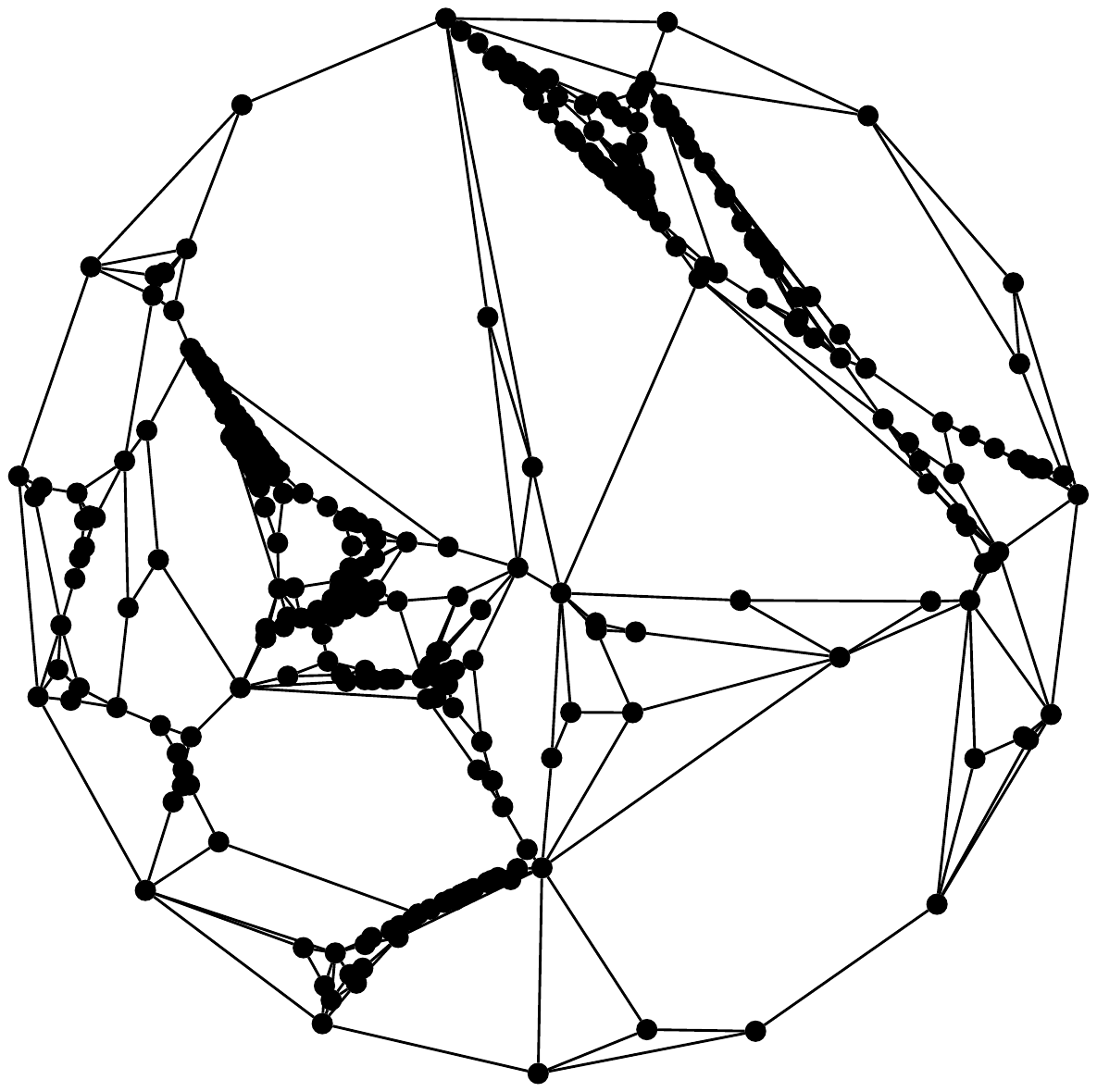} & 
        \includegraphics[width=.175\linewidth]{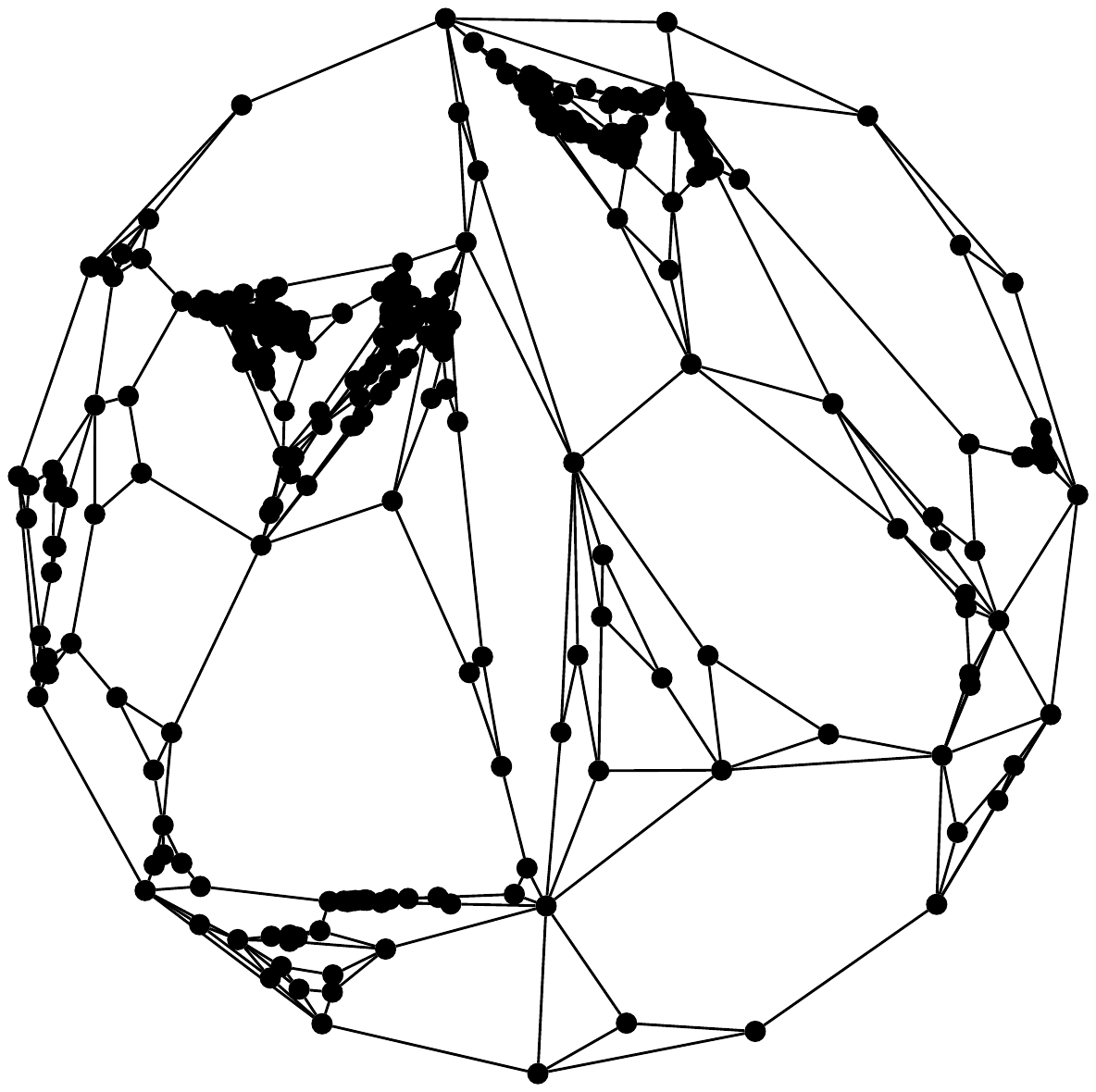} \\
        \hline
        & $\rho(\Gamma) =11035$ & $\rho(\Gamma) =328$  & $\rho(\Gamma) =252$ & $\rho(\Gamma) =214$ & $\rho(\Gamma) = 1116, r=2$ \\
        \hline
        \multirow{-10}{*}{\rotatebox[origin=c]{90}{$G(59,149)$}} &
        \includegraphics[width=.175\linewidth]{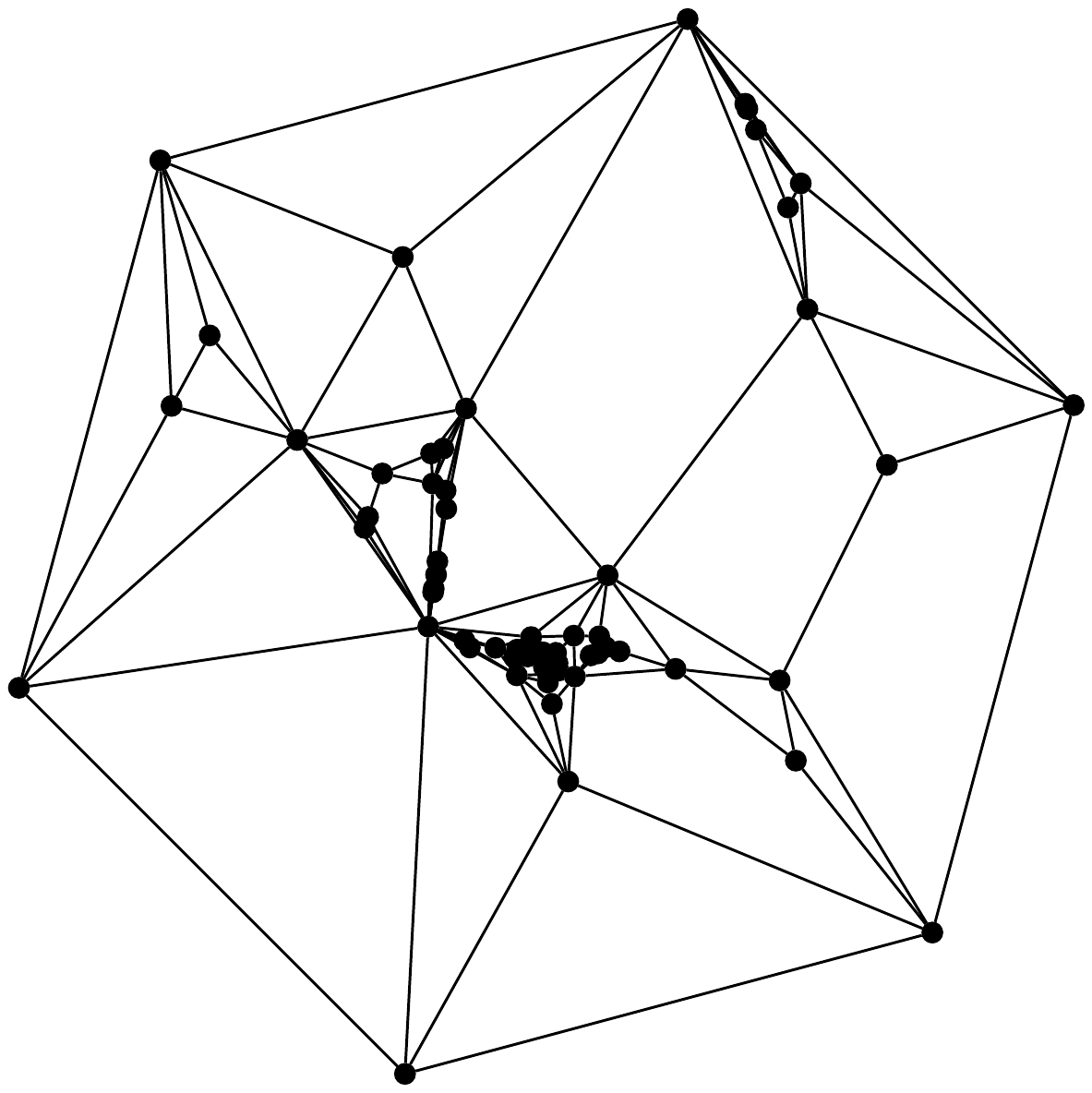} & 
        \includegraphics[width=.175\linewidth]{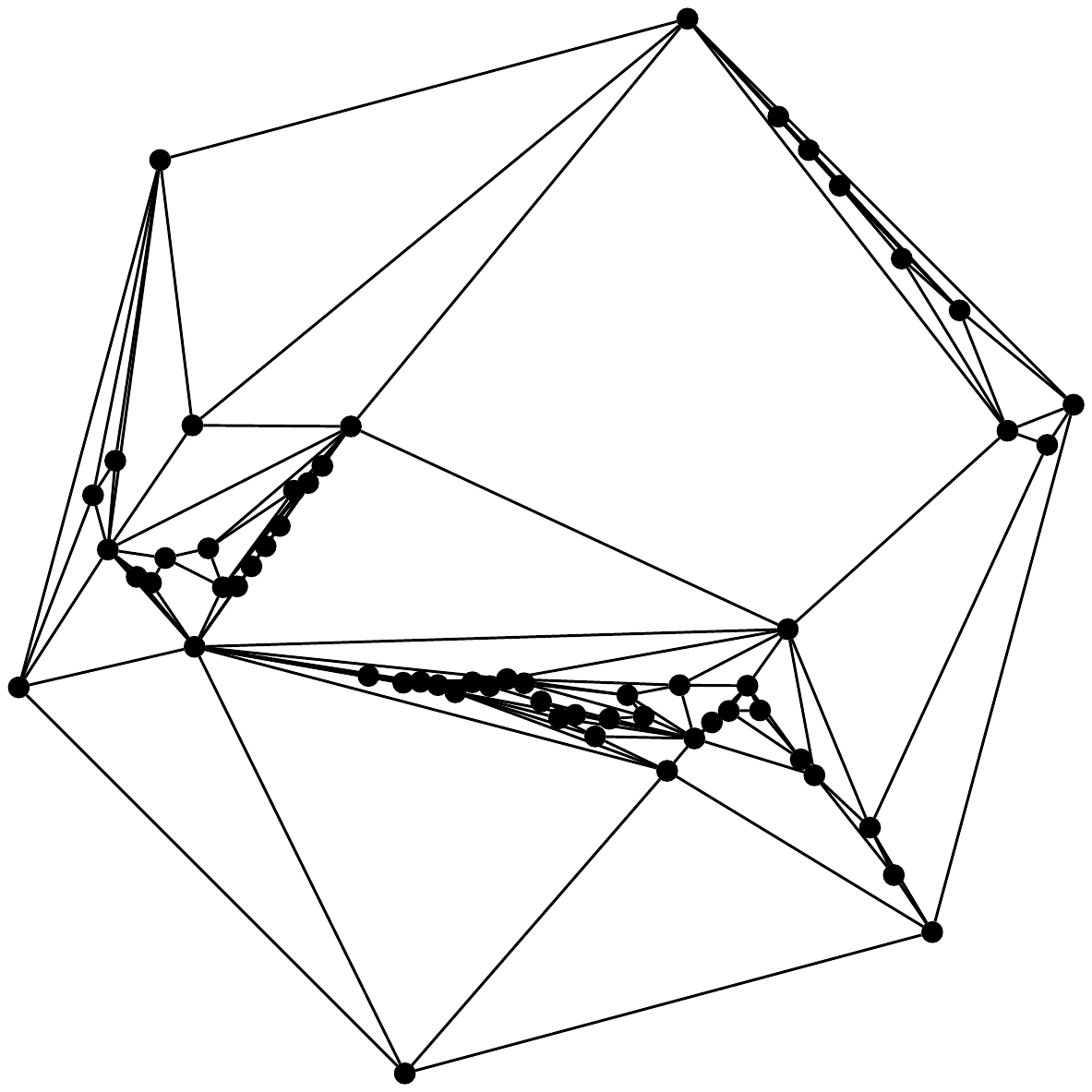} &  
        \includegraphics[width=.175\linewidth]{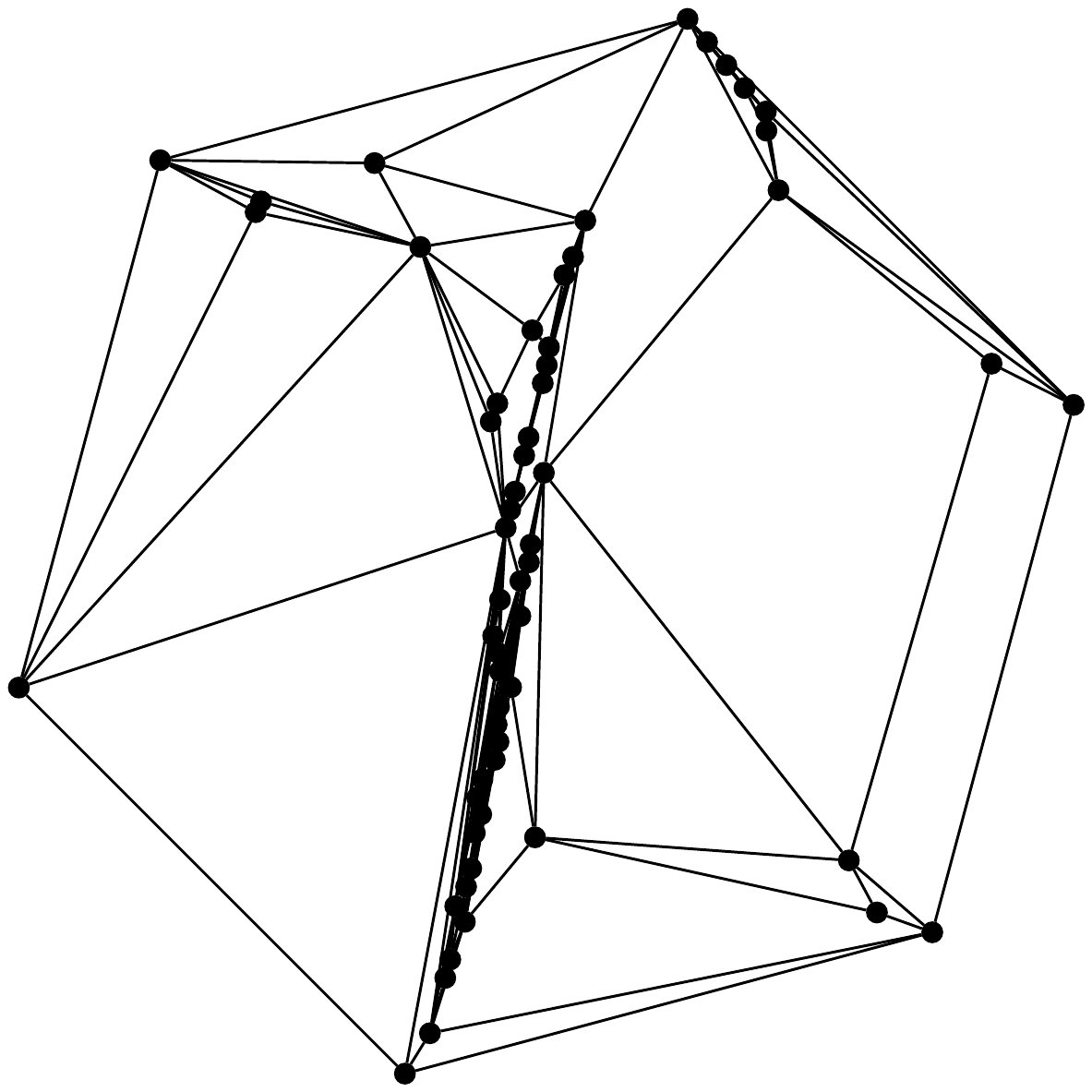} & 
        \includegraphics[width=.175\linewidth]{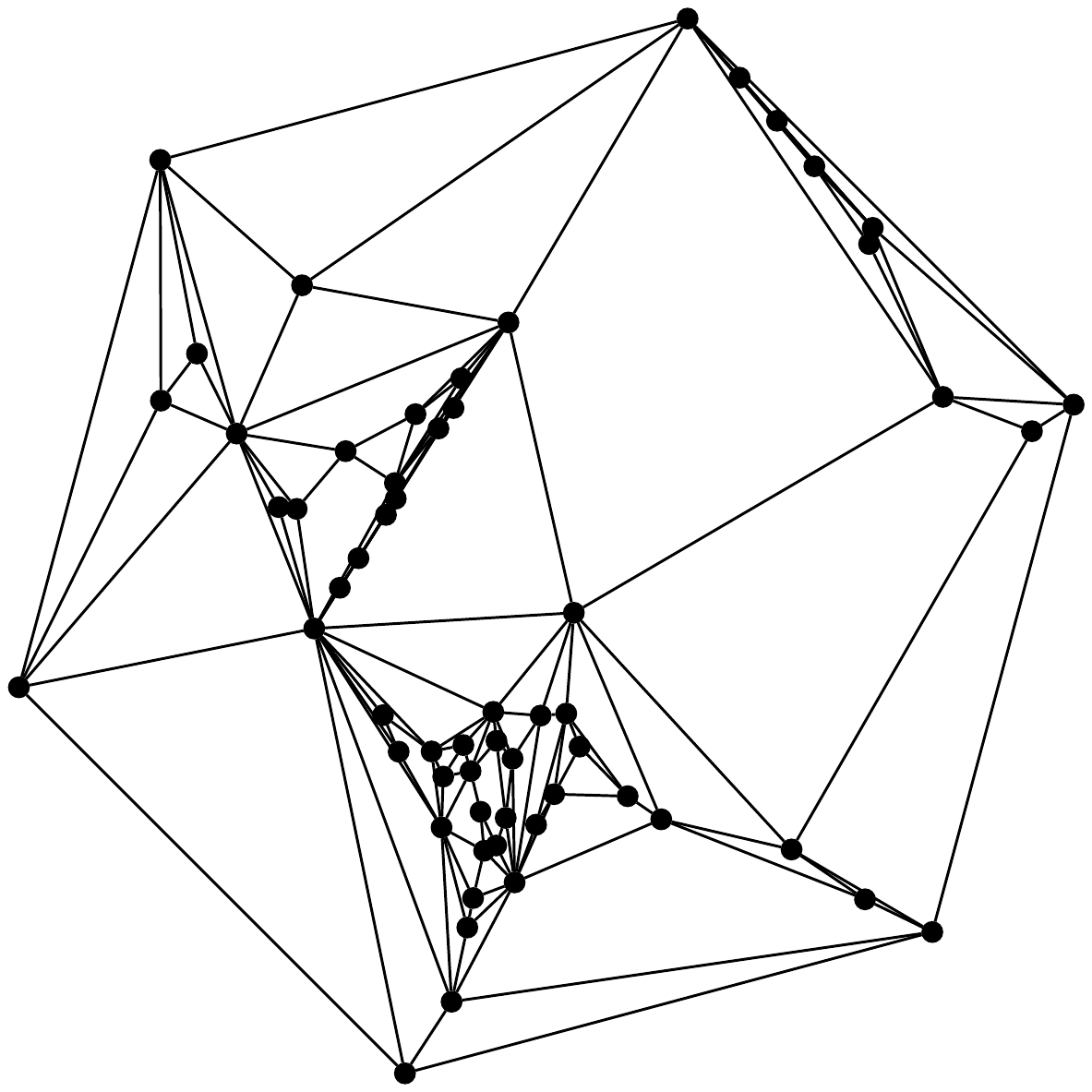} & 
        \includegraphics[width=.175\linewidth]{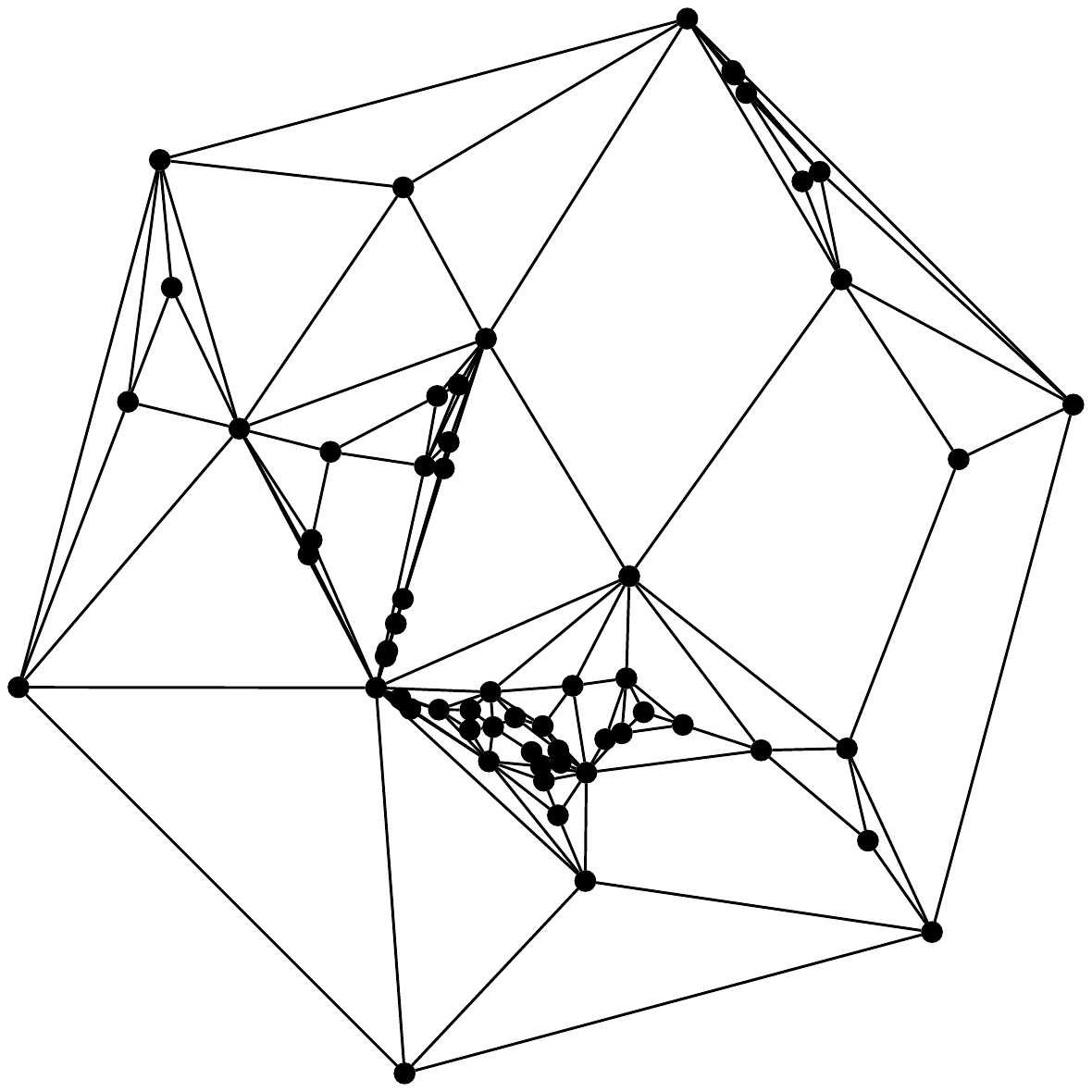} \\
        \hline
        & $\rho(\Gamma) =154$ & $\rho(\Gamma) =41$ & $\rho(\Gamma) =49$ & $\rho(\Gamma) =41$ & $\rho(\Gamma) = 125, r=2$ \\
        \hline
        
  \end{tabular}
\end{table}

\begin{table}[p]
\caption{Worst and best rotations for the graph of Fig.~\ref{fig:kaleidoscope2}.}
\label{tbl:kal2}
\begin{center}
      \begin{tabular}
            {ccccc} \hline & Tutte & $x$-spread & $y$-spread & $xy$-morph\\
            \hline 
        
            \multirow{-10}{*}{\rotatebox[origin=c]{90}{$\alpha = 60^\circ$}} &
            \includegraphics[width=.23\linewidth]{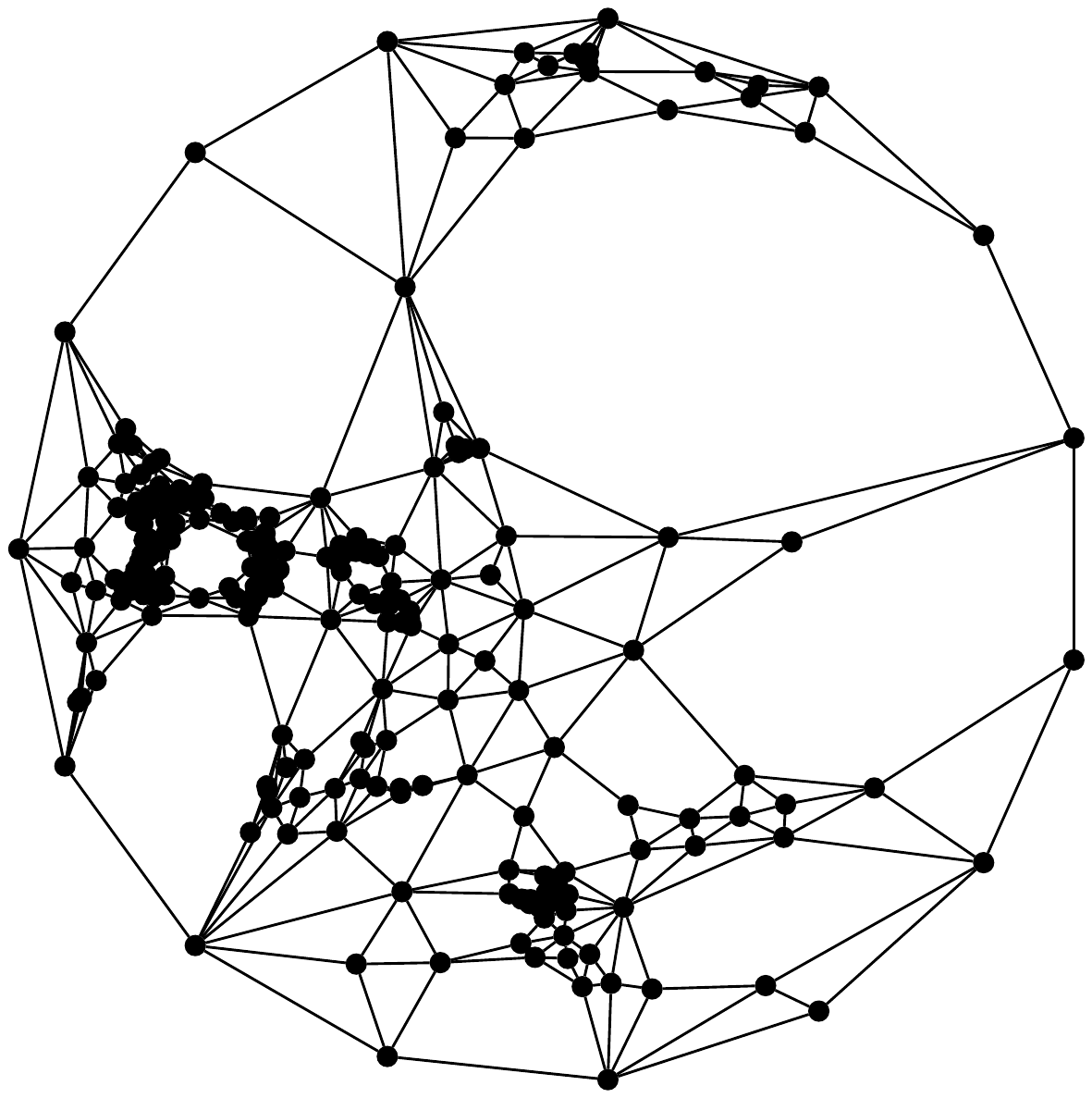} &
            \includegraphics[width=.23\linewidth]{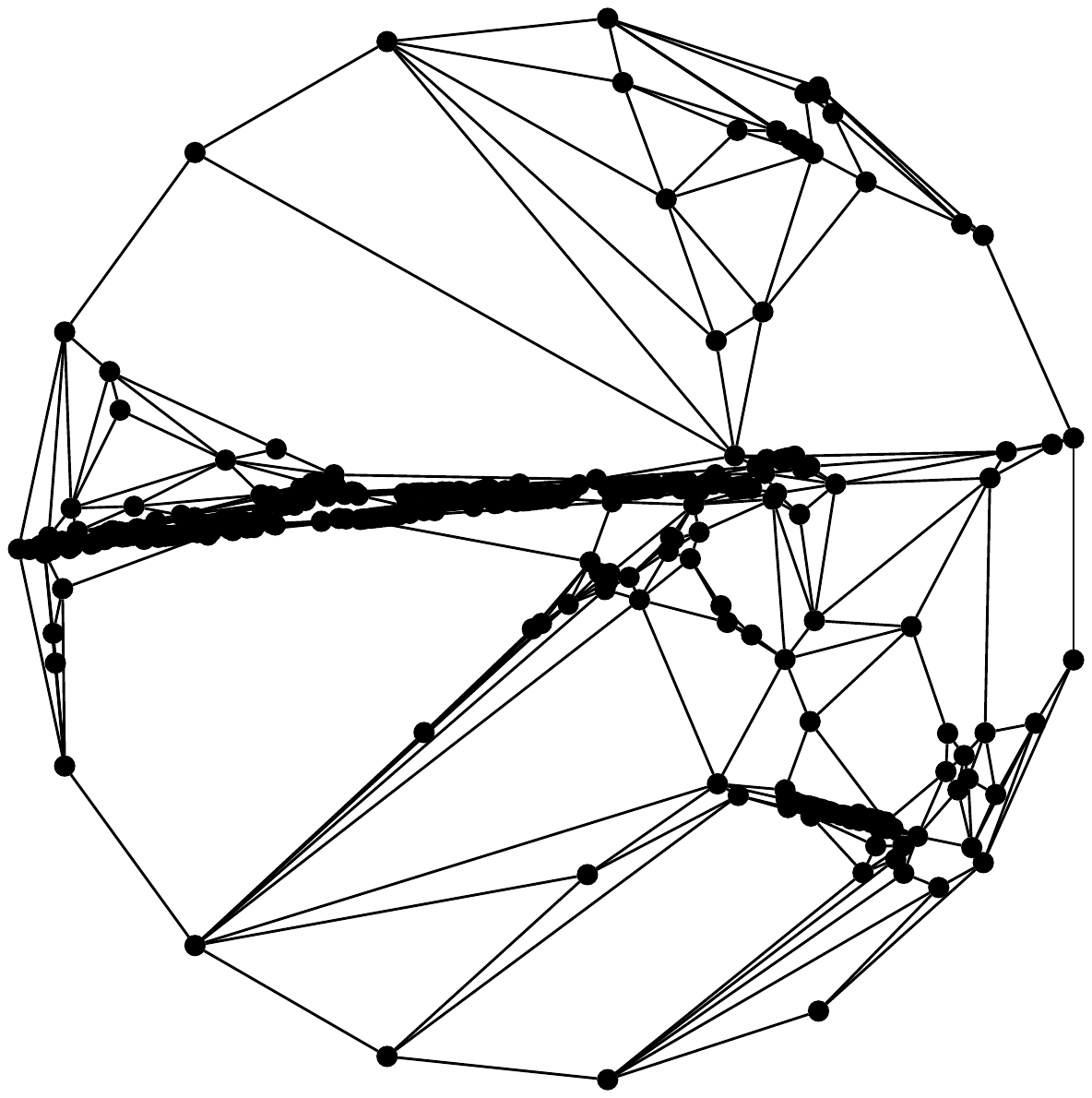} &  
            \includegraphics[width=.23\linewidth]{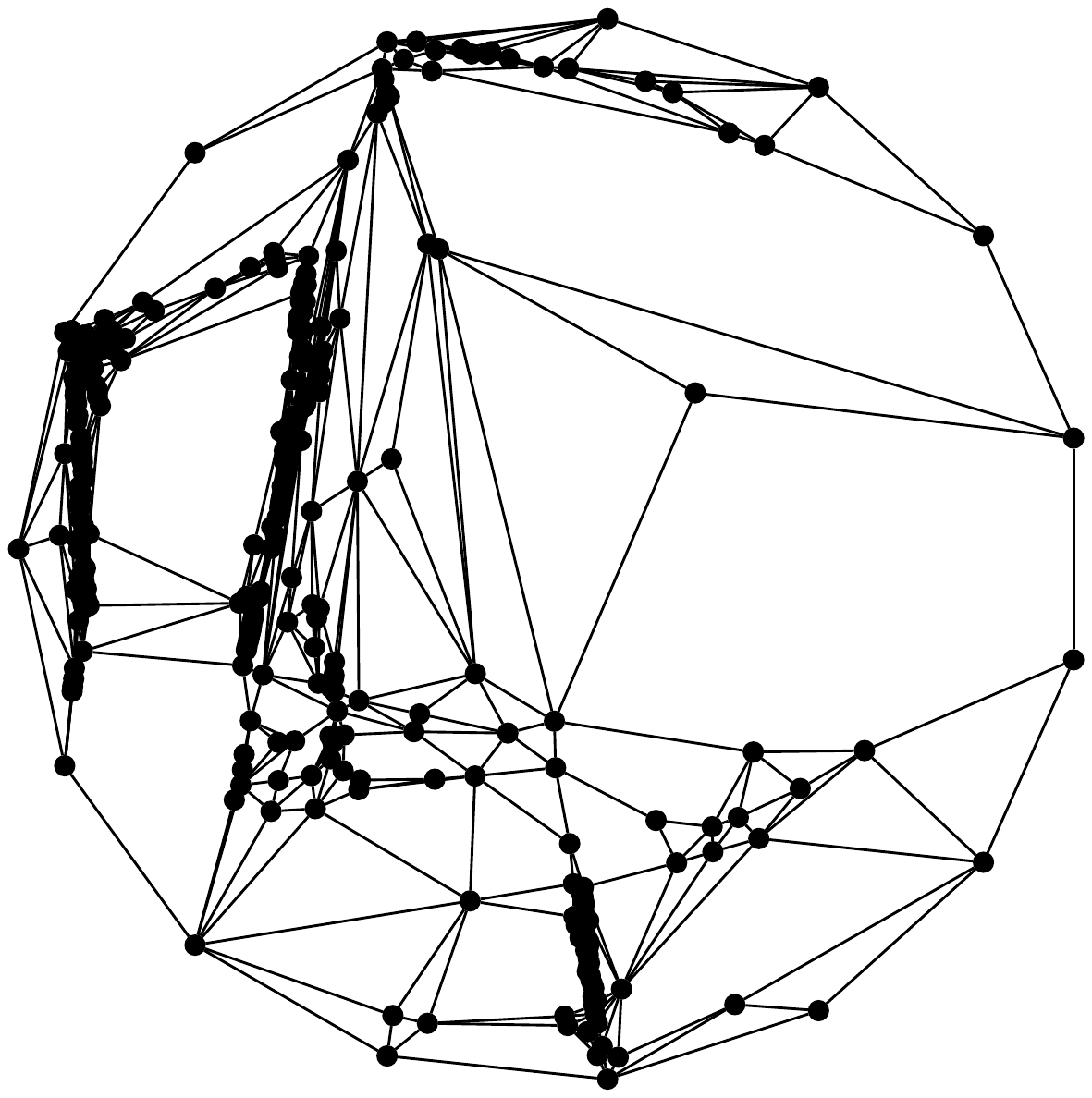} & 
            \includegraphics[width=.23\linewidth]{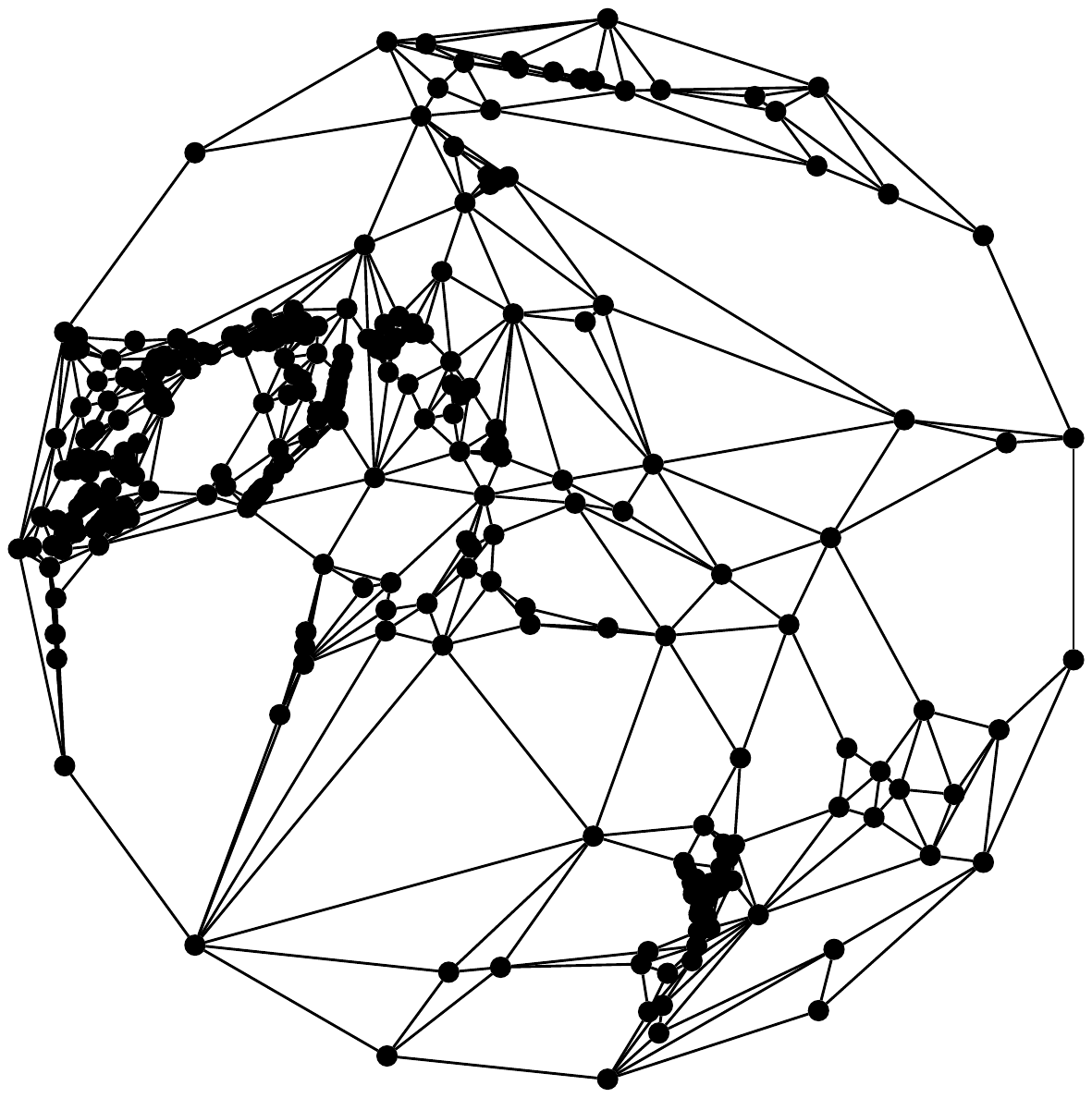} \\
            \hline
            \multirow{-10}{*}{\rotatebox[origin=c]{90}{$\alpha = 85^\circ$}} &
            \includegraphics[width=.23\linewidth]{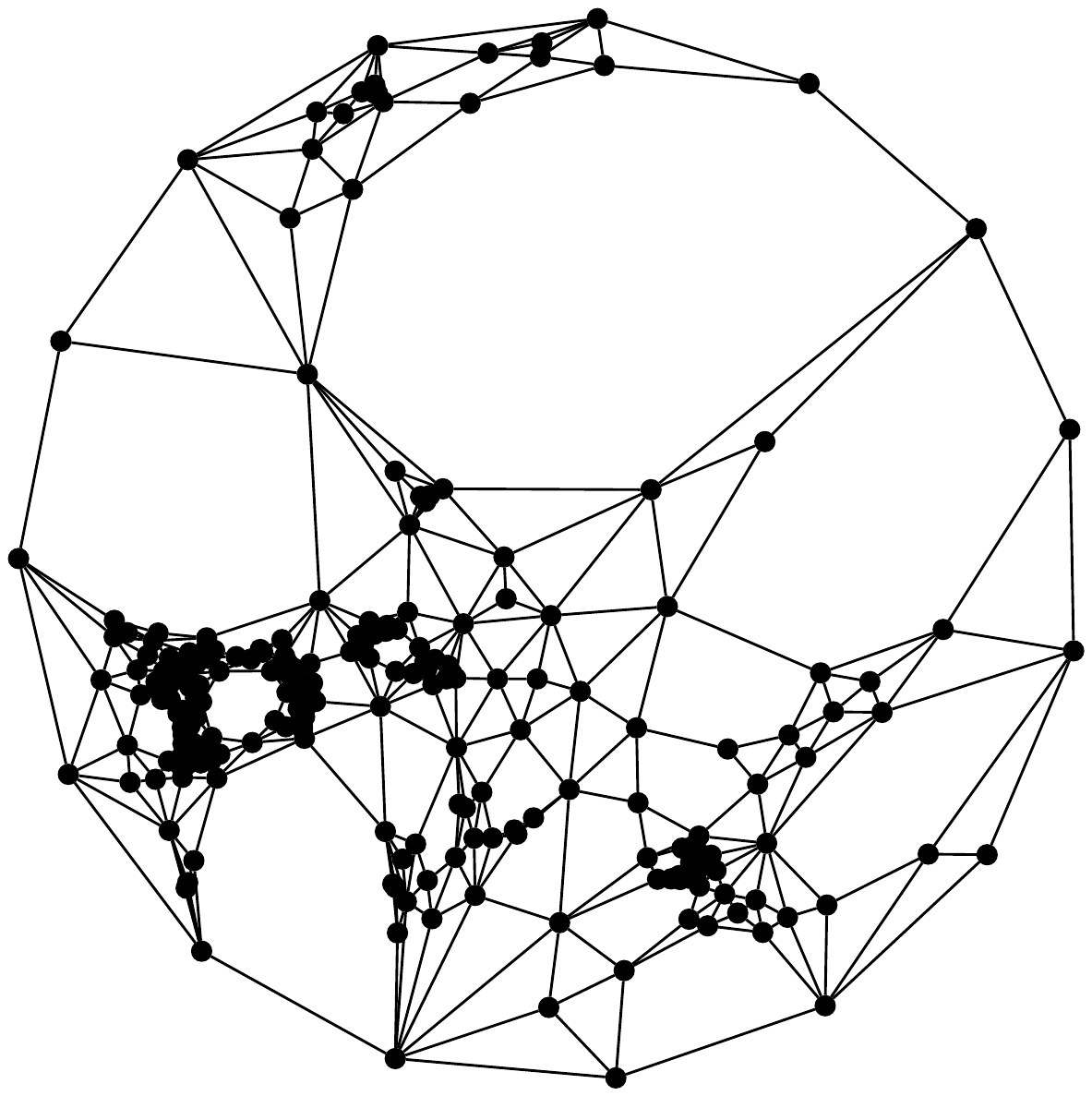} &
            \includegraphics[width=.23\linewidth]{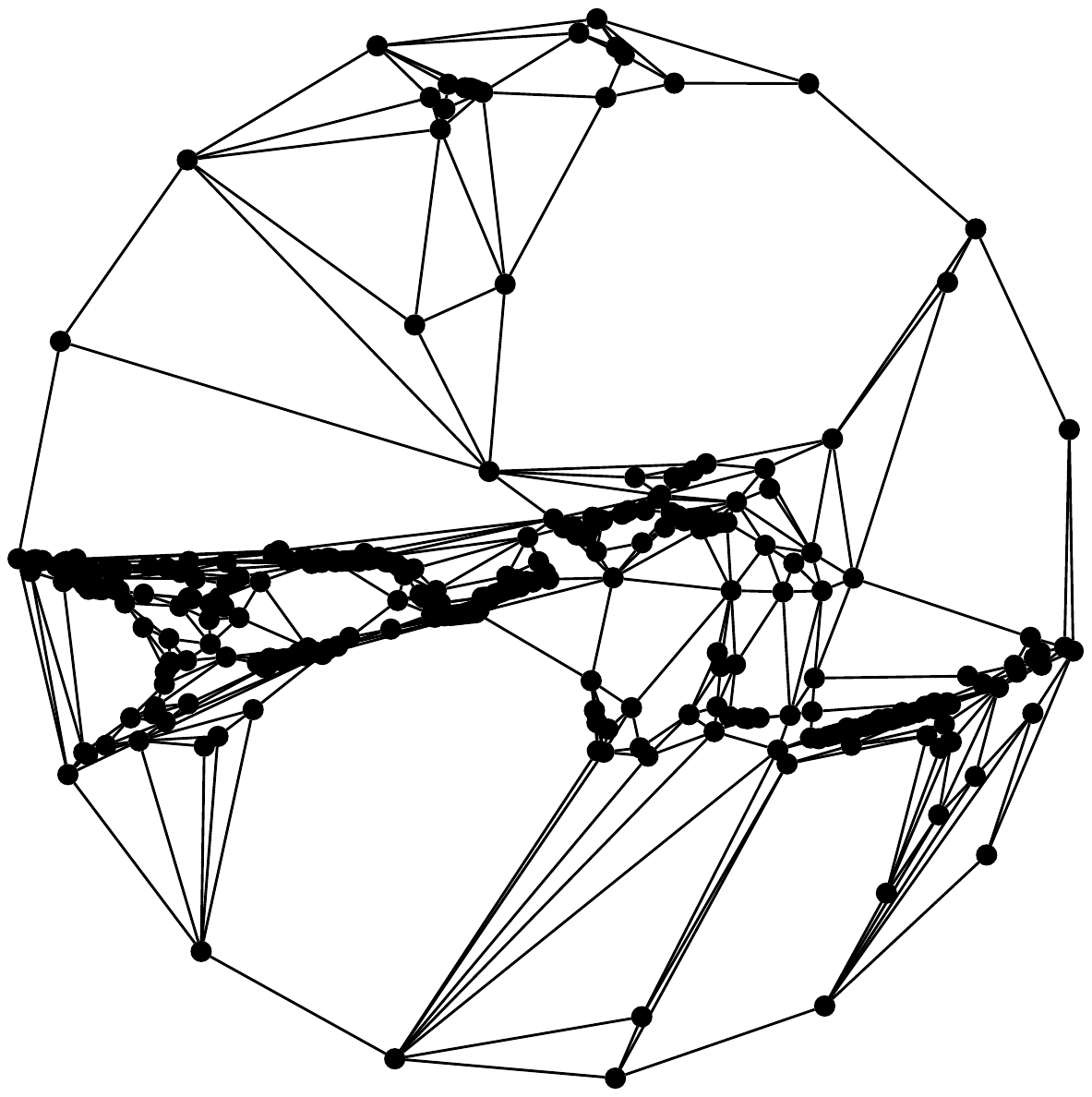} &  
            \includegraphics[width=.23\linewidth]{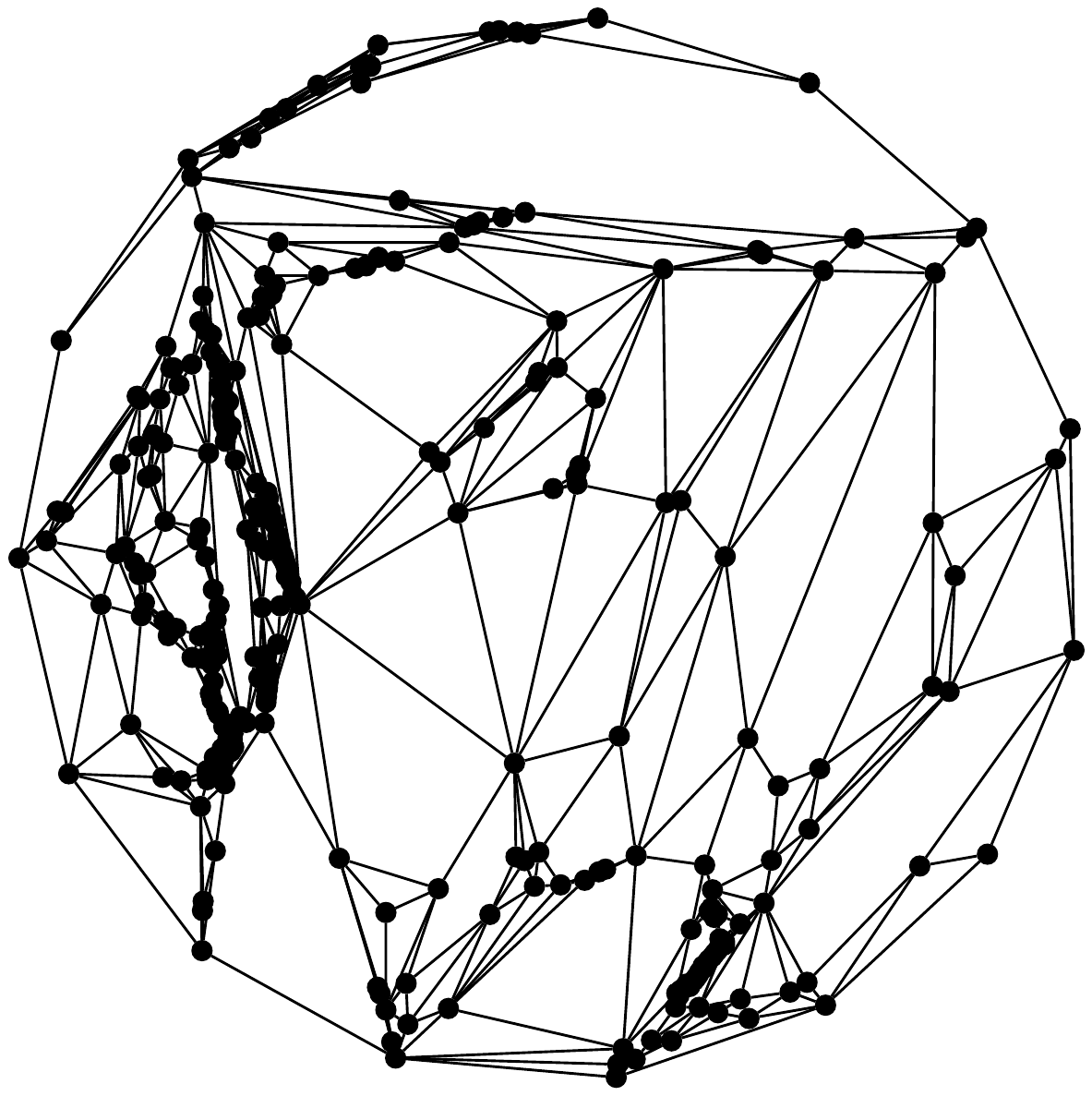} & 
            \includegraphics[width=.23\linewidth]{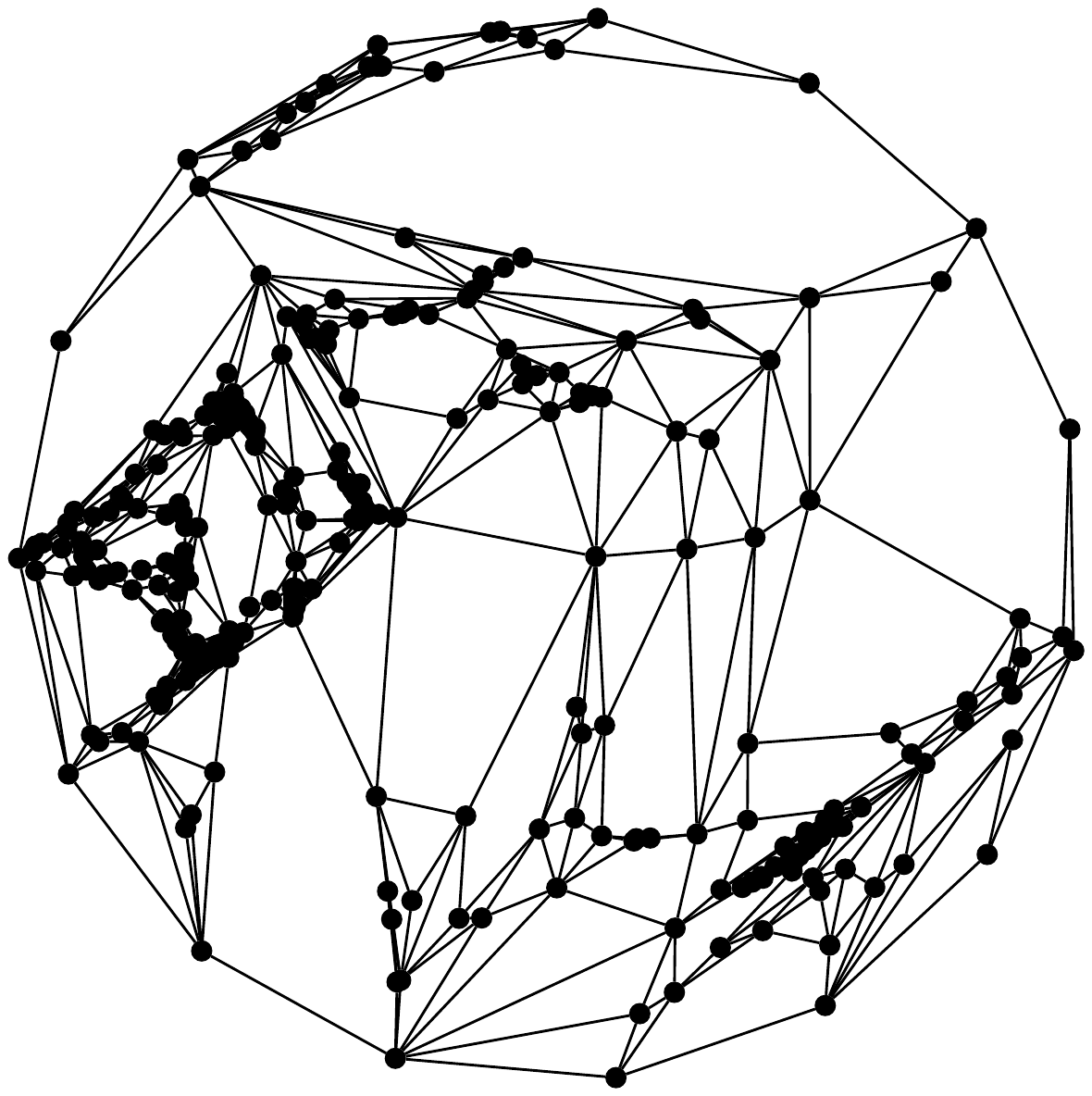} \\

      \end{tabular}
\end{center}
\end{table}

\section{Completely uniform vertex spacing}

The method we implemented for our experiments to spread $x$-coordinates uniformly fixes the outer face of the drawing to be a regular polygon. However, the coordinates of this polygon may not be exactly aligned with a system of $n$ completely uniformly spaced points along the $x$-axis. Additionally, after its initial unweighted Tutte drawing, our method cannot reorder the $x$-coordinates of the points, so when many interior points have $x$-coordinates between some two outer polygon vertices, and few interior points have $x$-coordinates between some other two outer polygon vertices, our method cannot ameliorate that imbalance. In this appendix we describe a method for constructing stress-graph embeddings with completely uniform vertex $x$-coordinate spacing, in the order given by an arbitrary $st$-ordering of the graph. Both the weights for this embedding and the placement for the outer face vertices can be found in linear time. In exchange, we lose control of the $y$-coordinates, even for the vertices of the outer face, making this method unsuitable for combining with other weights in an $xy$-morph.

\begin{theorem}
Let $G$ be an arbitrary three-connected planar graph, with one of its faces chosen to be the outer face, and with its vertices ordered in an arbitrary $st$-ordering respecting that choice of outer face. Then there exists a convex placement for the outer face vertices, and a system of positive weights for the interior edges, for which the stress-graph embedding for this outer face placement and system of weights gives each vertex an $x$-coordinate equal to the index of its position in the $st$-ordering. The outer face placement and system of weights can be constructed from $G$ in linear time.
\end{theorem}

\begin{proof}
The only difficulty is finding a convex placement for the outer face, respecting the fixed $x$-coordinates given by the $st$-ordering. After this is done, the weights for a stress-graph embedding can be chosen in exactly the same way as we did for the $x$-uniform drawing with a regular polygon as its outer face. If the outer face is a triangle, it is easy to choose $y$-coordinates for its vertices, giving it a convex embedding that respects the fixed $x$-coordinates, so from now on we assume that the outer face has more than two sides.

We first choose a convex sequence of slopes for the outer face edges, leaving two edges (the top and bottom) horizontal. The choice of which is to be the top and bottom edge is arbitrary, except that both edges must separate the leftmost and rightmost vertex in the outer face ordering, and they cannot both be incident to the leftmost or rightmost vertex. In this way, both the leftmost and rightmost vertex are part of a chain of one or more non-horizontal edges.

Next, we place the leftmost vertex of the outer face with its given $x$-coordinate and an arbitrary $y$-coordinate and place neighboring vertices along non-horizontal edges so that they have the given $x$-coordinate and the chosen slope with respect to their neighbors. In this way, the placement of the entire left chain can be determined. Symmetrically, we can place the entire right chain. However, this may not leave the top and bottom edges horizontal.

Finally, we apply a linear transformation to the $y$-coordinates of the right chain so that it extends over the same range of $y$ coordinates as the left chain. This will in general change the slopes of the edges in the right chain, but preserve their convexity, as well as preserving the $x$-coordinates of the points. After this transformation, the left and right chains can be connected to each other by horizontal edges, completing the placement of the outer face of $G$ in a convex polygon that respects the $x$-coordinates coming from the $st$-ordering of $G$.
\end{proof}
\end{appendix}
\end{document}